\documentclass[authoryear,preprint,review,3p,12pt]{elsarticle}

\usepackage{amssymb}
\usepackage{amsmath}
\usepackage{subcaption}
\usepackage{float}\usepackage{epsfig}
\usepackage{hyperref}
\usepackage{float}
\usepackage{bm}
\usepackage{multirow}
\usepackage{makecell}
\usepackage{ragged2e}
\usepackage[authoryear]{natbib}  

\usepackage{amsfonts}%
\usepackage{mathrsfs}%
\usepackage[title]{appendix}%
\usepackage{xcolor}%
\usepackage{textcomp}%
\usepackage{manyfoot}%
\usepackage{booktabs}%
\usepackage{algorithm}%
\usepackage{algorithmicx}%
\usepackage{algpseudocode}%
\usepackage{listings}%
\DeclareCaptionLabelFormat{new}{\textup{(}{\text{#2}}\textup{)}}
\usepackage{overpic}

\begin{document}
\captionsetup[sub]{font=normalsize,textfont=it}

\begin{frontmatter}



\title{Inflation and instabilities of a spherical magnetoelastic balloon} 


\author[inst1]{Shaikh Nadeem Karim}

\affiliation[inst1]{organization={Department of Applied Mechanics and Biomedical Engineering},
            addressline={Indian Institute of Technology Madras}, 
            city={Chennai},
            postcode={600036}, 
            country={India}}

\author[inst1]{Ganesh Tamadapu}

\begin{abstract}

This study explores the instabilities during the axisymmetric inflation of an initially spherical magnetoelastic balloon, modeled as a magnetizable Ogden material, under combined internal pressure and a non-uniform magnetic field generated by current-carrying coils. The nonlinear interplay of geometric and material effects leads to governing equations sensitive to bifurcations and instabilities. A coordinate singularity at the poles of the balloon is identified within the system of governing differential equations, which is resolved through an appropriate choice of field variables and L'Hôpital's rule. Stability analysis reveals that as inflation progresses, axisymmetry is broken through a supercritical pitchfork bifurcation, resulting in a pear-shaped equilibrium. This symmetry is later restored through a reverse subcritical pitchfork bifurcation, forming an isolated loop of pear-shaped solutions containing stable and unstable branches in the case of a six-parameter Ogden material model (SPOM). The onset of symmetry-breaking bifurcations is influenced by material parameters and magnetic field intensity, with critical values beyond which such bifurcations are suppressed. Both symmetry-preserving and pear-shaped configurations are stable under small asymmetric perturbations in both magnetic and non-magnetic cases. Snap-through transitions between pear-shaped and axisymmetric configurations are also observed. 
\end{abstract}




\end{frontmatter}


\section{Introduction} 
The inflation mechanics of a polymeric membrane is a classical and intriguing problem, as it involves both material and geometric nonlinearity. Understanding the mechanics of thin-walled structures is important in the design across scales, from nano-scale structures \citep{Yao2011} to large-scale space structures \citep{space_book}. The control of pressure in inflatable structures has emerged as a trending strategy for agile and rapid actuation of soft robots \citep{Combution_robot}. Inflatable polymer structures are also being utilized in the development of advanced surgical tools for minimally invasive procedures \citep{surgical_tools} and in bio-electronics applications \citep{Bio-elect-application}. Many biological materials like cell membranes, tissues, arterial walls, etc. can be modeled as an inflatable polymeric membrane \citep{Membranes_in_Bio}.         

The inflation behavior of elastomeric and gel balloons involves both geometric and material nonlinearity, making the governing equations prone to various bifurcations. These mathematical bifurcations manifest as different physical instabilities. The inflation of thin-walled polymeric structures has been extensively studied, with a focus on explaining the instabilities that arise during the process. Among the most notable is the limit point instability, characterized by a non-monotonic relationship between pressure and stretch \citep{GT_PhysRevE}.

Other prominent phenomena include wrinkling, where localized buckling occurs in the membrane due to compressive stresses, and symmetry breaking or shape bifurcations, where an inflated shape transitions into another, such as a spherical balloon becoming pear-shaped. The wrinkling of pressurized cylindrical and hemispherical balloons, using the relaxed strain energy approach, is analyzed in \cite{AMITPATIL_JMPS}. The energy-based stability criteria used to study the bifurcation from a spherical to a pear-shaped balloon is discussed in \cite{FU201433}. Additionally, the symmetry-breaking bifurcation in toroidal membranes is explored in \cite{ROYCHOWDHURY2018328}. Numerical methods such as finite element analysis have also been employed to investigate the inflation mechanics of ellipsoidal balloons and their post-bifurcation equilibrium curves in  \cite{Wang_IJNM2018}.

It is intriguing to explore the effects of external fields, such as magnetic, electric, and chemical potential, on the instabilities in the inflation mechanics of thin-walled structures made of active materials like dielectrics, magneto-elastics, and gels. The influence of a direct current (DC) electric field on the inflation and shape bifurcation instabilities of spherical and cylindrical inflatable structures has been investigated in \cite{Cai_JAM2015} and \cite{GHOSH_JMPS2021}, respectively.

The inflation mechanics of a toroidal magneto-elastic membrane, subject to a non-uniform magnetic field generated by a current-carrying coil placed at the center of the undeformed membrane, is examined in \cite{REDDY_IJNM2017}. Additionally, the effect of a magnetic field produced by a dipole, located along the axis of symmetry, on the inflation and wrinkling of a circular magneto-elastic plate is studied in \cite{Barham_ActaM2007} and \cite{SAXENA_IJNM2019}.

Recently, an interesting phenomenon known as the delayed burst caused by the coupling between diffusion and deformation has been documented by  \cite{TengLi_JMPS2019} in spherical gel balloons made from neo-Hookean material. A similar delayed burst was also observed in torus-shaped gel balloons, as reported in \cite{TAMADAPU_IJNM2022}, where the study used the Gent and neo-Hookean material models and introduced a novel phenomenon termed the delayed short burst specifically for Gent materials.

Existing literature on thin-walled inflatable structures made of active materials primarily focuses on configurations such as a cylindrical membrane with a magnetic field generated by a current-carrying coil placed along the cylinder's axis \citep{REDDY_IJSS2018}, a magneto-elastic toroidal membrane with a current-carrying coil at its center, and a circular plate under the influence of a magnetic field created by a dipole \citep{Barham_ActaM2007, SAXENA_IJNM2019}. However, the problem of axisymmetric inflation mechanics for initially spherical magneto-elastic balloons under the influence of a magnetic field generated by the current-carrying coil(s) appears to be unexplored in the literature.

Existing stability analyses primarily focus on the stability of spherical configurations in initially spherical polymeric balloons and the emergence of pear-shaped configurations \citep{CHEN_IJNM1991, FU201433}. The existence of isola bifurcations (closed loops of non-spherical solutions) was demonstrated in \cite{CHEN_IJNM1991}, while a theoretical framework describing isolas, where the loop shrinks to a point termed the ``isola center'' as a critical parameter is approached, was presented in \cite{birth_of_isola}. Isola curves have been identified experimentally and numerically in various contexts, including the buckling of spherical elastic shells \cite{shperical_shell_buckling1}, axisymmetric buckling of hollow hemispheres \cite{hemisphere_buckling2}, and the buckling of elastic arches \cite{arch_buckling3}. Symmetry-breaking bifurcations in toroidal membranes \citep{ROYCHOWDHURY2018328} and instabilities in fluid-loaded cylindrical membranes \citep{AMITPATIL_Royal_soc} have also been studied. However, the effects of magnetic fields on bifurcations and instabilities in initially spherical magnetoelastic balloons remain largely unexplored.

The literature thus seems incomplete in two significant areas: First, the axisymmetric inflation mechanics of initially spherical magneto-elastic balloons, and second, the influence of magnetic fields on the bifurcations and instabilities of these spherical magneto-elastic balloons. In this paper, we address and explore these two important questions. 

This study investigates the axisymmetric inflation mechanics of an initially spherical magnetoelastic balloon subjected to a non--uniform magnetic field generated by current-carrying coil(s). A variational approach is used to derive a system of non--linear coupled ordinary differential equations that account for the Ogden strain energy density, magnetic energy for weakly magnetizable materials, and pressure work. Stability is analyzed by introducing axisymmetric and asymmetric perturbations that satisfy the geometric boundary conditions of the inflation problem, focusing on the onset of symmetry-breaking bifurcations and post--bifurcation deformations. As the balloon inflates, the symmetric solution branch undergoes spontaneous symmetry breaking at a critical pressure via a supercritical pitchfork bifurcation, resulting in shapes that retain axial symmetry but lack symmetry in the $Y^1-Y^2$ plane. Remarkably, the axisymmetry is restored at a lower pressure through a reverse subcritical pitchfork bifurcation, forming an isolated loop of pear-shaped solutions.

\section{Kinematics of axisymmetric deformation} 
\label{sec: Kinematics of deformation} 
\begin{figure}[h] 
    \centering
    \includegraphics[width=\linewidth]{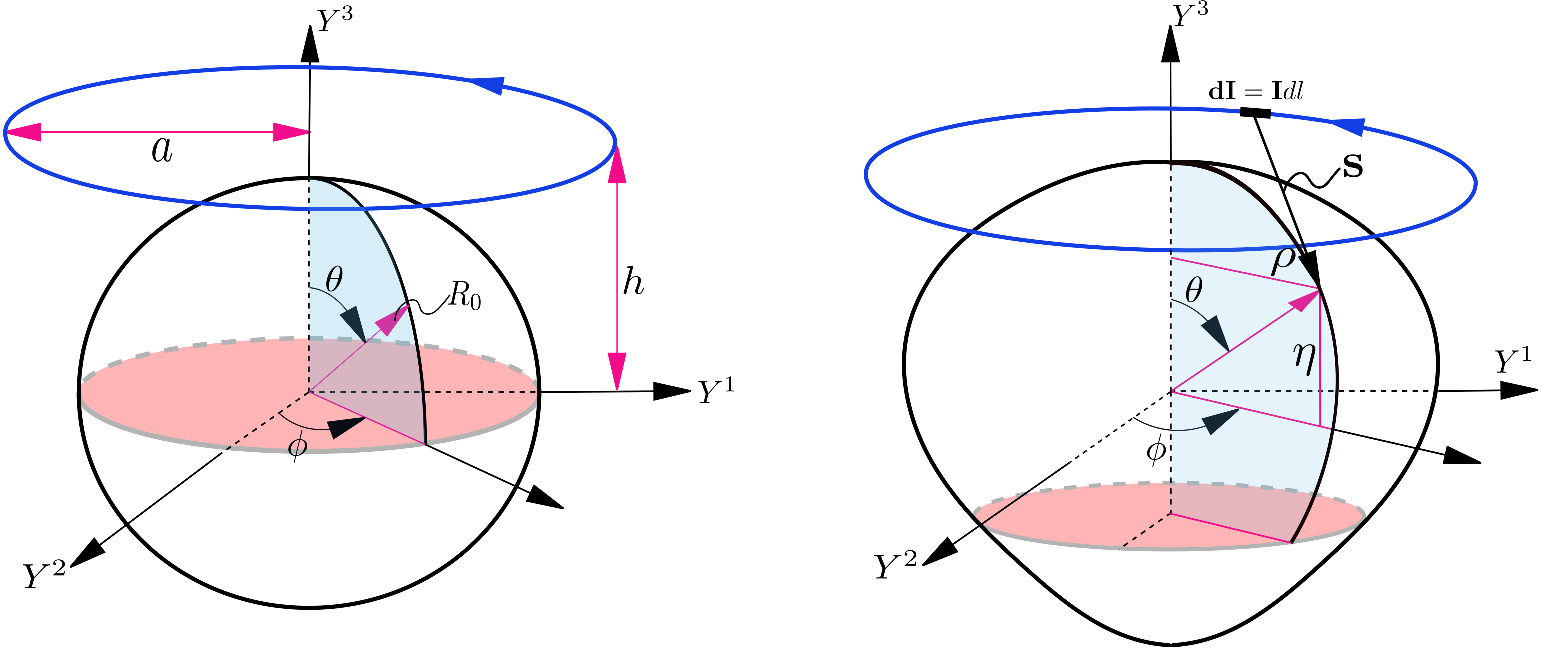}
    \caption{Undeformed and deformed configuration of initially spherical balloon undergoing axisymmetric deformation under the influence of  magnetic field produced by a current carrying coil.} 
    \label{fig:enter-label}
\end{figure}
The finite inflation of a spherical balloon made of nonlinear magneto-elastic, incompressible polymeric material is studied in this work. Let the radius of the balloon is $R_0$ and thickness $T_0$ in the undeformed configuration. The deformed configuration is assumed to be axisymmetric (about the axis $Y^3$ as shown in Fig.~\ref{fig:enter-label}). A spherical coordinate system, $\theta$, $\phi$, and $\zeta$, respectively, in the meridional, circumferential, and normal (perpendicular to the surface in the thickness direction) is used to formulate the finite inflation problem. The position vector $\mathbf{X}$ of a point on the mid-surface ($\zeta=0$) of undeformed configuration and the position vector $\mathbf{x}$ of the same point in deformed configuration are expressed as   
\begin{align}
\begin{split}
  \mathbf{X}(\theta,\phi)&= R_0 \sin{\theta} \cos{\phi} \, \mathbf{E}_1 + R_0 \sin{\theta} \sin{\phi} \, \mathbf{E}_2 + R_0 \cos{\theta} \, \mathbf{E}_3, \\ 
  \mathbf{x}(\theta,\phi)&= \bar{\rho} \cos{\phi} \, \mathbf{e}_1 + \bar{\rho} \sin{\phi} \, \mathbf{e}_2 + \bar{\eta} \, \mathbf{e}_3,   
  \label{Position vec} 
\end{split}   
\end{align}
where $\mathbf{E}_i$ and $\mathbf{e}_i$ are the orthonormal bases of Euclidean coordinate systems $Y^i$ and $y^i$, respectively. The deformation mapping $\mathcal{X}$ is given as, 
\begin{equation}
    \mathbf{x}(\theta,\phi) = \mathcal{X} \left( \mathbf{X}(\theta,\phi) \right). 
\end{equation}

The bases vectors $\mathbf{G}_i$ and $\mathbf{g}_i$ for the undeformed and deformed mid-surface respectively, are given as,

\begin{equation}
    \mathbf{G}_i = \frac{\partial \mathbf{X}}{\partial X^i}, \quad \mathbf{g}_i = \frac{\partial \mathbf{x}}{\partial X^i}  \quad \text{where} \, \left(X^1,X^2\right) = \left(\theta , \phi \right).  
    \label{tangentC vec} 
\end{equation} 

The surface normal vector $\mathbf{N}$ in the undeformed configuration at point $\mathbf{x}$ and the surface normal vector $\mathbf{n}$ in the deformed configuration at point $\mathbf{x}$ are given, respectively, as,  

\begin{equation}
\mathbf{N} = \frac{ \mathbf{G}_1 \times \mathbf{G}_2}{|\mathbf{G}_1 \times \mathbf{G}_2 |} \quad \text{and} \quad \mathbf{n} = \frac{ \mathbf{g}_1 \times \mathbf{g}_2}{|\mathbf{g}_1 \times \mathbf{g}_2 |}. 
\label{current normal vec}
\end{equation}

The vector triads ($\mathbf{G}_1$, $\mathbf{G}_2$, $\mathbf{N}$) and ($\mathbf{g}_1$, $\mathbf{g}_2$, $\mathbf{n}$) form the bases in the curvilinear coordinate system ($\theta$, $\phi$, $\zeta$) for the undeformed and deformed configurations, respectively. The quantities $G_{ij} = \mathbf{G}_i \cdot \mathbf{G}_j$ and $g_{ij} = \mathbf{g}_i \cdot \mathbf{g}_j$ for $i,j=\{1,2,3\}$ represent the covariant metric tensor components for the undeformed and deformed configurations, respectively. The metric tensors $\mathbf{G}$(undeformed) and $\mathbf{g}$(deformed) are expressed as,  

\begin{equation}
\mathbf{G} = R_0^2\,\mathbf{G}_1 \otimes \mathbf{G}_1 +  R_0^2 \sin^2\theta \,\mathbf{G}_2 \otimes \mathbf{G}_2,  
\label{Gmetric} 
\end{equation} 
\begin{equation}
\mathbf{g} = \left(\bar{\rho}_{,\theta}^2 + \bar{\eta}_{,\theta}^2 \right)\, \mathbf{g}_1 \otimes \mathbf{g}_1 +  \bar{\rho}^2\,\mathbf{g}_2 \otimes \mathbf{g}_2. 
\label{gmetric} 
\end{equation} 
Here, $\lambda_3$ is the principal stretch ratio in $\zeta$ direction. The components of the inverse of the metric tensors $\mathbf{G}$ and $\mathbf{g}$ are given as  $G^{-1}_{\alpha \beta} = G^{\alpha \beta}$, and $g^{-1}_{\alpha \beta}  = g^{\alpha \beta}$, respectively.  The deformation gradient tensor $\bf{F}$, which is a linear mapping between the deformed and the undeformed configurations, is given as  $\bf{F} = g_{\beta} \otimes G^{\beta}$. The left Cauchy-Green tensor $\mathbf{C} = \mathbf{F}^T\mathbf{F} = g_{\alpha \beta} \mathbf{G^\alpha} \otimes \mathbf{G^\beta} =g_{\alpha \beta} G^{\alpha \gamma} \mathbf{G}_{\gamma} \otimes \mathbf{G}^\beta $ can be written as 
\begin{align}
\mathbf{C} =\frac{\bar{\rho}_{,\theta}^2 + \bar{\eta}_{,\theta}^2 }{R_0^2}\, \mathbf{G}_1\otimes\mathbf{G}^1 + \frac{\bar{\rho}^2}{R_0^2 \sin^2\theta}\, \mathbf{G}_2\otimes\mathbf{G}^2. 
\end{align}
In-plane principal stretches in $\theta$ and $\phi$ directions are given as, 
\begin{equation}
    \lambda_\theta = \frac{ \sqrt{\bar{\rho}_{,\theta}^2 + \bar{\eta}_{,\theta}^2} }{R_0}, \quad \lambda_\phi = \frac{\bar{\rho}}{R_0 \sin\theta}.    
\end{equation} 
The out-of-plane stretch normal to the mid-surface is $\lambda_3=\text{det}\left( \mathbf{G}\right)/\text{det}\left( \mathbf{g}\right)$.
\section{Total potential energy} 
The total energy potential for a weakly magnetizable nonlinear magneto-elastic membrane inflation \citep{Barham_2008}, can be written as follows,  
\begin{equation}
\label{Total Energy}
E_{\text{total}} = T_0\int_A \Phi_{\text{e}}\left( \mathbf{C} \right)\,{\rm d} A  -  T_0\int_A \Phi_{\text{m}}\left( \mathbf{C}, \mathbf{H}\right)\,{\rm d} A -
\int_v W_p\,{\rm d}v, 
\end{equation}
where $T_0$ is the undeformed spherical balloon thickness, $A$ is the mid-surface area of the undeformed balloon, and $v$ is the enclosed volume of the deformed configuration. Here, $\Phi_{\text{e}}$ in (\ref{Total Energy}) is the elastic strain energy per unit reference volume, $\Phi_{\text{m}}$ is the magnetisation energy per unit reference volume, and 
$W_p$ is related to the pressure energy. 

In this work, the material is assumed to follow incompressible Ogden  material model for the elastic part of the potential, which is given as, 

 
\begin{equation}
    \Phi_{\text{e}} = \sum_{i=1}^{3} \frac{\bar{\mu_i}}{\alpha_i} \left( \lambda_1^{\alpha_i} + \lambda_2^{\alpha_i}  + \lambda_3^{\alpha_i} - 3 \right), \label{energy Ogden}
\end{equation}
where $\bar{\mu}_i$ are the material constants and $\alpha_i$ are the shape parameters related to shear modulus ($\mu$) as $2\mu=\sum_{i=1}^3\bar{\mu}_i\alpha_i$.

The magnetic energy density used is given as, 
\begin{equation}
    \Phi_\text{m}\left( \mathbf{C} , \mathbf{H} \right) =  \frac{1}{2} \mu_0 \chi \mathbf{h}_a\cdot\mathbf{h}_a,    
    \label{mag energy} 
\end{equation}
where $\mu_0$ is the permeability of vacuum, $\mathbf{h}_a$ is the applied magnetic field, and $\chi$ is magnetic susceptibility per unit deformed volume. In this formulation, the self-generated magnetic field is assumed to be negligible compared to the applied magnetic field \citep{Barham_2008}. The applied magnetic field $\mathbf{h}_a$ is a nonuniform magnetic field generated by a DC current-carrying circular coil of radius $\bar{a}$ placed at height $h$ from equatorial plane $Y^1 - Y^2$ such that the coil axis coincides with the axis of symmetry of deformation i.e. axis $Y^3$. The applied field $\mathbf{h}_a$ is derived using Biot-Saverts's law as follows,  
\begin{equation}
    {\rm d}\mathbf{h}_a  = \frac{I}{4\pi} \frac{{\rm d}\mathbf{l} \times \mathbf{s} }{|\mathbf{s}|^3}, 
    \label{Biot Savert} 
\end{equation}
where $I$ is current in the coil, ${\rm d}\mathbf{l}$ is a differential length element on the wire at an angle $\phi_i$ from $Y^1$ axis, $\mathbf{s}$ is the position vector of a point on the mid-surface of the balloon from ${\rm d} \mathbf{l}$ as shown in Fig.~\ref{fig:enter-label}. Vectors $\mathbf{s}$ and ${\rm d} \mathbf{l}$ are obtained as follows, 
\begin{equation}
    {\rm d}\mathbf{l} = -{\rm d}l\sin\phi_i\,\mathbf{e}_1 + {\rm d}l\cos\phi_i\,\mathbf{e}_2,  \quad  {\rm d}l = \bar{a}\,{\rm d}\phi_i,   
    \label{vec dl}
\end{equation}
\begin{equation}
    \mathbf{s} = \left( \bar{\rho} \cos \phi - \bar{a}\cos\phi_i \right) \, \mathbf{e}_1 + \left( \bar{\rho} \sin \phi - \bar{a}\sin\phi_i \right) \, \mathbf{e}_2 + \left( \bar{\eta} - \bar{h} \right)\, \mathbf{e}_3, 
    \label{vec s}
\end{equation} 
\begin{equation}
    {\rm d}\mathbf{l}\times \mathbf{s} = {\rm d}\phi_i \left[ \left(\bar{\eta} -\bar{h}\right)  \bar{a} \cos \phi_i \, \mathbf{e}_1 + \left(\bar{\eta} -\bar{h}\right)\bar{a} \sin \phi_i \, \mathbf{e}_2 + \left( \bar{a}^2 - \bar{\rho}\bar{a}\cos\left( \phi_i - \phi \right)  \right) \, \mathbf{e}_3  \right].  
    \label{di cross ds} 
\end{equation}
From (\ref{Biot Savert}), (\ref{vec s}) and (\ref{di cross ds}), the  applied magnetic field $\mathbf{h}_a$ can be written as, 
\begin{align}
    \mathbf{h}_a = \frac{I}{4\pi} 
     \int_{0}^{2\pi} \frac{ \left(\left(\bar{\eta} -{\bar h} \right) \bar{a} \cos \phi_i \right) \, \mathbf{e}_1 + ( \left(\bar{\eta} - {\bar h}\right) \bar{a} \sin \phi_i) \, \mathbf{e}_2 + \left( \bar{a}^2 - \bar{\rho}\bar{a}\cos\left( \phi_i - \phi \right)  \right) \, \mathbf{e}_3}{ \left( \bar{\rho}^2 + \left(\bar{\eta} -\bar{h}\right)^2 + \bar{a}^2 - 2\bar{\rho}\bar{a}\cos\left( \phi_i - \phi \right) \right)^{3/2} } \, {\rm d} \phi_i. 
    \label{ha calc} 
\end{align}  
Introducing the non-dimensional quantities  $\left( {\rho} , {\eta} , {a} , h \right) = \left( \bar{\rho} , \bar{\eta} , \bar{a} , \bar{h} \right)/R_0$, the applied magnetic field $\mathbf{h}_a$ can be rewritten as, 
\begin{align}
    \mathbf{h}_a = \frac{I}{4\pi R_0} 
     \int_{0}^{2\pi} \frac{ ( \left({\eta} - h \right) {a} \cos \phi_i ) \, \mathbf{e}_1 + ( \left({\eta} -h\right) {a} \sin \phi_i) \, \mathbf{e}_2 + \left( {a}^2 - {\rho}{a}\cos\left( \phi_i - \phi \right)  \right) \, \mathbf{e}_3}{ \left( {\rho}^2 + \left({\eta} - h \right)^2 + {a}^2 - 2{\rho}{a}\cos\left( \phi_i - \phi \right) \right)^{3/2} } \,  {\rm d} \phi_i.
    \label{ha calcnd} 
\end{align}  
The magnetic energy $\Phi_{\text{m}}$ can be written as, 
\begin{align}
    \Phi_{\text{m}} = \frac{1}{2} \mu_0 \chi \mu \mathcal{M} \mathcal{H}^2, 
\end{align}  
where
$
    \mathcal{M} ={I^2}/{16\pi^2 R_0^2 \mu }
    \label{M nond}
$
and 
\begin{equation}
    \mathcal{H} = \left| \int_{0}^{2\pi} \frac{ ( \left({\eta} - h \right) {a} \cos \phi_i ) \, \mathbf{e}_1 + ( \left({\eta} -h\right) {a} \sin \phi_i) \, \mathbf{e}_2 + \left( {a}^2 - {\rho}{a}\cos\left( \phi_i - \phi \right)  \right) \, \mathbf{e}_3}{ \left( {\rho}^2 + \left({\eta} - h \right)^2 + {a}^2 - 2{\rho}{a}\cos\left( \phi_i - \phi \right) \right)^{3/2} } \, {\rm d} \phi_i \right| 
    \label{H cali} 
\end{equation}

The third term in \eqref{Total Energy} represents the internal energy due to the presence of gas at gauge pressure $\bar{P}$, where the integral is performed over the deformed/inflated volume, and expressed as  

\begin{equation} 
    \int_v W_p \, {\rm d}v = \int^{2\pi}_0 \int^{\pi}_0 \left( \frac{1}{3} P\,\mathbf{x}\cdot \mathbf{n} \sqrt{g} \right) \,  {\rm d}\theta\,{\rm d}\phi = - \int^{2\pi}_0 \int^{\pi}_0 \left( \frac{1}{2} P \rho^2 \eta_{,\theta}  \right) \, {\rm d}\theta\,{\rm d}\phi  
\label{P_energy1}
\end{equation} 
where $\mathbf{x}$ is the position vector given in \eqref{Position vec} and $g=\text{det}\left( \mathbf{g}\right)$ is the determinant of the deformed metric tensor. 

\section{Governing equations} 
The total potential energy $\bar{E}_T$ for axisymmetric inflation of a magneto-elastic polymeric balloon can be written as, 
\begin{equation}
    \bar{E}_{T} = \int_0^{2\pi} \int_0^{\pi} \left[ \left( \Phi_{\text{e}} - \Phi_{\text{m}}\right)T_0 \sqrt{G}  + \frac{1}{2} \bar{P}\bar{\rho}^2\bar{\eta}_{,\theta}\right]   {\rm d}\theta \,{\rm d}\phi, 
    \label{E_T}
\end{equation}
where $G =\text{det}(\mathbf{G})$ is the determinant of the undeformed metric tensor. Substituting  $\Phi_{\text{m}}$ from  (\ref{mag energy}) in \ref{E_T},   

\begin{equation}
\label{var energy1} 
    \bar{E}_T = \int_0^{2\pi} \int_0^{\pi} \left[ \left( {\Phi}_{\text{e}} -\frac{1}{2} \mu_0 \chi \mu \mathcal{M} \mathcal{H}^2 \right)T_0R_0^2 \sin \theta  + \frac{1}{2}\bar{P}{\bar{\rho}}^2{\bar{\eta}}_{,\theta}\right]\,{\rm d}\theta \,{\rm d}\phi,  
\end{equation}
 $\bar{E}_T$ is a functional in terms of primary field variables and independent variable $\theta$ i.e. $\bar{E}_T = \bar{E}_T\left(\bar{\rho}, \bar{\rho}_{,\theta}, \bar{\eta}, \bar{\eta}_{,\theta}, \theta \right)$. A primary field variable $\bar{\rho}$ is replaced with $\lambda_2$ as the new primary field variable by doing the following substitutions, 
\begin{equation}
    \bar{\rho} = \lambda_2 R_0 \sin \theta, \quad \lambda_{2,\theta} \sin\theta = v,  \quad \bar{w} = \bar{\eta}_{,\theta}, \quad \lambda_1 = \frac{ \sqrt{ \left(\lambda_2\cos\theta + v\right)^2R_0^2 + \bar{w}^2 } }{R_0},
    \label{subs1}
\end{equation} 
The total potential energy is non-dimensionalized using $\mu^*=\mu/2$, 
\begin{equation}
 \left({\rho}, {\eta} , w, {a} \right)= \frac{\left(\bar{\rho}, \bar{\eta}, \bar{w}, \bar{a} \right)}{R_0}\quad {\hat\Phi} = \frac{\Phi_{\text{e}}}{\mu^*}\quad {P} = \frac{\bar{P}R_0}{T_0 \mu^*}, \quad E_T = \frac{\bar{E}_T}{\mu^*}, \quad   \mu_i = \frac{\bar{\mu}_i}{\mu^*},  
 \label{nd1} 
\end{equation} 
where $\bar{\mu_i}$ and $\mu_i$ are dimensional and non-dimensionalised material parameters for Ogden material model given in (\ref{energy Ogden}).   

Using equations \eqref{nd1} and \eqref{subs1} total potential energy can be written as, 

\begin{equation} 
\label{Total Energy nd}
    E_T = \int_0^{2\pi} \int_0^{\pi} \left[ \left( \hat{\Phi} -\frac{1}{2} \mu_0 \chi  \mathcal{M} \mathcal{H}^2 \right)\sin\theta + \frac{1}{2}{P}\lambda_2^2{{\eta}}_{,\theta}\sin^2\theta\right]   {\rm d}\theta \,{\rm d}\phi,
\end{equation}  

\begin{equation*} 
    E_T = \int_0^{2\pi} \int_0^{\pi} \hat{\Pi} \,   {\rm d}\theta \,{\rm d}\phi,
\end{equation*}
Total potential energy functional now becomes $E_T = E_T\left( \lambda_2 , v, \eta, w , \theta \right)$. Using the principle of minimum potential energy, equations of equilibrium are obtained as 
\begin{align}
    \label{equilibrium} 
    \frac{\partial \hat{\Pi}}{\partial \lambda_2} - \frac{\rm d}{\rm d \theta} \left( \frac{\partial \hat{\Pi}}{\partial v} \sin\theta \right) =0, \qquad  \frac{\partial \hat{\Pi}}{\partial \eta} -  \frac{\rm d }{\rm d \theta} \frac{\partial \hat{\Pi}}{\partial w} = 0. 
\end{align}
The boundary conditions for axisymmetric inflation of the balloon are given as
\begin{align} \label{BC1}
    \rho\left( 0 \right) = 0, \quad \rho\left( \pi \right) = 0, \quad \eta_{,\theta} \left( 0 \right) = 0, \quad \eta_{,\theta} \left( \pi \right) = 0.
\end{align} 
Substituting \eqref{subs1} in \eqref{equilibrium} converts a set of two second-order ordinary differential equations \eqref{equilibrium} to a set of four first-order ordinary differential equations given as 
\begin{align}
   & \lambda_{2,\theta} =  \frac{v}{\sin\theta},\label{gdef1}\\   
   & v_{,\theta} = F_1\left( v, \eta, w, \lambda_2 , \theta; P \right),\label{gdef2}\\
    & w_{,\theta} = F_2\left( v,\eta,w , \lambda_2 ,  \theta; P \right), \label{gdef3}\\
    & \eta_{,\theta} = w \label{gdef4} 
\end{align} 
with the boundary conditions in terms of new field variables as,  
\begin{equation}\label{bc}
 v(0)=0,\quad w(0)=0,\quad v(\pi)=0,\quad w(\pi)=0.
\end{equation} 
Due to the coordinate singularity at the poles ($\theta=0$ and $\theta=\pi$), the right hand side of the equations \eqref{gdef1} -\eqref{gdef4} attain an indeterminate $0/0$ form and are thus valid only for $\theta\neq 0 $ and $\theta \neq \pi$. Applying L'Hospital's rule to indeterminate form, set equations valid for the poles are given as, $\lambda_{2,\theta} = 0 $, $v_{,\theta} = 0$, $w_{,\theta} = \left( -\mu_o \chi  \mathcal{M} \mathcal{H} \mathcal{H}_{,\eta} -P\lambda_2^2\cos\theta  \right)/\left(2\Phi_{,ww} \right) $, $\eta_{,\theta}=0$. (see \ref{sec: l2 equation} for details)        

\section{Solution procedure} 
Following a similar procedure used in \cite{roychowdhury2015inflating,roychowdhury2015response,TAMADAPU_IJNM2022,TAMADAPU_Euro_A} and \cite{TAMADAPU_IJSS_rim}, the two-point boundary value problem (TPBVP) given by \eqref{gdef1}-\eqref{bc} for the axisymmetric inflation of a magneto-elastic balloon is solved as follows. For the plane of symmetric solutions one would expect the stretches at the poles should be same. First, $\lambda_2$ value at $\theta=0$ is chosen for a fixed value of $\mathcal{M}$ with pressure $P$ and $\eta(0)(=\eta(\pi))$ as unknown quantities to be determined using shooting method. The ode set \eqref{gdef1} is integrated from $\theta=0$ to $\pi/2$ and $\theta=\pi$ to $\pi/2$ with $P$ and $\eta(0)=\eta(\pi)$ as guess values. The correct values of pressure and $\eta$ at the poles are obtained by shooting method with a cost function from the matching conditions at $\theta=\pi/2$ as given below: 

\begin{align*}
     &S\left(P,\eta(0),\lambda_2(\pi),\eta(\pi);\lambda_2(0)\right) =\left[\{\lambda_2^f\left(\pi/2\right)-\lambda_2^b\left(\pi/2\right)\}^2 + \{v^f\left(\pi/2\right)-v^b\left(\pi/2\right)\}^2\right.\nonumber\\
     &\left.\qquad\qquad\qquad\qquad+\{w^f\left(\pi/2\right)-w^b\left(\pi/2\right)\}^2+\{\eta^f\left(\pi/2\right)-\eta^b\left(\pi/2\right)\}^2 \right]^{1/2}.\
 \end{align*}
 Here, superscripts $f$ and $b$ represent the forward ($\theta=0$ to $\pi/2$) and backward ($\theta=\pi$ to $\pi/2$) integration solutions, respectively.
 The shooting method involves searching for local minima of the cost function that satisfy $|S|<10^{-8}$ for input values of $\lambda_2(0)$ and $\mathcal{M}$, which yields an exact solution for the 4-tuple parameter $\left[P,\eta(0),\lambda_2(\pi), \eta(\pi) \right]$. The shooting method is implemented through the Nelder–Mead search technique (\cite{nelder_simplex}) using the MATLAB fminsearch function.

 Now the stretch value $\lambda_2\left( 0\right)$ from the correct 4-tuple $\left[ P,\eta(0),\lambda_2(\pi), \eta(\pi) \right]$ is given a small increment $\delta$. For this new value of stretch $\lambda_2\left( 0\right) + \delta$ at the pole, shooting method is implemented again to get the next correct 4-tuple.   

\section{Cauchy stresses} 
Following the work of \cite{Barham_2008}, the Cauchy stresses for the incompressible magneto-elastic material are given as follows: 
\begin{align}
    {\bf \sigma} = \frac{\partial \Phi_{\text{e}} }{\partial \mathbf{F}}\mathbf{F}^T + \frac{\partial}{\partial \mathbf{F}}\left( \frac{1}{2} \chi \mu_0 |\mathbf{h}_a|^2 \right)\mathbf{F}^T  + \mu_0 \left( \mathbf{h}_a \otimes \mathbf{h}_a - \frac{1}{2} |\mathbf{h}_a|^2 \mathbf{I} \right) + \mu_0\mathbf{h}_a\otimes\mathbf{m} -q\mathbf{I},   
\end{align}
where, $(.)_\mathbf{F} = \frac{\partial(.)}{\partial\mathbf{F}}$ denotes the partial derivative with respect to the deformation gradient $\mathbf{F}$, and $q$ is the constraint pressure, serving as a Lagrange multiplier for the constrained minimization of the total energy potential $E_T$.
\begin{align}
    \left(\frac{\partial \Phi_{\text{e}} }{\partial \mathbf{F}}\mathbf{F}^T\right)_\theta = \lambda_1\frac{\partial \Phi_{\text{e}}}{\partial \lambda_1}, \quad  \left(\frac{\partial \Phi_{\text{e}} }{\partial \mathbf{F}}\mathbf{F}^T\right)_\phi = \lambda_2\frac{\partial  \Phi_{\text{e}} }{\partial \lambda_2},  \quad \left(\frac{\partial \Phi_{\text{e}} }{\partial \mathbf{F}}\mathbf{F}^T\right)_\zeta = \lambda_3\frac{\partial  \Phi_{\text{e}}}{\partial \lambda_3}.
\end{align}

\begin{align*}
\begin{split}
   &\left[ \frac{\partial}{\partial \mathbf{F}}\left( \frac{1}{2} \chi \mu_0 |\mathbf{h}_a|^2 \right)\mathbf{F}^T \right]_\theta =  \frac{1}{2} \chi \mu_0 \frac{\partial}{\partial \lambda_1}\left(  |\mathbf{h}_a|^2 \right)\lambda_1 = 0 \\
   & \left[ \frac{\partial}{\partial \mathbf{F}}\left( \frac{1}{2} \chi \mu_0 |\mathbf{h}_a|^2 \right)\mathbf{F}^T \right]_\phi =  \frac{1}{2} \chi \mu \mathcal{M} \frac{\partial}{\partial \lambda_2}\left(  \mathcal{H}^2 \right)\lambda_2 
\end{split}  
\end{align*}
The dimensional stress components are given by 
\begin{align*}
    \mathbf{\bar{\sigma}}_\theta &= \left(\lambda_1\frac{\partial  \Phi_{\text{e}}}{\partial \lambda_1} -  (q\mathbf{I})_\theta \right) + (1+\chi) \mu\mathcal{M} (h_a)_\theta^2  - \frac{1}{2} \mu \mathcal{M} \mathcal{H}^2\\  
    \mathbf{\bar{\sigma}}_\phi &= \left(\lambda_2\frac{\partial \Phi_{\text{e}}}{\partial \lambda_2} -  (q\mathbf{I})_\phi \right) +   \frac{1}{2} \lambda_2 \chi \mu \mathcal{M} \frac{\partial}{\partial \lambda_2}\left(  \mathcal{H}^2 \right) - \frac{1}{2}\mu\mathcal{M} \mathcal{H}^2\\
    \mathbf{\bar{\sigma}}_\zeta &= \left(\lambda_3\frac{\partial \Phi_{\text{e}}}{\partial \lambda_3} -  (q\mathbf{I})_\zeta \right) +  (1+\chi) \mu\mathcal{M} (h_a)_\zeta^2  - \frac{1}{2}\mu \mathcal{M} \mathcal{H}^2  
\end{align*} 
Constraint pressure $q$ is calculated by setting the mechanical part of the stress in the thickness direction $\mathbf{\sigma}_\zeta$ to zero and is given by $q = \lambda_3 \frac{\partial \Phi_{\text{e}}}{\partial  \lambda_3 }$.
The expressions for the dimension less stress components (with shear modulus $\mu$) are given by  
\begin{align}
    \mathbf{\sigma}_\theta &= \left(\lambda_1\frac{\partial \Phi}{\partial \lambda_1} -  \lambda_3 \frac{\partial \Phi }{\partial \lambda_3 }   \right) + (1+\chi) \mathcal{M} (h_a)_\theta^2  - \frac{1}{2}\mathcal{M} \mathcal{H}^2, \\
    \mathbf{\sigma}_\phi &= \left(\lambda_2\frac{\partial \Phi}{\partial \lambda_2} -  \lambda_3\frac{\partial \Phi }{ \partial \lambda_3 } \right) +   \frac{1}{2} \lambda_2 \chi  \mathcal{M} \frac{\partial}{\partial \lambda_2}\left(  \mathcal{H}^2 \right) - \frac{1}{2}\mathcal{M} \mathcal{H}^2.
\end{align} 

 \section{Stability analysis}   
 \subsection{Kinematics of asymmetric deformation} 
Assuming general asymmetric deformation, the deformed position vector can be written as,
\begin{align}
    \mathbf{x}(\theta,\phi) = \Bar{\rho}(\theta,\phi) \cos\left[\beta(\theta,\phi)\right] \,\mathbf{e}_1 + \Bar{\rho}(\theta,\phi) \sin\left[\beta(\theta,\phi)\right] \,\mathbf{e}_2 + \Bar{\eta}(\theta,\phi) \, \mathbf{e}_3.       
\end{align} 
The covariant metric tensor associated with asymmetrically deformed configuration is given as, 
\begin{align} 
\begin{split}
    \mathbf{g_{\text{asm}}} =&\left( \Bar{\rho}^2 \beta_{,\theta}^2 + \Bar{\eta}_{,\theta}^2 + \Bar{\rho}_{,\theta}^2\right)\,\mathbf{g}_1\otimes \mathbf{g}_1 + \left( \Bar{\rho}^2 \beta_{,\theta}\beta_{,\phi} + \Bar{\eta}_{,\theta}\Bar{\eta}_{,\phi} + \Bar{\rho}_{,\theta}\Bar{\rho}_{,\phi}\right)\,\mathbf{g}_1\otimes \mathbf{g}_2 \\ 
    &+ \left( \Bar{\rho}^2 \beta_{,\theta}\beta_{,\phi} + \Bar{\eta}_{,\theta}\Bar{\eta}_{,\phi} + \Bar{\rho}_{,\theta}\Bar{\rho}_{,\phi}\right)\,\mathbf{g}_2\otimes \mathbf{g}_1 
    +\left( \Bar{\rho}^2 \beta_{,\phi}^2 + \Bar{\eta}_{,\phi}^2 + \Bar{\rho}_{,\phi}^2\right)\,\mathbf{g}_2\otimes \mathbf{g}_2. 
\end{split}    
\end{align}
The principal stretches $\Lambda_1$, $\Lambda_2$ for asymmetric deformation are eigenvalues of $\mathbf{C} =\mathbf{G}^{-1}\mathbf{g}$. The out-of-plane stretch normal to the mid-surface $\Lambda_3$ is given as, 
$\Lambda_3 =\text{det}\left( \mathbf{G}\right)/\text{det}\left( \mathbf{g_{\text{asm}}}\right)$. 
 
 The stability of the axisymmetric solution of the governing equations given by \eqref{gdef1}-\eqref{gdef4} is analyzed by introducing both axisymmetric as well as asymmetric perturbations. The primary field variables $\lambda_2$ and $\eta$ which locate a material point in the deformed configuration are added with perturbation as, 

 \begin{align} 
 \label{perturbed vars} 
     &{\lambda}^*_{2}(\theta,\phi) = \lambda^{(0)}_2(\theta) + \epsilon\,\lambda_{2}^{(1)}(\theta)+ \epsilon\,\lambda_{2}^{(2)}(\theta)\cos\phi+\mathcal{O}(\epsilon^2),\\
     &{\eta}^*(\theta,\phi) = \eta^{(0)}(\theta) + \epsilon\,\eta^{(1)}(\theta) + \epsilon\,\eta^{(2)}(\theta)\cos\phi+\mathcal{O}(\epsilon^2),\\
     &\beta^*(\theta,\phi) = \theta+  \epsilon\,\beta^{(2)}(\theta)\sin\phi+\mathcal{O}(\epsilon^2),
 \end{align}
where $\epsilon$ is a book-keeping parameter for the perturbation, ${\lambda}^*_{2}, \eta^*$ and $\beta^*$ are perturbed primary field variables, $\lambda_2^{(0)}$ and $\eta^{(0)}$ are solutions of the axisymmetric inflation problem, $\lambda_2^{(1)}$, $\eta^{(1)}$, are introduced perturbations to study the plane of symmetry breaking solution, $\lambda_2^{(2)}$, $\eta^{(2)}$ and $\beta^{(2)}$ are introduced perturbations to study asymmetric solution. The membrane is perturbed from the axisymmetric solution to the plane of symmetry breaking or complete asymmetric solution to study the linear stability problem and to find the possible existence of the plane of symmetry breaking or asymmetric solution branch.    

The perturbations introduced for the primary field variables are given in terms of Legendre's polynomial functions are given as,
\begin{equation} 
\label{Legendre vars}
    \lambda_{2}^{(i)}(\theta) =  \sum_{k=1}^N a^{i,k} P_k\left(\cos\theta\right),  \; \eta^{(i)}(\theta) = \sum_{k=1}^N b^{i,k} P_k\left(\cos\theta\right), \; \beta^{2}(\theta)=\sum_{k=1}^N c^{2,k} P^1_k\left(\cos\theta\right),\;i=1,2.
\end{equation}  
Using \eqref{perturbed vars}-\eqref{Legendre vars} perturbed left Cauchy-Green tensor $\Tilde{C}$ and the perturbed total energy potential ${E}^*_T$ are calculated. The perturbed total energy potential ${E}^*_T$ is expressed as,  
\begin{align}
    {E}^*_T = {E}^*_T\left( \lambda^{*}_2 , \eta^{*},\beta^*\right).
\end{align} 
Note that $\phi$ dependent perturbations are absent in the above equations for the plane of symmetry breaking solutions. 
In genereal, the perturbed total energy potential is expanded in terms of $\epsilon$ about the axisymmetric equilibrium solution up to second order in $\epsilon$ as, 
\begin{align}
{E}^*_T\left( \lambda^{*}_2 , \eta^{*},\beta^* \right) = &E^*_{T}\left(\lambda^{(0)}_2 , \eta^{(0)} \right) + \epsilon\,\delta{E}^*_T\left(\lambda_2^{(0)} , \eta^{(0)}   \right) + \epsilon^2 \delta^2 {E}^*_T\left(\lambda_2^{(0)} , \eta^{(0)}   \right) + \mathcal{O}(\epsilon^3)\nonumber\\ = &E^*_{T_0} + \epsilon{E^*_{T_1}} + \epsilon^2{E^*_{T_2}} + \mathcal{O}(\epsilon^3)
\end{align}  
The governing equations \eqref{gdef1}-\eqref{gdef4} for the axisymmetric deformation are derived using the principle of minimum potential, hence the first variation of total energy potential is zero ($E^*_{T_1}=0$). The stability of the perturbed configuration is determined by the positive definiteness of the second variation of total energy potential ($E^*_{T_2}$). The perturbed total energy potential can be written in terms of coefficients of Legendre polynomials (see equation \eqref{Legendre vars}) as, \begin{align}
    {E}^*_T = {E}^*_T\left( z^{i,1},z^{i,2},\hdots\right),\quad z^{i} = \{a^{i,k},b^{i,2},c^{2,k}\}, \quad k=1,2,3,\hdots,N,
\end{align}
where $i=1$ and $i=2$ correspond to plane of symmetry breaking and asymmetric perturbations, respectively.
The positive definiteness of the second variation $E^*_{T_2}$ is given by the Legendre-Hadamard or ellipticity condition \cite{Truesdell},  
\begin{align} 
\label{stability condition}
    \sum_{p=1}^{(i+1)N}\sum_{q=1}^{(i+1)N}\frac{\partial^2 {E^*_T} }{\partial z^{i,p}\,\partial z^{i,q}}\bigg|_{z^i=0} \delta z^{i,p}\,\delta z^{i,q}> 0, \quad i=1\,\text{or}\,2.
\end{align} 
The violation of the Legendre-Hadamard condition for a chosen axisymmetric deformed configuration indicates a possible instability. This implies the second order partial derivative term (usually called as Hessian matrix) in \eqref{stability condition} needs to be positive definite. Therefore, loosley speaking all the eigenvalues of the Hessian matrix need to be positive for the stability of a given equilibrium configuration around which perturbation are carried. 
At least one negative eigenvalue indicates the existence of the symmetry-breaking solutions.

\section{Results and discussions} 

\begin{table}[ht]
\centering  
\begin{tabular}{c c c c c c c } 
 \hline
 Ogden model & $\alpha_1$ & $\alpha_2$ & $\alpha_3$ & $\mu_1$ & $\mu_2$ & $\mu_3$ \\ [0.5ex] 
 \hline\hline
  SPOM1  & 1.3 & 4 & -2 &3.00508  &0.0145 &-0.0177  \\ 
 \hline
 SPOM2 & 1.73 & 4 & -2 &2.25815 &0.0145 &-0.0177  \\
 \hline
 FPOM1 & 1.3 & 0 & -2 &3.04969 &0 &-0.0177  \\
 \hline
 FPOM2 & 1.73 & 0 & -2 &2.29168 &0 &-0.0177  \\
 \hline
 TPOM1 & 1.3 & 0 & 0 &3.07692 &0 &0 \\
 \hline
 TPOM2 & 1.98 & 0 & 0 &2.0202 &0 &0 \\ [1ex] 
\hline
\end{tabular}
\caption{SPOM - Six parameter Ogden material, FPOM - Four parameter Ogden material, TPOM - Two parameter Ogden material.} 
\label{Table:Materials}
\end{table}
Calculations are performed for the dimensionless material parameters given in Tab.~\ref{Table:Materials}. The parameters $\mu_i$ and $\alpha_i$ are chosen such that $\sum_{i=1}^3\mu_i\alpha_i=4$.

\subsection{Deformation in the absence of magnetic field} 
\label{sec: no mag}
\begin{figure}[ht!]
\centering
\begin{overpic}[width=0.48\linewidth,trim=0in 0in 0in 0in]{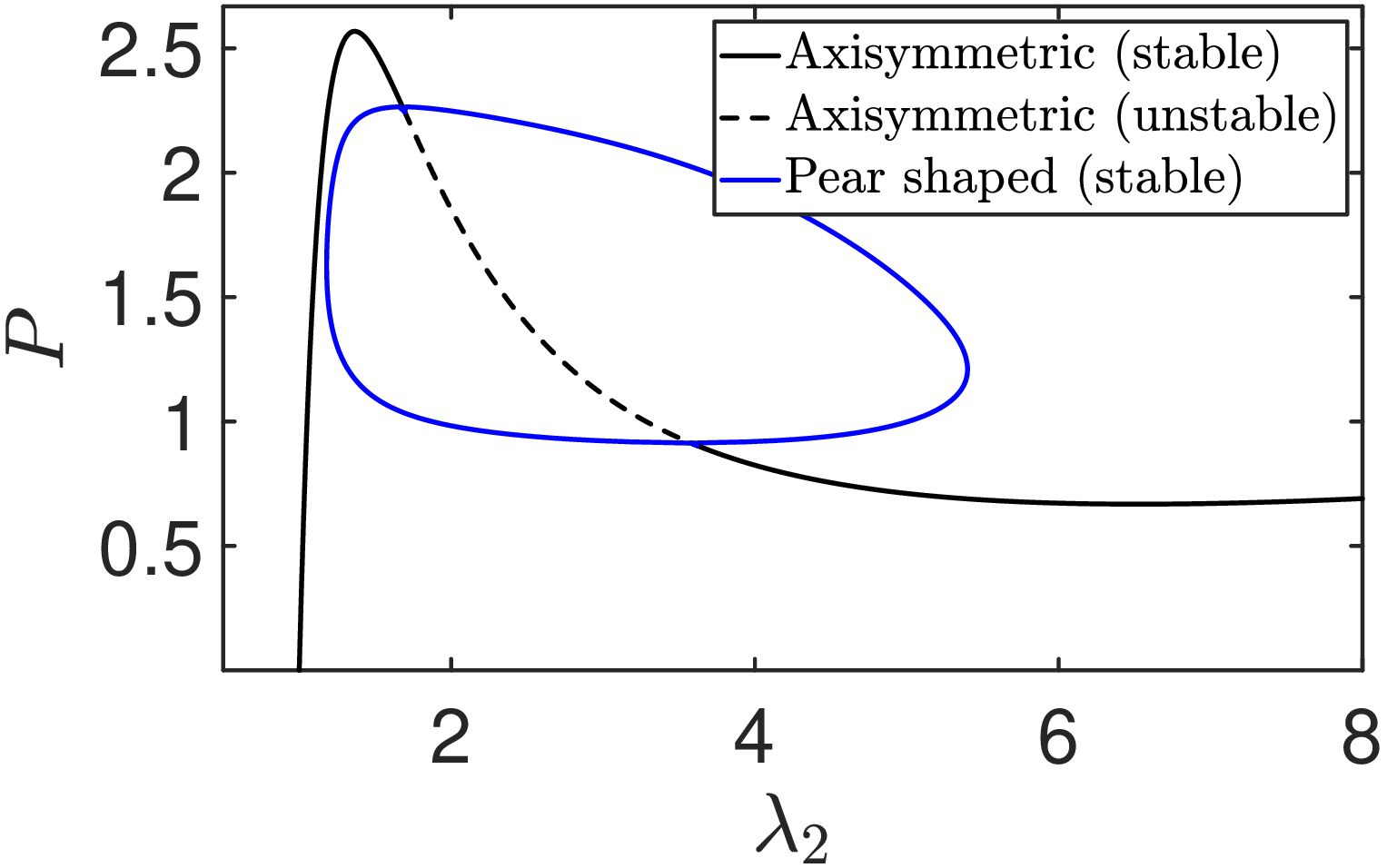}
\put(-3,62){\subcaptiontext*[1]{}}
\put(29,56){\color{black}\small $P_{cr1}$}
\put(47,33){\color{black}\small $P_{cr2}$}
\end{overpic}\hspace{0.1in}
\begin{overpic}[width=0.48\linewidth,trim=0in 0in 0in 0in]{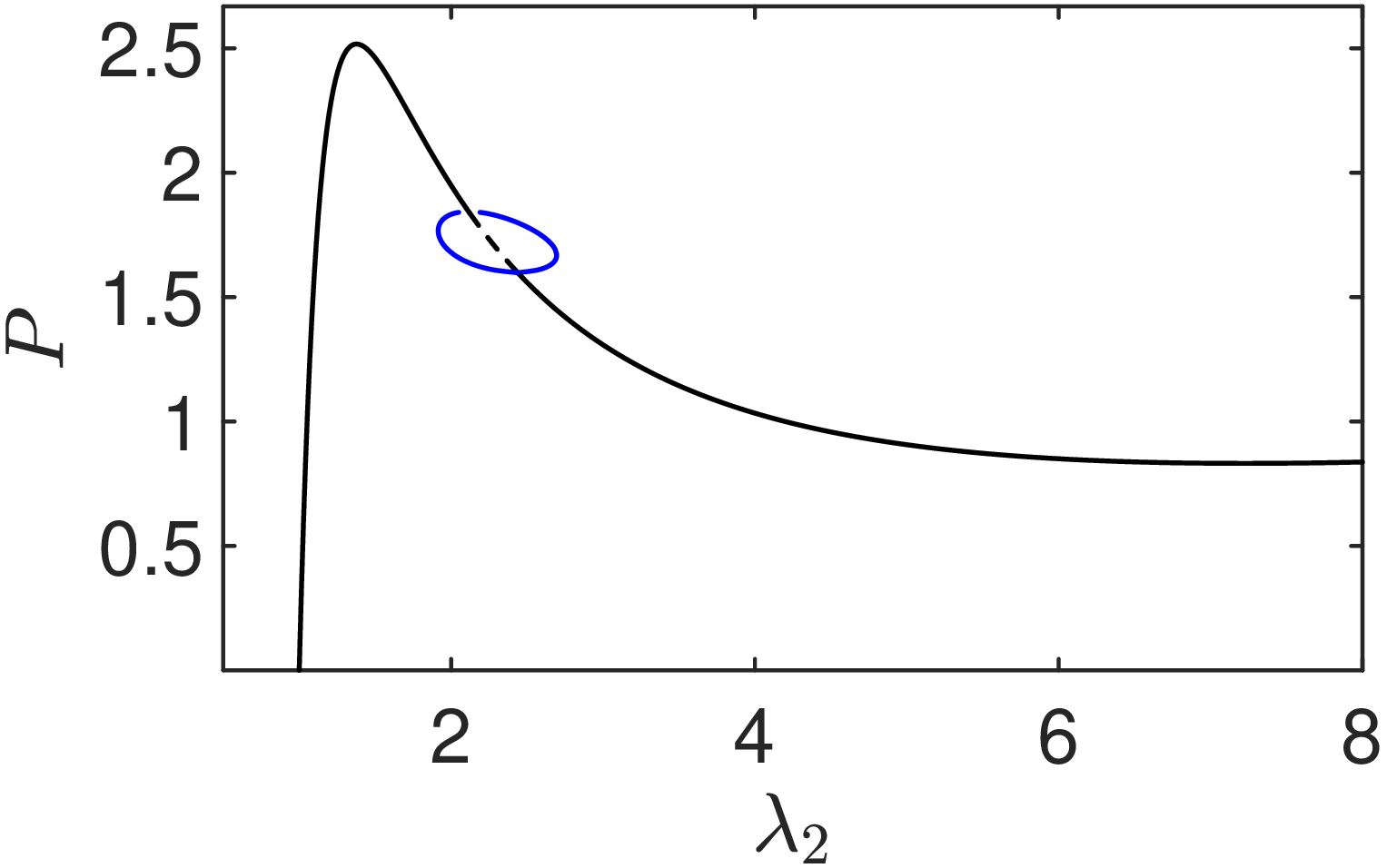}
\put(-3,62){\subcaptiontext*[2]{}}
\end{overpic} \\ 
\vspace{0.1in}
\begin{overpic}[width=0.48\linewidth,trim=0in 0in 0in 0in]{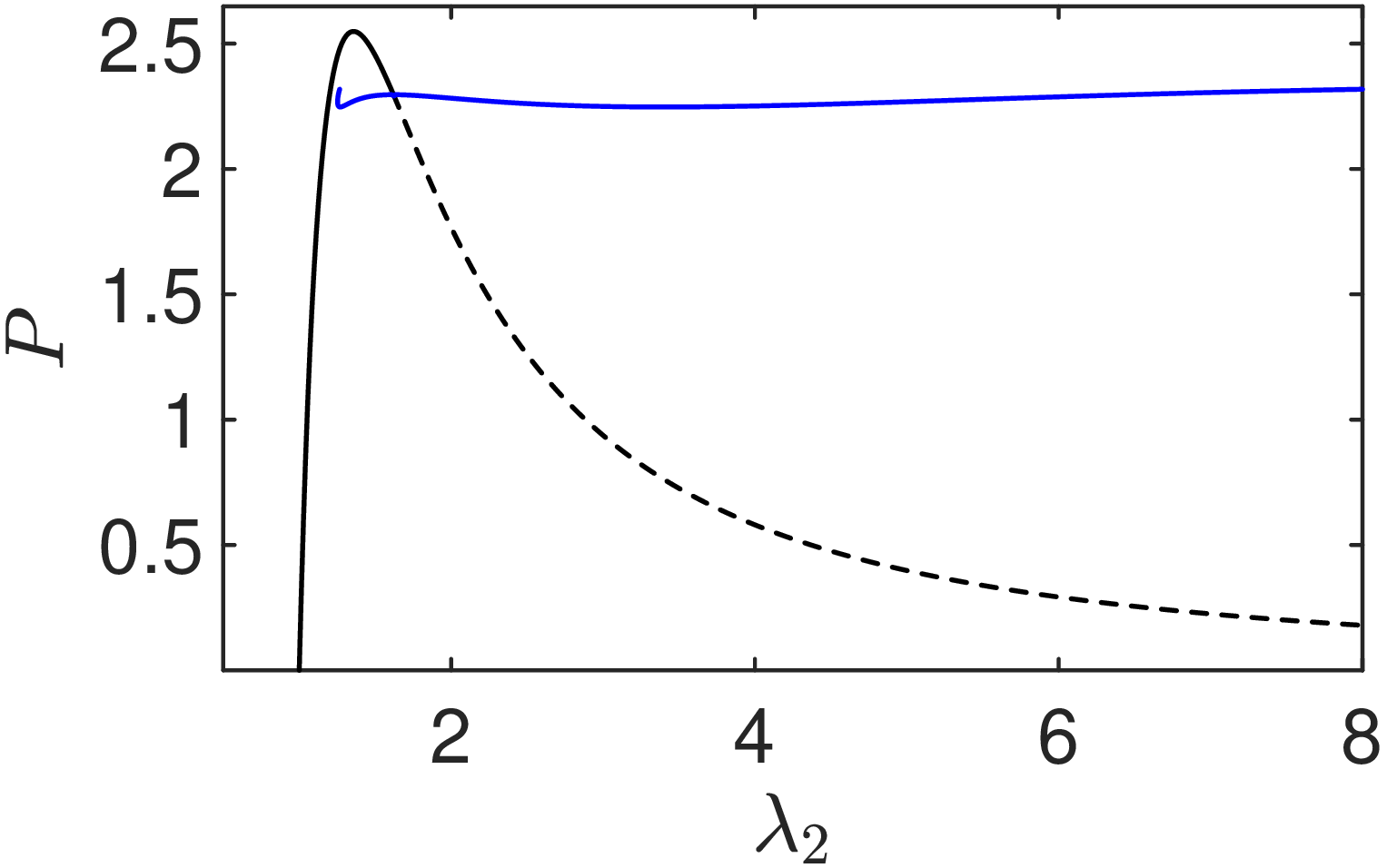}
\put(-3,62){\subcaptiontext*[3]{}}
\end{overpic}\hspace{0.1in}
\begin{overpic}[width=0.48\linewidth,trim=0in 0in 0in 0in]{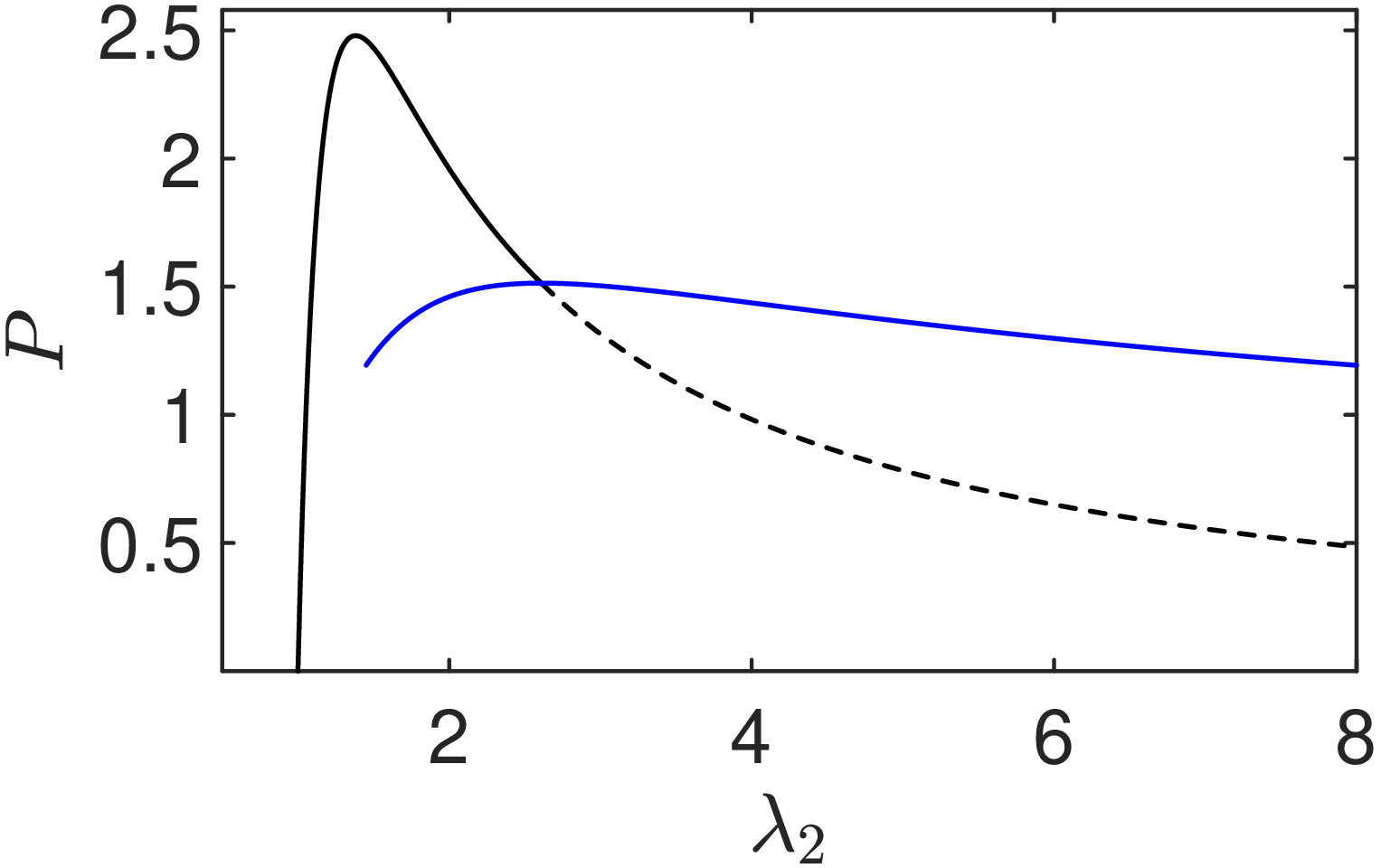}
\put(-3,62){\subcaptiontext*[4]{}}
\end{overpic}

\caption{$P$ vs $\lambda_2$ at the pole ($\theta=0$, $\pi$) for the axisymmetric and pear--shaped deformations of the balloon in the absence of magnetic field. (a) SPOM1 and (b) SPOM2 material parameters with isola bifurcations where axisymmetric shape retains its stability after some stretch by reverse supercritical pitchfork bifurcation (c) TPOM1 and (d) TPOM2 with no isola bifurcation. } 
\label{fig:M0} 
\end{figure}
Consider the case with zero applied magnetic field ($\mathcal{M} = 0 $). The spherical elastomeric balloon is inflated by controlling the pressure inside the balloon. The inflated configuration remains spherical until the deformation reaches a stretch value where, a bifurcation point is observed beyond the limit point pressure stretch. The solutions observed after the bifurcation point are plane of symmetry breaking also called in the literature as pear-shaped configurations. Such pear-shaped solutions on either side of the axisymmetric curves represent stretches at the north and south poles and vice versa. While the bifurcation of the spherical equilibrium into a pear-shaped equilibrium is studied in \cite{FU201433} and \cite{CHEN_IJNM1991}, it is important to revisit the specific impact of material parameters on the post-bifurcation solutions, particularly in the context of Ogden type material model. 

Fig.~\ref{fig:M0} demonstrates the effect of material parameters on the post-bifurcation (pear-shaped) equilibrium curve with the Ogden material model. For both SPOM and FPOM parameters (see Tab.~\ref{Table:Materials}), an isola bifurcation exists in the form of an isolated closed loop representing pear-shaped solutions as shown in Figs.~\ref{fig:M0}(a) and \ref{fig:M0}(b). The local bifurcations show supercritical and reverse supercritical pitchfork normal behavior at the top ($P_{cr1}$) and bottom ($P_{cr2}$) bifurcation points, respectively. As $\alpha_1$ increases, the isola solution curve shrinks to the isola center and eventually disappears beyond a specific critical value of $\alpha_1$. On the other hand, for TPOM parameters, there is no isola bifurcation observed (see Figs.~\ref{fig:M0}(c) and \ref{fig:M0}(d)). However, supercritical pitchfork bifurcation is observed with hardening and softening type nonlinearity as shown in Figs.~\ref{fig:M0}(c) and ~\ref{fig:M0}(d), respectively. Also, the pressure at which the plane of symmetry breaking bifurcation occurs decreases with increasing $\alpha_1$.  

There is a critical value of $\alpha_1$ (not shown in figures) for the SPOM, FPOM, and TPOM models beyond which pear-shaped solutions no longer exist. For the values of $\alpha_1$ beyond this threshold, the spherical equilibrium remains stable, and the system does not transition to a non-spherical configuration.

\subsection{Effect of magnetic field on balloon inflation: single coil arrangement}   
\label{sec:single coil} 

\begin{figure}[htb!]
\centering
\begin{overpic}[width=0.48\linewidth,trim=0in 0in 0in 0in]{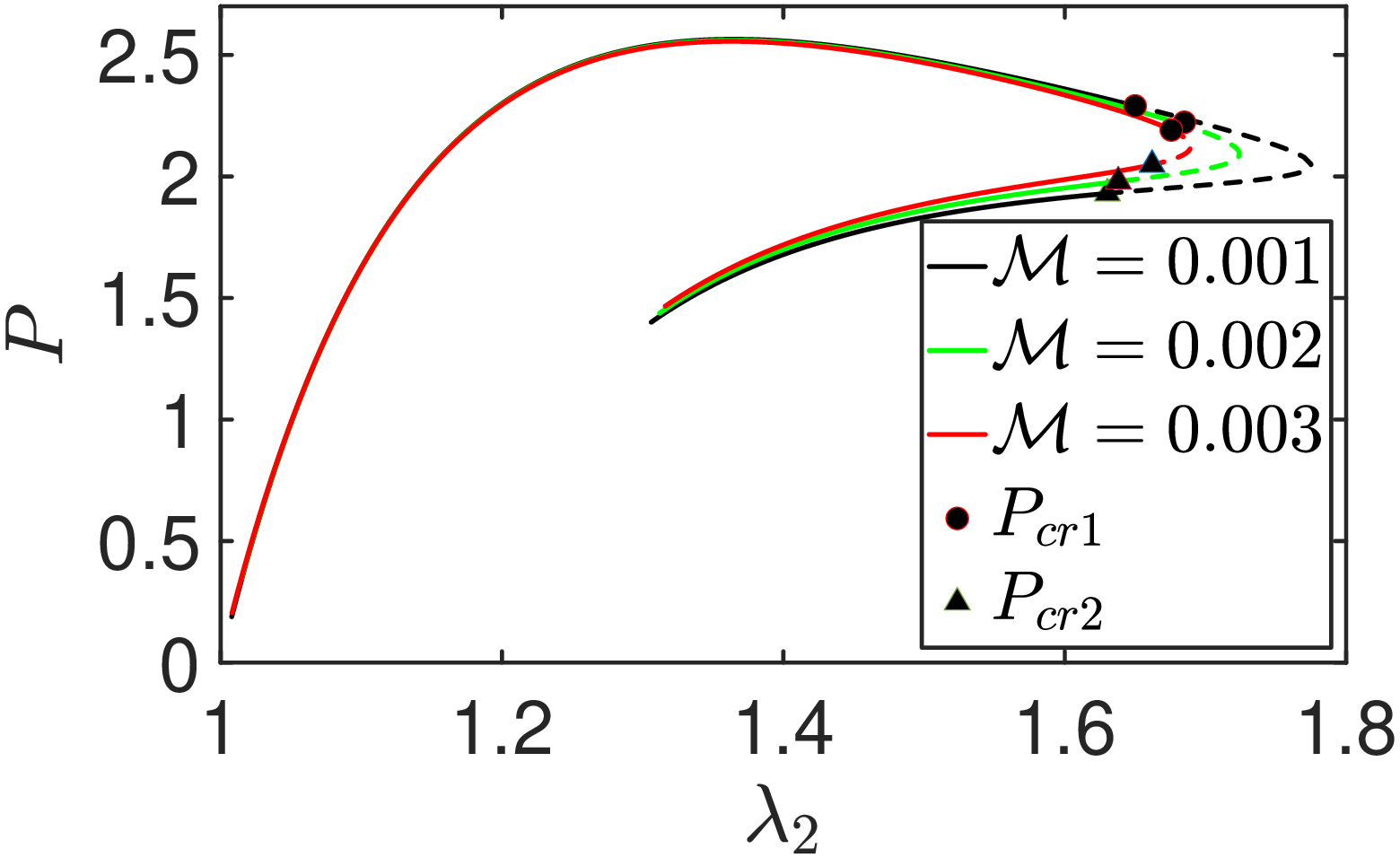}
\put(-3,62){\subcaptiontext*[1]{}}
\end{overpic} \hspace{0.1in}
\begin{overpic}[width=0.48\linewidth,trim=0in 0in 0in 0in]{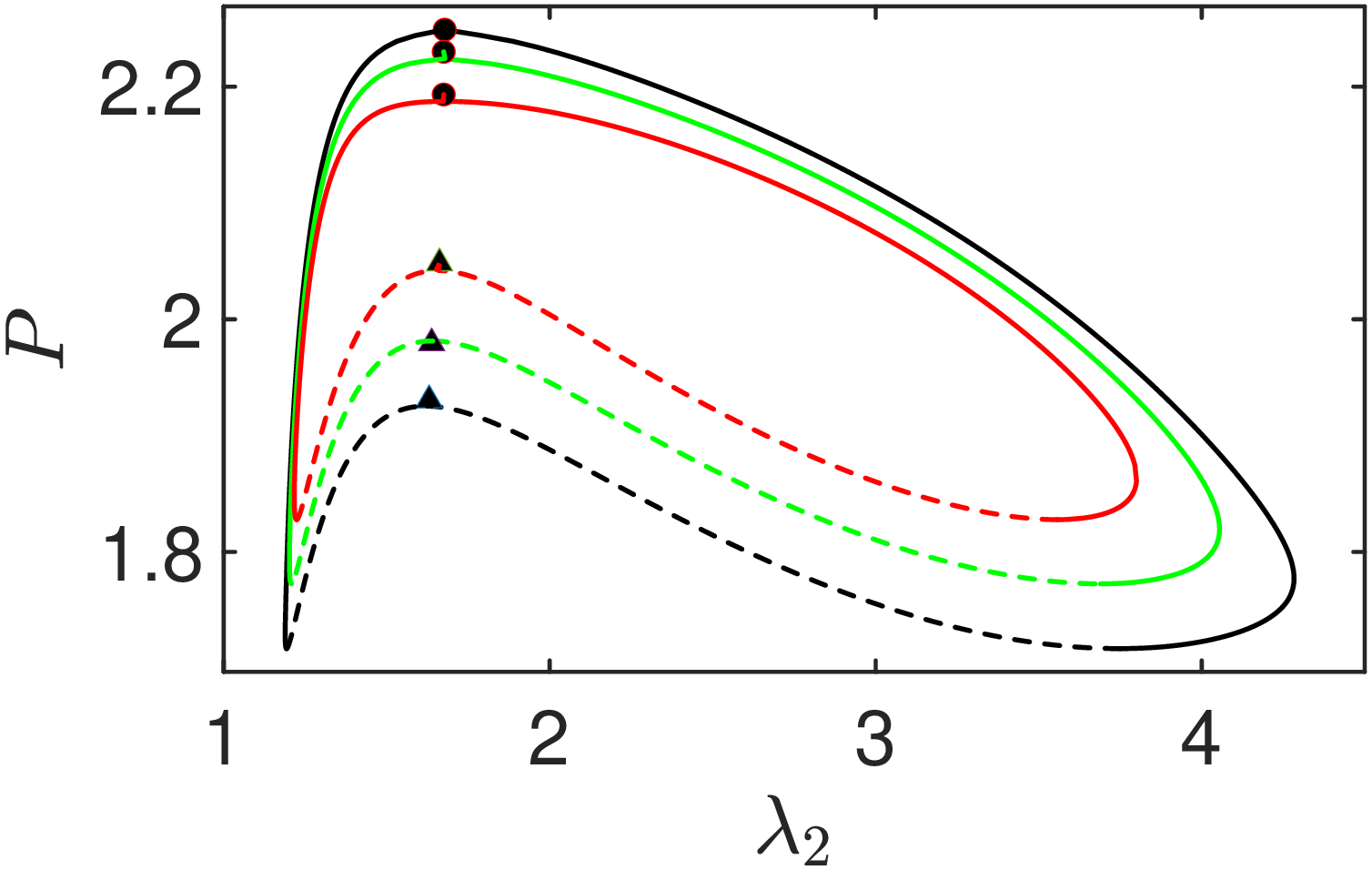}
\put(-3,62){\subcaptiontext*[2]{}}
\end{overpic} \\ 
\vspace{0.1in}

\begin{overpic}[width=0.48\linewidth,trim=0in 0in 0in 0in]{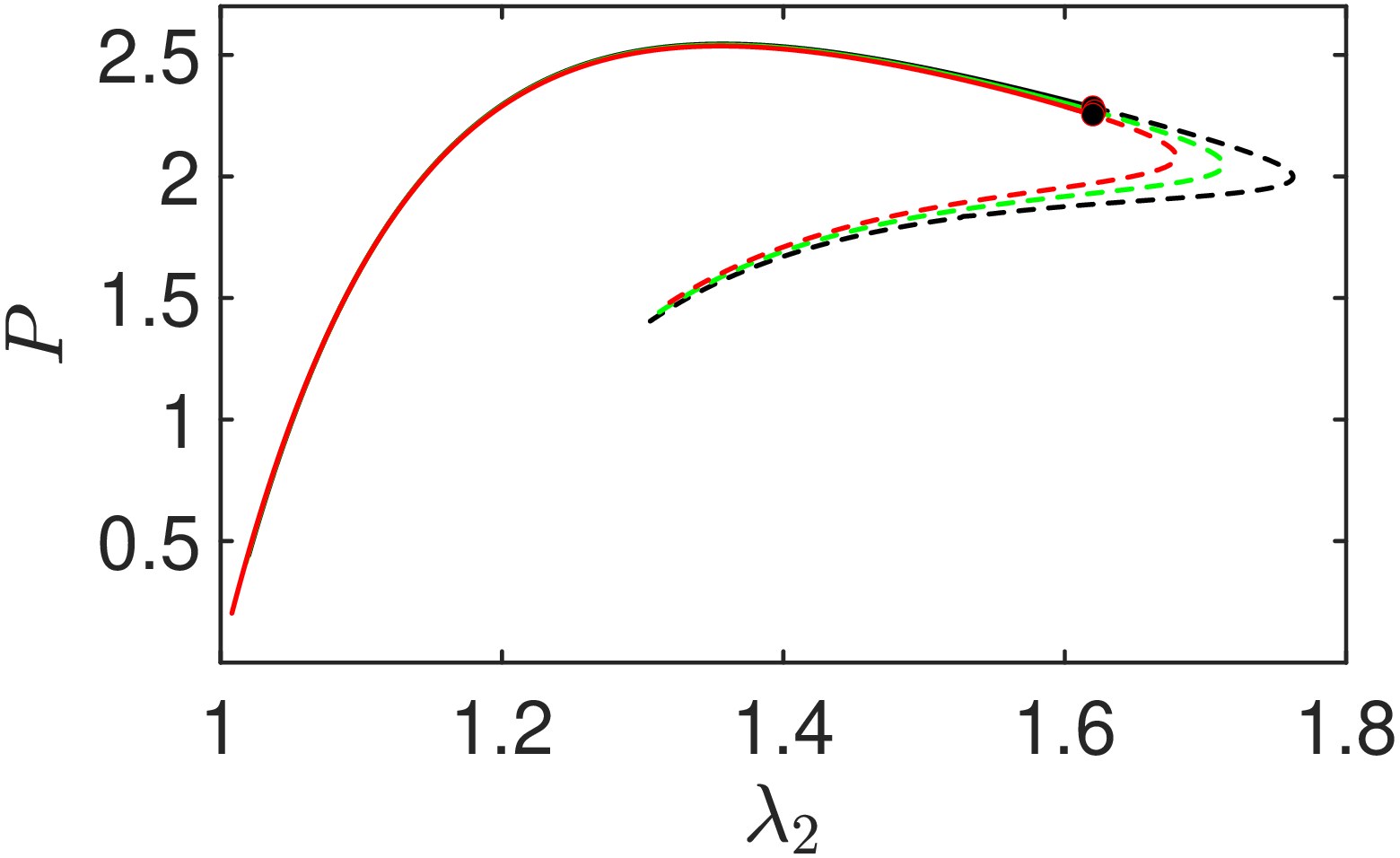} 
\put(-3,62){\subcaptiontext*[3]{}}
\end{overpic}\hspace{0.1in}
\begin{overpic}[width=0.48\linewidth,trim=0in 0in 0in 0in]{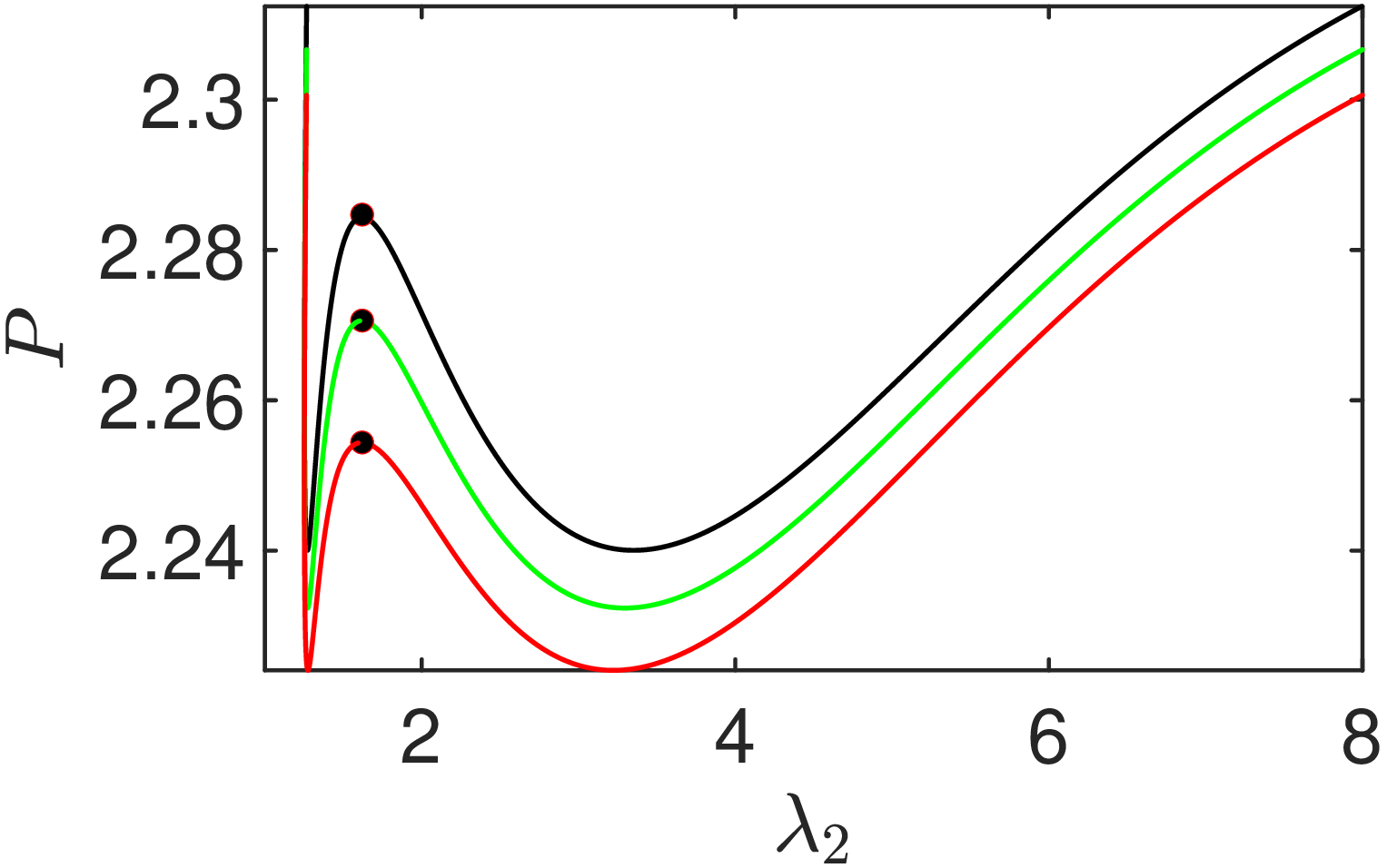}
\put(-3,62){\subcaptiontext*[4]{}}
\end{overpic}

\caption{ $P$ vs $\lambda_2$ at pole $\theta=0$ (a) SPOM1 plane of symmetry  curve. (b) SPOM1 pear--shaped curve (c) TPOM1 plane of symmetry curve. (d) TPOM1 pear--shaped curve. The solid line represents a stable equilibrium and the dotted line represents unstable equilibrium.}
\label{fig:Magdiff} 
\end{figure}

\begin{figure}[htb!]
\centering
\begin{overpic}[width=0.48\linewidth,trim=0in 0in 0in 0in]{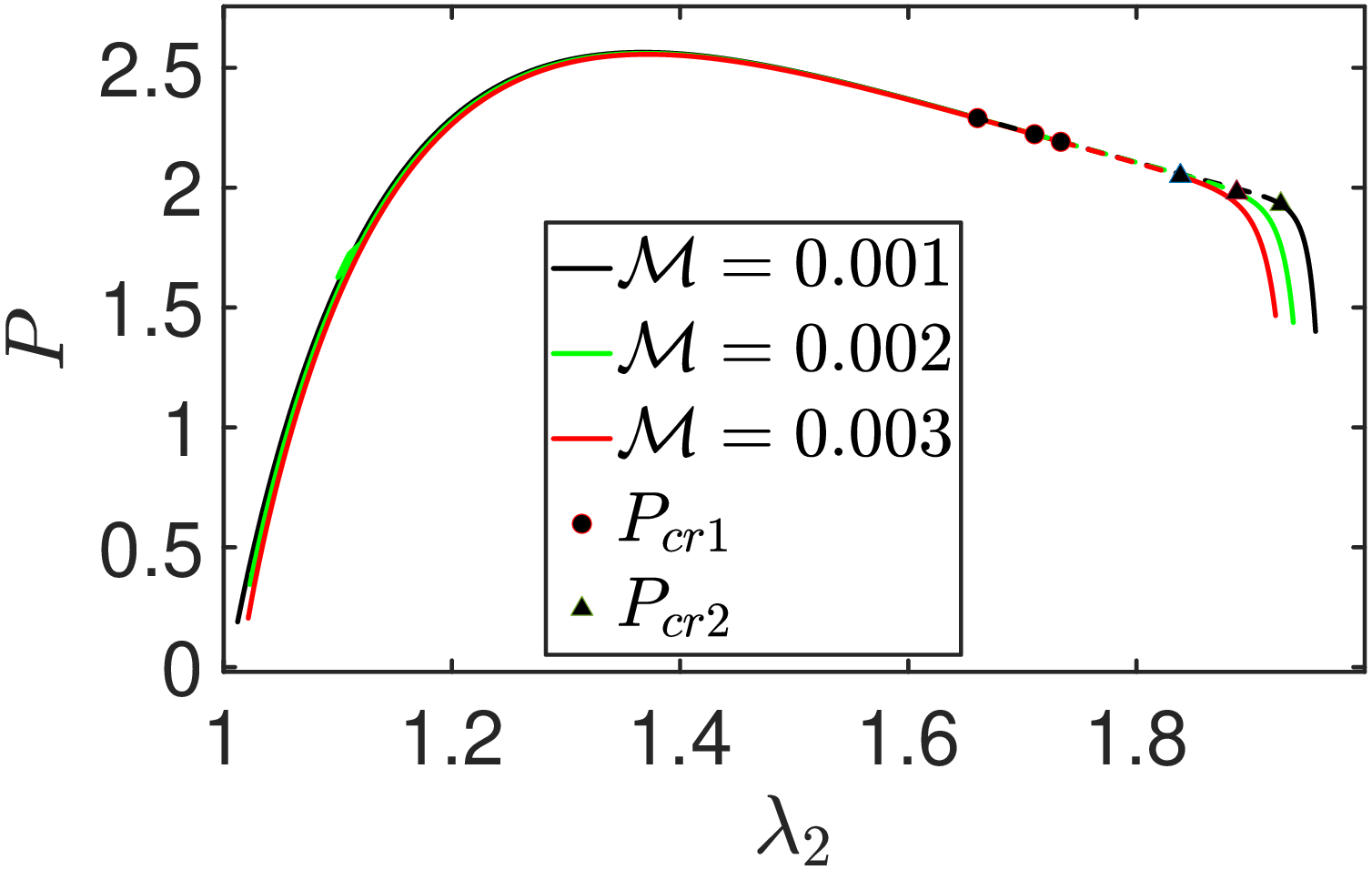}
\put(-3,62){\subcaptiontext*[1]{}}
\end{overpic}\hspace{0.1in}
\begin{overpic}[width=0.48\linewidth,trim=0in 0in 0in 0in]{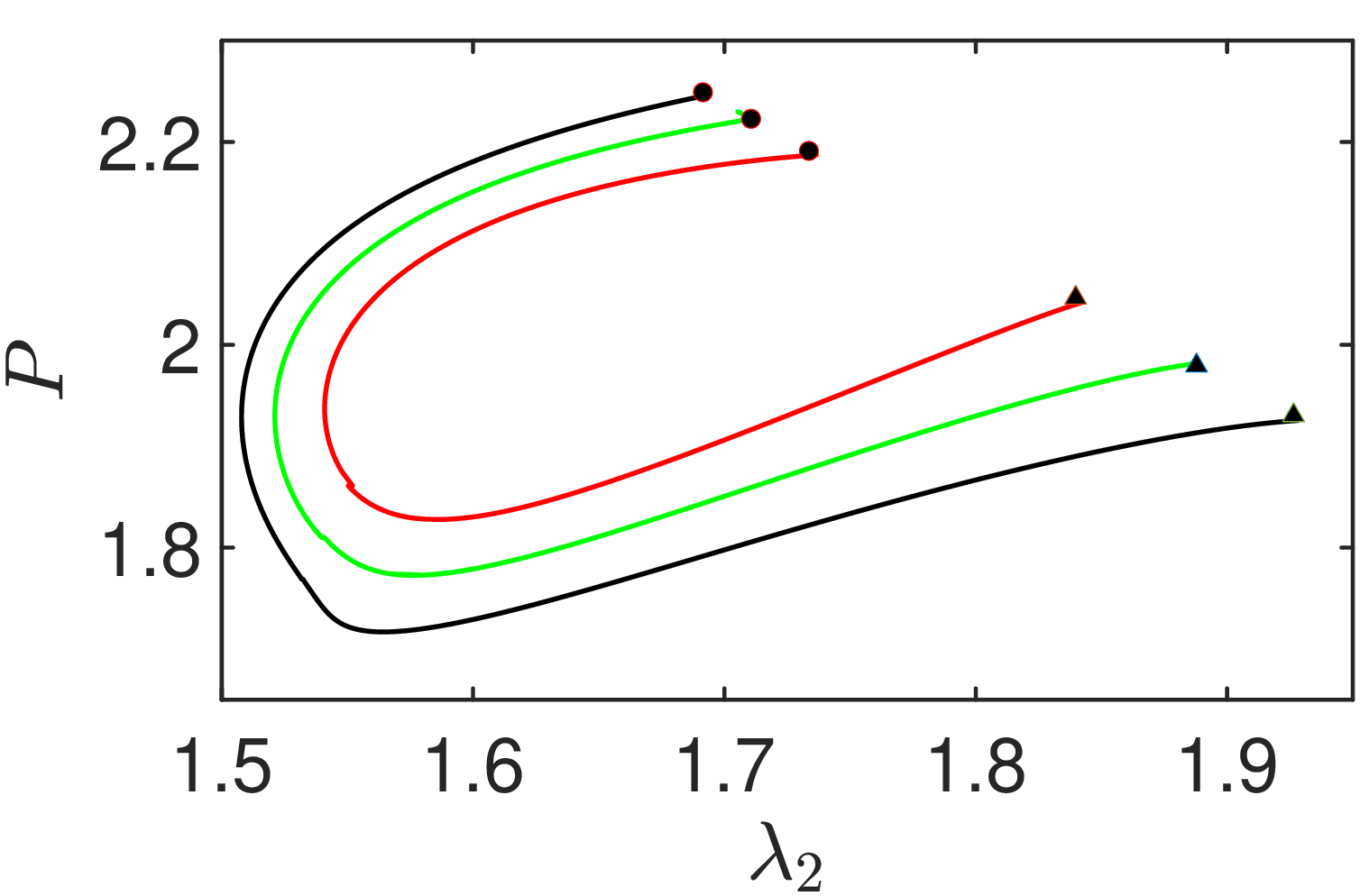}
\put(-3,62){\subcaptiontext*[2]{}}
\end{overpic} \\ 
\vspace{0.1in}
\begin{overpic}[width=0.48\linewidth,trim=0in 0in 0in 0in]{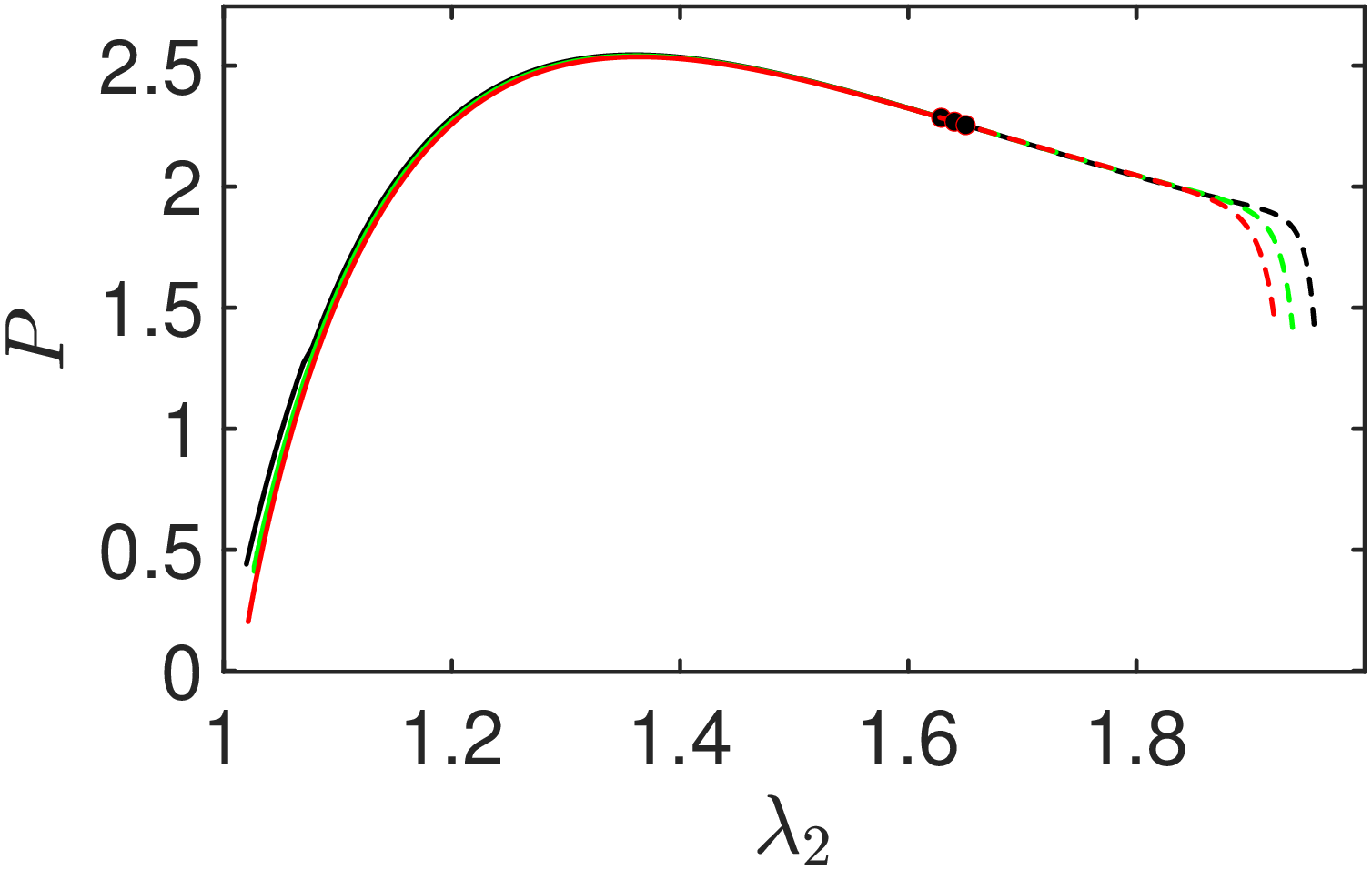}
\put(-3,62){\subcaptiontext*[3]{}}
\end{overpic}\hspace{0.1in}
\begin{overpic}[width=0.48\linewidth,trim=0in 0in 0in 0in]{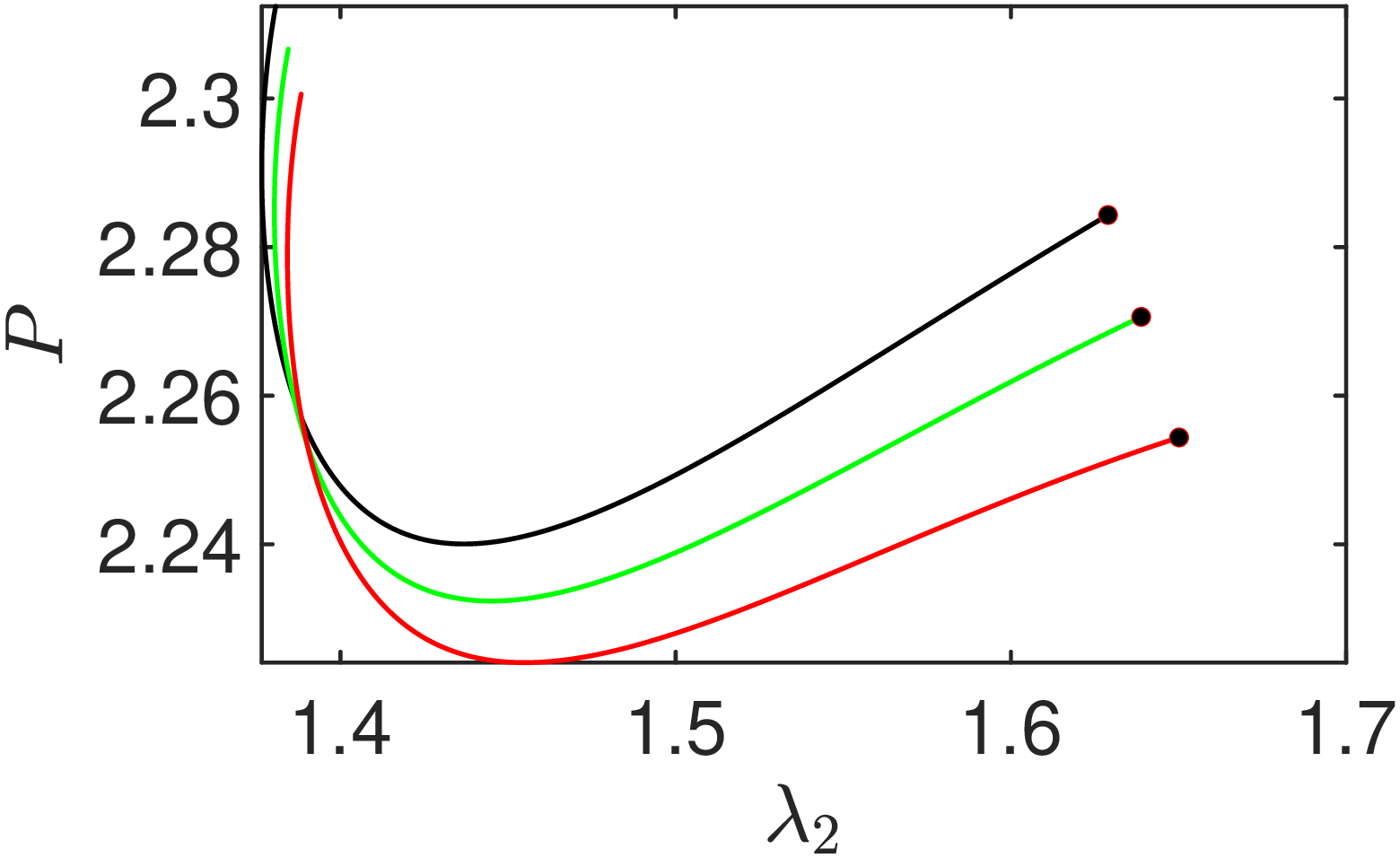}
\put(-3,62){\subcaptiontext*[4]{}}
\end{overpic}
\caption{ $P$ vs $\lambda_2$ at equator $\theta=\pi/2$  (a) SPOM1 plane of symmetry curve. (b) SPOM1 pear-shaped curve. (c) TPOM1 plane of symmetry curve. (d) TPOM1 pear-shaped curve. The solid line represents a stable equilibrium and the dotted line represents unstable equilibrium.}
\label{fig:Magdiff_eq}
\end{figure}
\begin{figure}[bt]
\centering
\begin{overpic}[width=0.48\linewidth,trim=0in 0in 0in 0in]{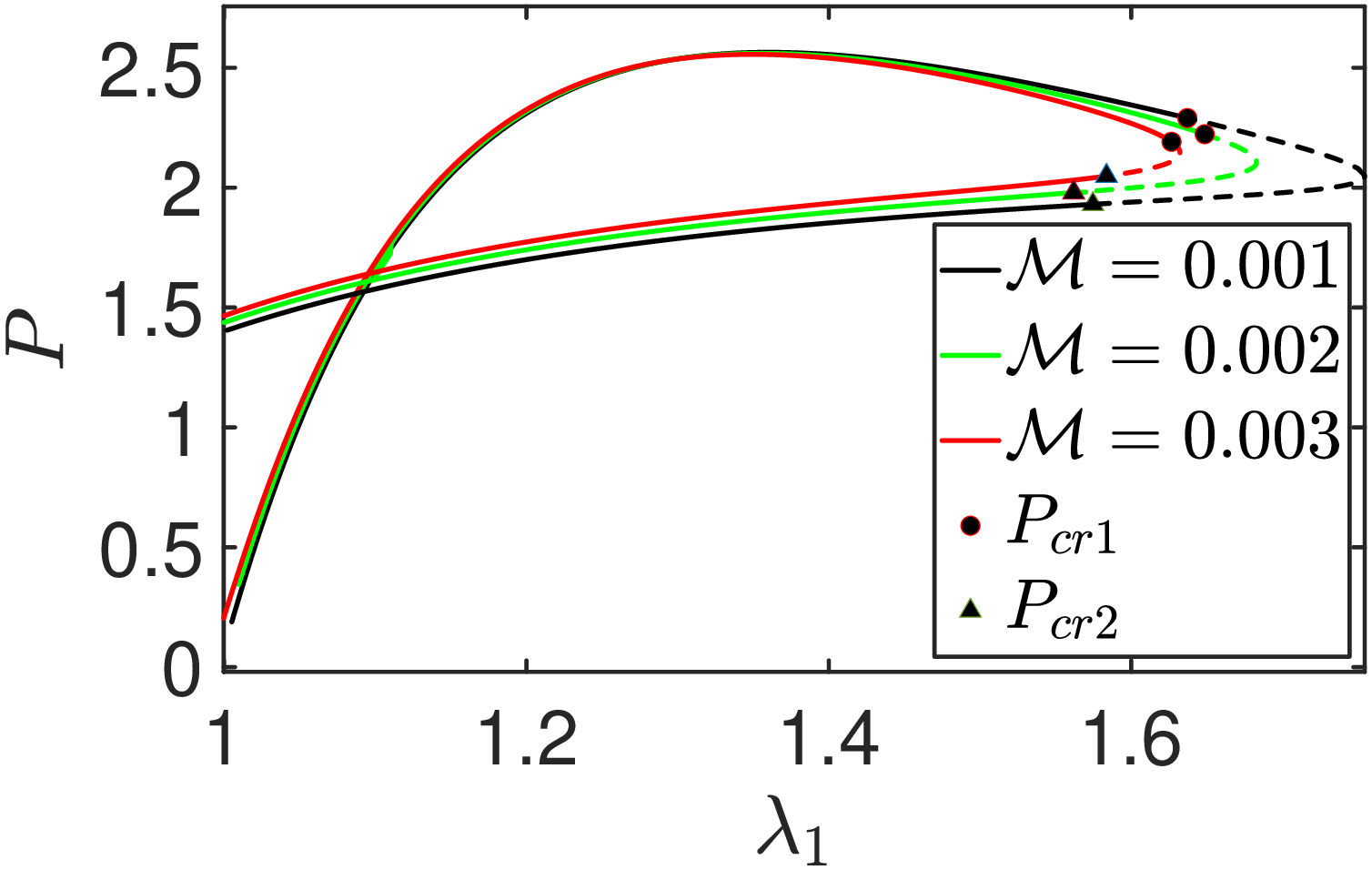}
\put(-3,62){\subcaptiontext*[1]{}}
\end{overpic}\hspace{0.1in}
\begin{overpic}[width=0.48\linewidth,trim=0in 0in 0in 0in]{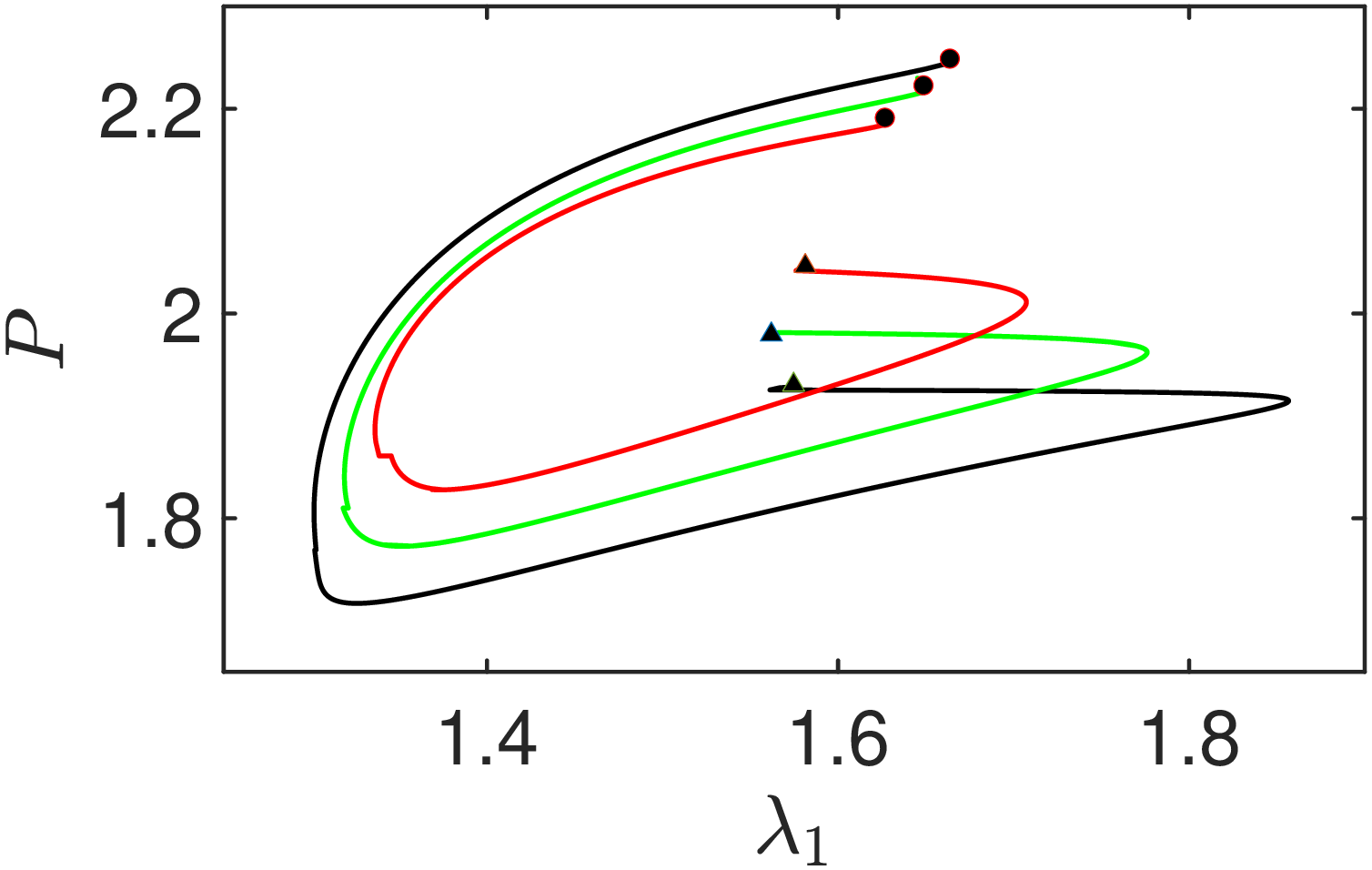}
\put(-3,62){\subcaptiontext*[2]{}}
\end{overpic} \\ 
\vspace{0.1in}
\begin{overpic}[width=0.48\linewidth,trim=0in 0in 0in 0in]{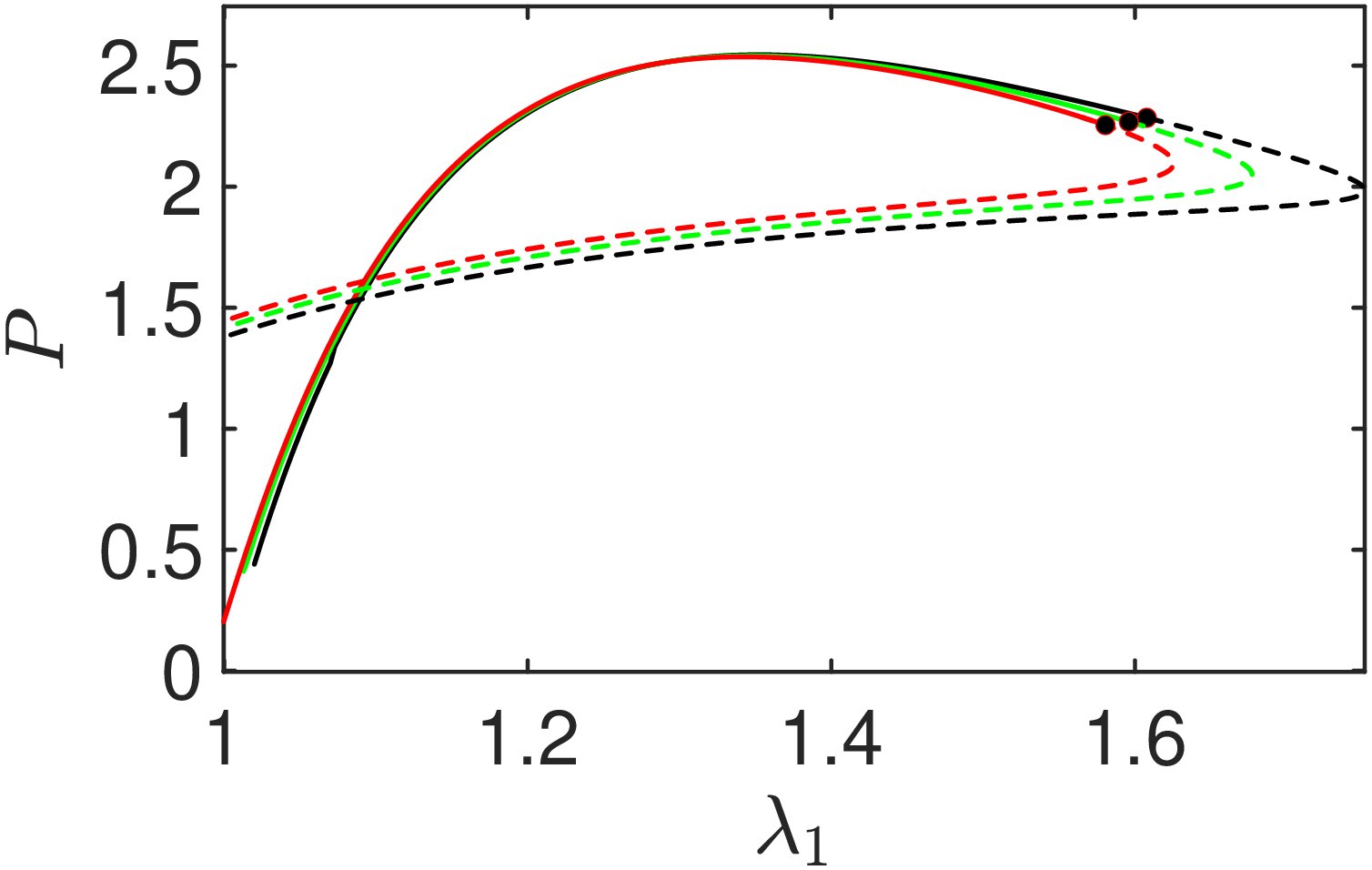}
\put(-3,62){\subcaptiontext*[3]{}}
\end{overpic}\hspace{0.1in}
\begin{overpic}[width=0.48\linewidth,trim=0in 0in 0in 0in]{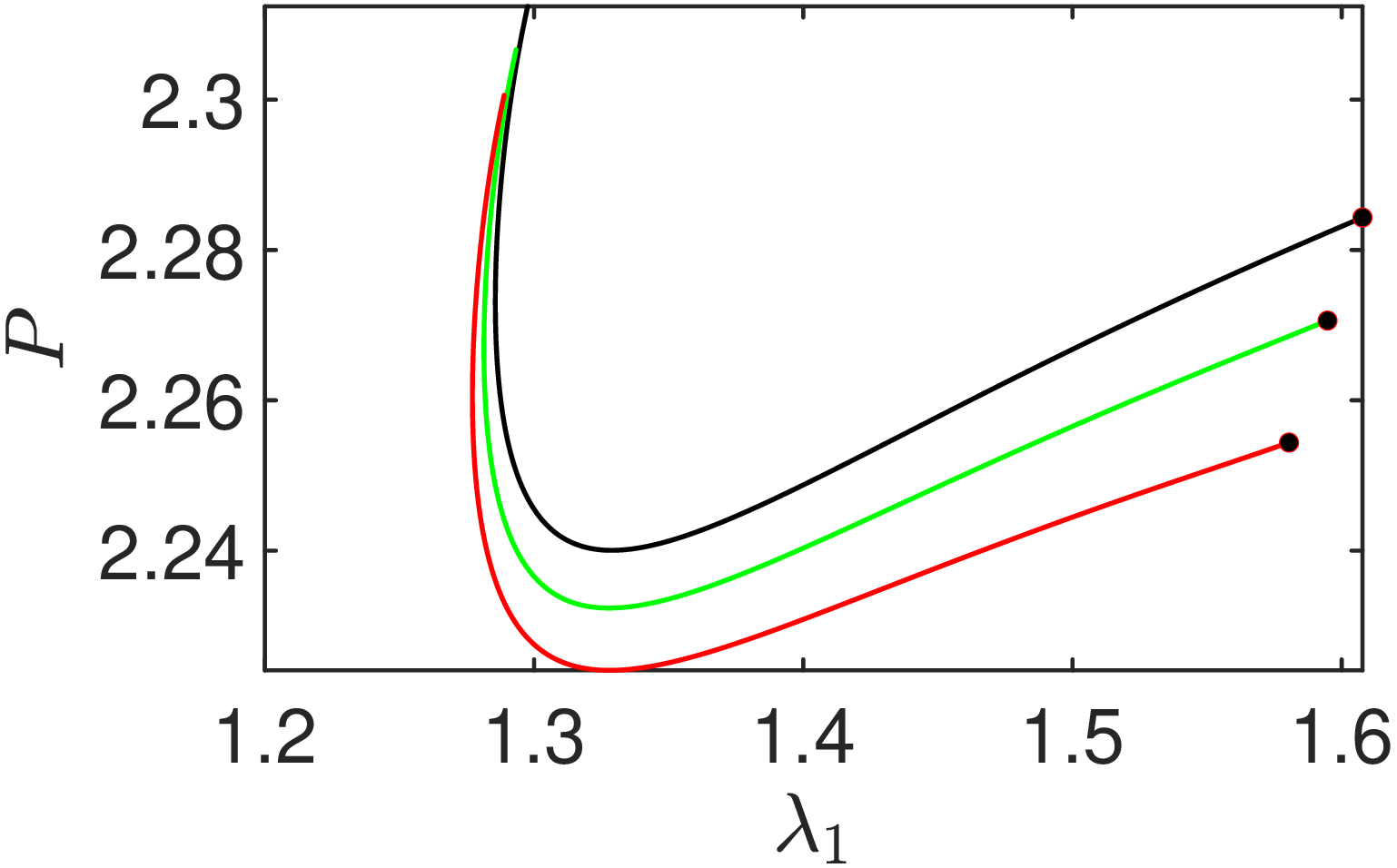}
\put(-3,62){\subcaptiontext*[4]{}}
\end{overpic}

\caption{ $P$ vs $\lambda_1$ at equator $\theta=\pi/2$ (a) SPOM1 plane of symmetry curve. (b) SPOM1 pear--shaped curve. (c) TPOM1 plane of symmetry curve. (d) TPOM1 pear--shaped curve. The solid line represents a stable equilibrium and the dotted line represents unstable equilibrium.} 
     \label{fig:Magdiff_eq-l1}
\end{figure}

This section explores the effects of non-uniform applied magnetic field created by a current-carrying coil on the deformation mechanics of an incompressible magnetoelastic balloon. At smaller stretches, the influence of the magnetic field is negligible but becomes increasingly pronounced as the stretches grow larger. This occurs because the magnetic field intensity diminishes as the distance between the wire and the membrane increases. A similar observation is reported in \cite{REDDY_IJNM2017} for a toroidal membrane.

As the applied magnetic field intensity increases, the limit point pressure of the primary (plane of symmetry) curve decreases. Notably, at smaller stretch ratios, the primary curve does not have an equilibrium state at a pressure lower than a certain critical value on the left of the limit point on the primary curve. The smallest pressure in the pre-limit point region of the primary curve is commonly referred in the literature as “magnetic limit point.” At larger stretches, the primary curve exhibits a “turning point” after the limit point pressure, and this turning point occurs at progressively smaller stretches with increasing magnetic field intensity as shown in Fig.~\ref{fig:Magdiff}(a). For SPOM and FPOM parameters, another interesting feature observed with the introduction of magnetic field is the existence of multiple plane of symmetric solutions for a given stretch at the pole. 

In Figs.~\ref{fig:Magdiff}(a) and \ref{fig:Magdiff}(b), for the SPOM parameters, the first critical pressure, $P_{cr1}$ (filled circle), decreases with increasing magnetic field intensity, while the second critical pressure, $P_{cr2}$ (filled triangle), increases. Similar results are observed for the FPOM parameters (not shown in figures). Though isolated pear-shaped solutions are observed with the magnetic field in Fig.~\ref{fig:Magdiff}(b), part of the solution curve on the isolated loop is marked with dashed lines representing instability. This was not observed in the absence of magnetic field as entire isolated curve in Fig.~\ref{fig:M0}(a) is found to be stable. The detailed discussion on stability can be found in the later sections. 

In the case of TPOM parameters, there is only a single critical pressure where bifurcation occurs, and this bifurcation happens at progressively lower pressures as the magnetic field strength increases, as shown in Fig.~\ref{fig:Magdiff}(c). This behavior mirrors the effect of increasing $\alpha_1$, as discussed in Section \ref{sec: no mag}. Beyond a critical magnetic field intensity ($\mathcal{M}=0.004$ for the SPOM model and $\mathcal{M}=0.0078$ for the TPOM model), the bifurcated curve no longer exists (not shown in figure), and the plane of the symmetry curve remains entirely stable. It is interesting to observe that in the presence of magnetic field, the pear-shaped curve has a limit point on the pressure stretch curve which is different from the bifurcation point. 
Figs.~\ref{fig:Magdiff_eq} and \ref{fig:Magdiff_eq-l1} show $P$ versus $\lambda_2$ and $\lambda_1$, respectively, at the equator for both axisymmetric and plane of symmetry breaking cases. After the turning point $\lambda_2$ at the equator nearly remains constant, as shown in Fig.~\ref{fig:Magdiff_eq}.  However, $\lambda_1$ at the equator continues to reduce after the turning point as shown in Fig.~\ref{fig:Magdiff_eq-l1}. When $\lambda_1$ at the equator drops below one, this corresponds to the region where the stress $\sigma_{\theta}$ becomes negative, which indicates an impending wrinkling condition. Under the influence of a magnetic field, wrinkled configurations may appear at lower pressures along the axisymmetric branch, following the second critical point ($P_{cr2}$), and at the magnetic limit point.  
\subsection{Effect of magnetic field on balloon inflation: two coil arrangement}

\begin{figure}[ht!]
\centering
\begin{overpic}[width=0.48\linewidth,trim=0in 0in 0in 0in]{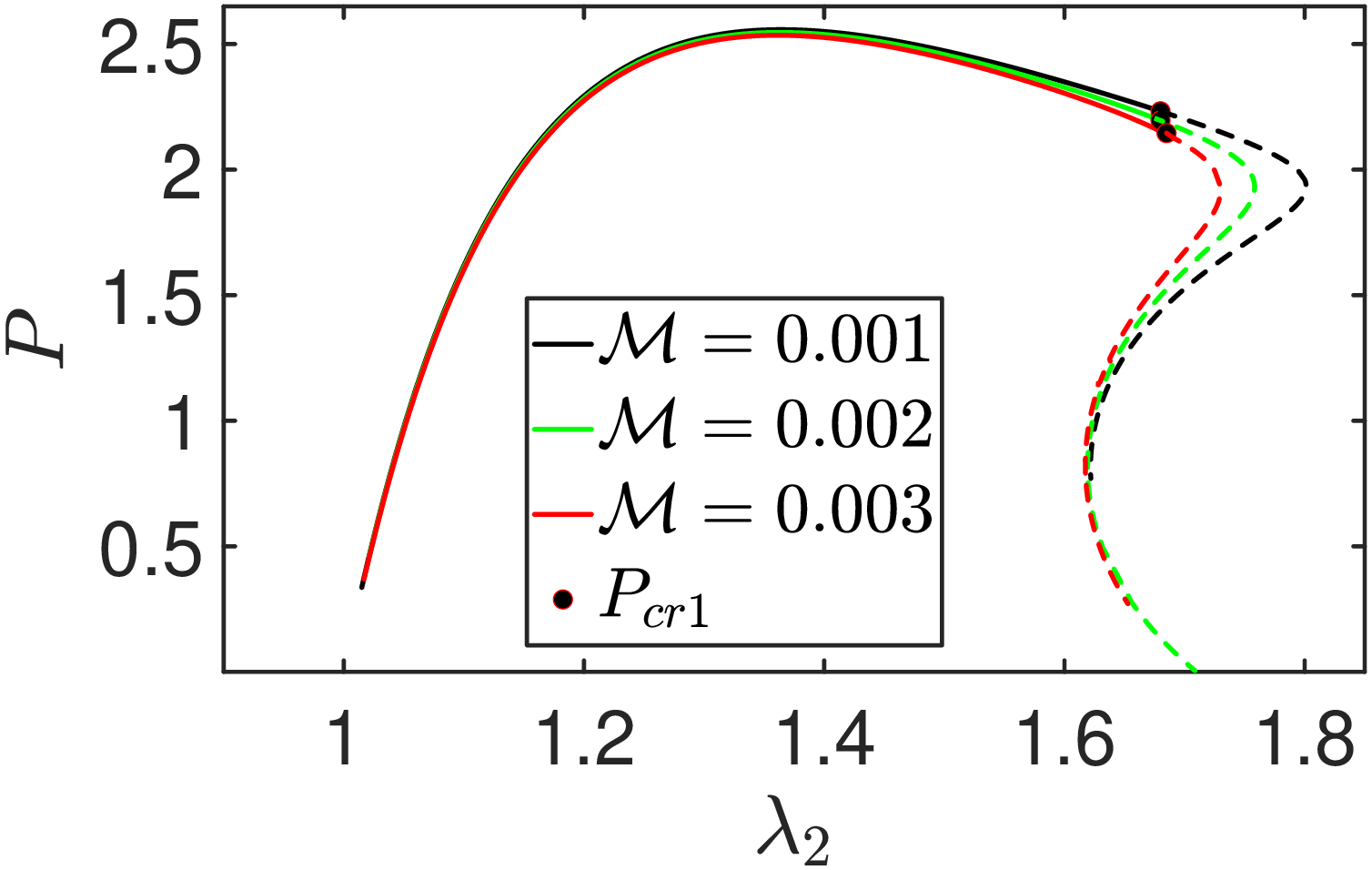}
\put(-3,62){\subcaptiontext*[1]{}}
\end{overpic}\hspace{0.1in}
\begin{overpic}[width=0.48\linewidth,trim=0in 0in 0in 0in]{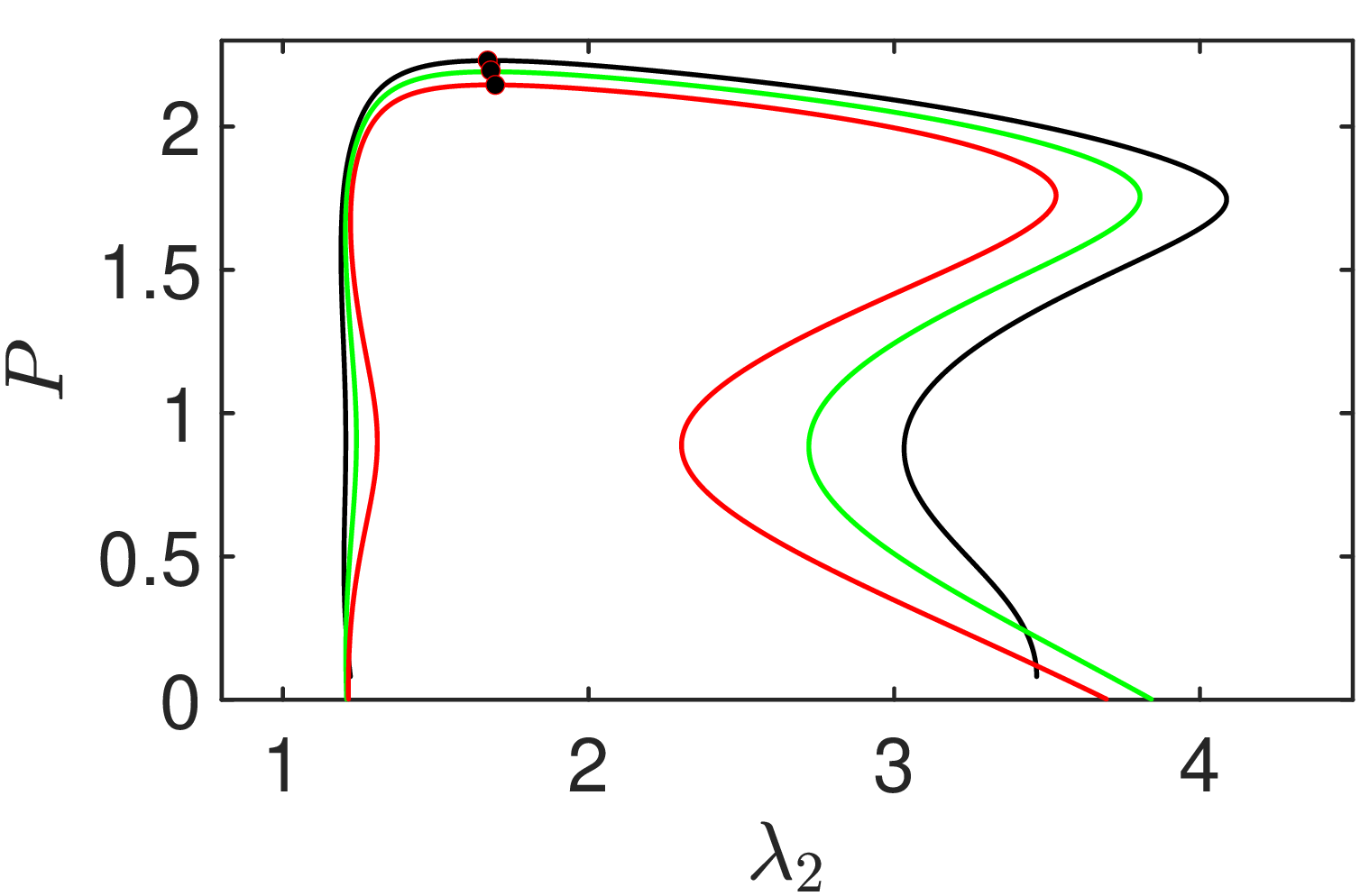}
\put(-3,62){\subcaptiontext*[2]{}}
\end{overpic} \\ 
\vspace{0.1in}
\begin{overpic}[width=0.48\linewidth,trim=0in 0in 0in 0in]{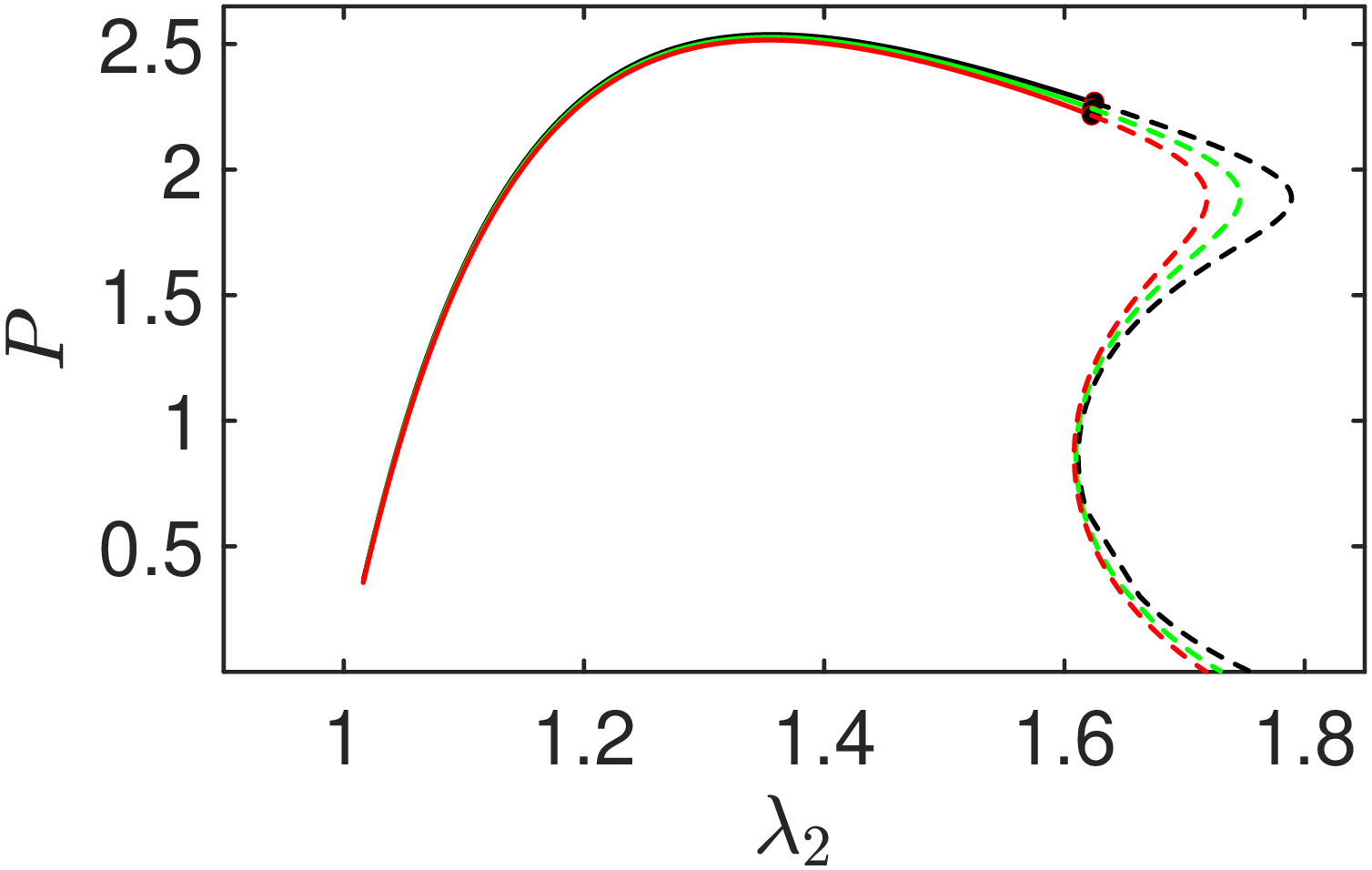}
\put(-3,62){\subcaptiontext*[3]{}}
\end{overpic}\hspace{0.1in}
\begin{overpic}[width=0.48\linewidth,trim=0in 0in 0in 0in]{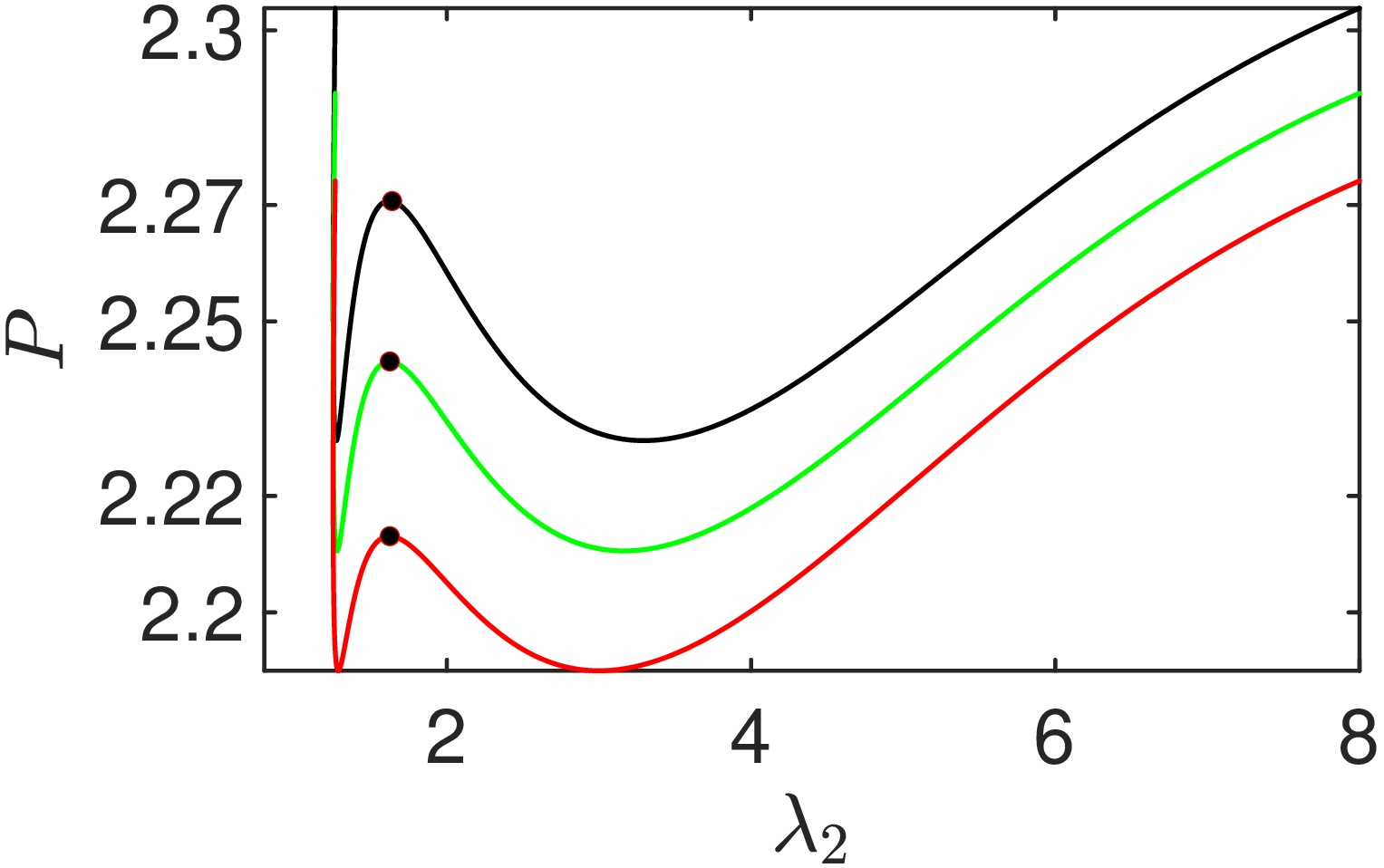}
\put(-3,62){\subcaptiontext*[4]{}}
\end{overpic}

\caption{ $P$ vs $\lambda_2$ at pole $\theta=0$ for two coil arrangement (a) SPOM1 plane of symmetry curve. (b) SPOM1 pear--shaped curve. (c) TPOM1 plane of symmetry curve. (d) TPOM1 pear--shaped curve. The solid line represents a stable equilibrium and the dotted line represents unstable equilibrium.} 
\label{fig:Magdiff_pole-l2_TC}
\end{figure}

This section examines the effects of a non-uniform magnetic field generated by two coils positioned at heights $h=1$ and $h=-1$ relative to the equator of the undeformed balloon, each with a radius $a=1.7321$. The coil at $h=1$ carries a counterclockwise current, while the coil at $h=-1$ carries a clockwise current. The observations for this configuration are consistent with those in Section \ref{sec:single coil}: increasing the magnetic field intensity reduces the limit point pressure of the primary (plane of symmetry) curve, the turning point shifts to smaller stretches with increasing field intensity, and the smallest pressure in the pre-limit point section, known as the “magnetic limit point,” marks the critical threshold for equilibrium at lowest pressure, as shown in Fig.~\ref{fig:Magdiff_pole-l2_TC}(a). 

\begin{figure}[ht!]
\centering
\begin{overpic}[width=0.48\linewidth,trim=0in 0in 0in 0in]{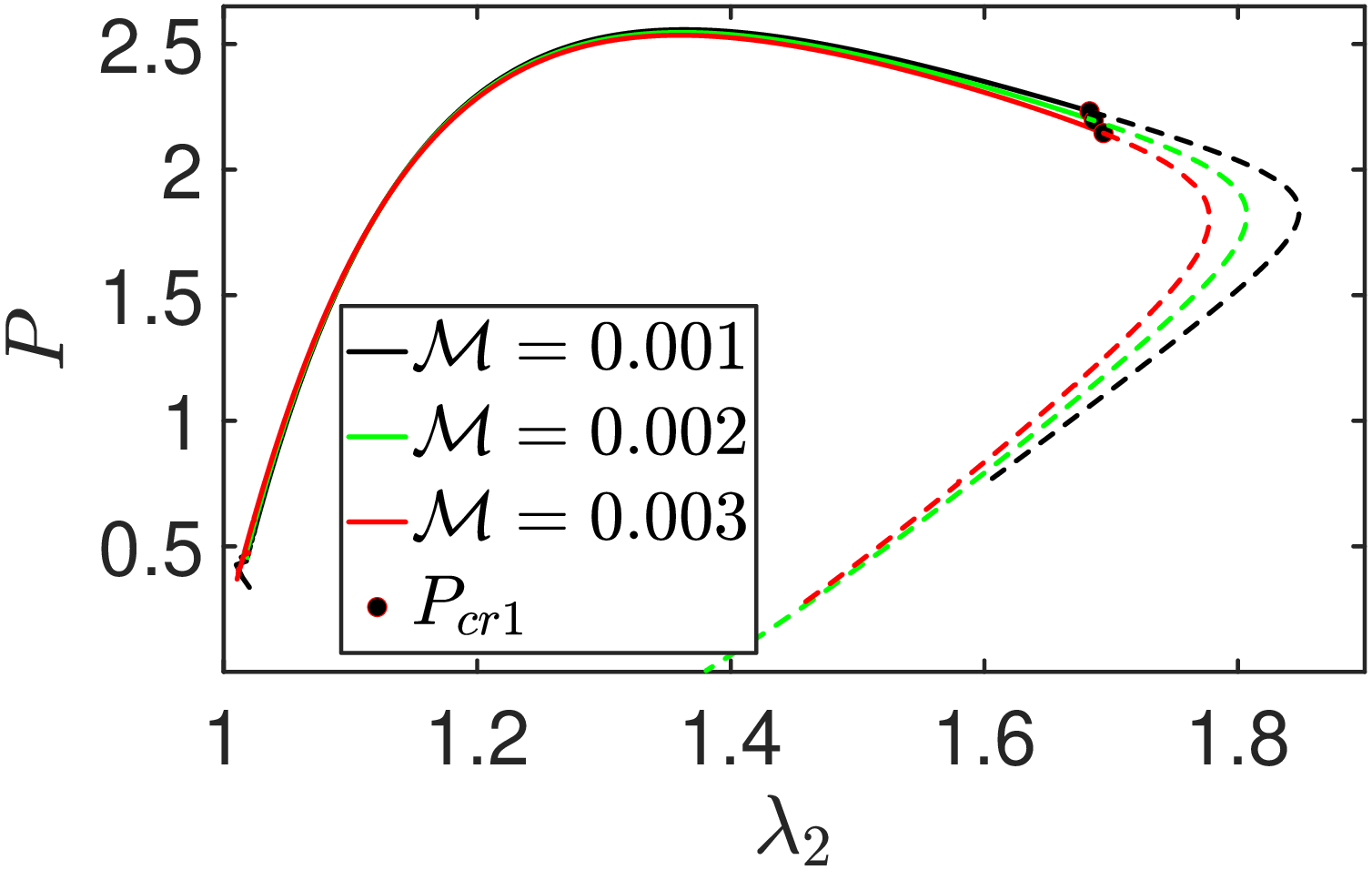}
\put(-3,62){\subcaptiontext*[1]{}}
\end{overpic}\hspace{0.1in}
\begin{overpic}[width=0.48\linewidth,trim=0in 0in 0in 0in]{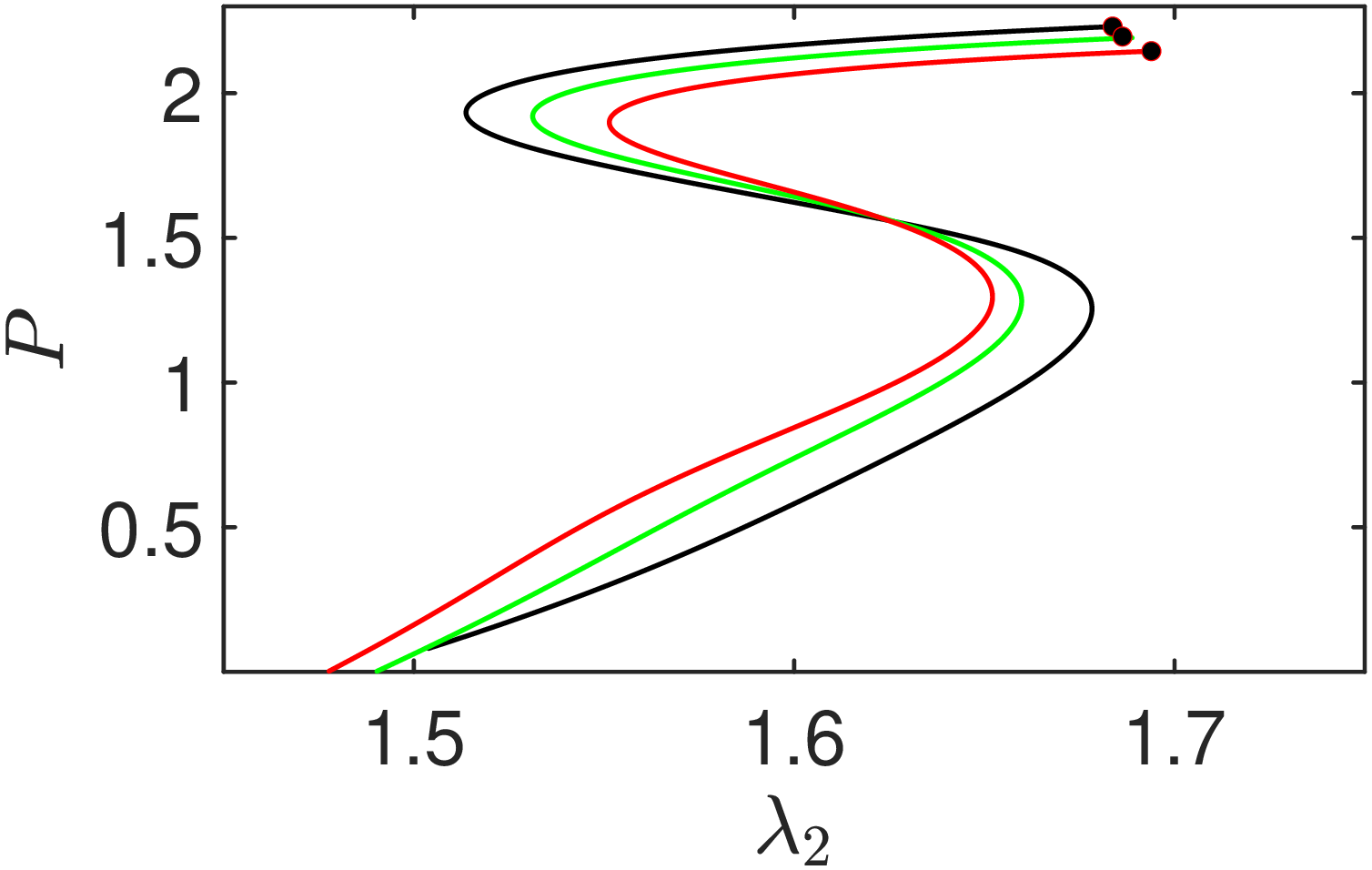}
\put(-3,62){\subcaptiontext*[2]{}}
\end{overpic} \\ 
\vspace{0.1in}
\begin{overpic}[width=0.48\linewidth,trim=0in 0in 0in 0in]{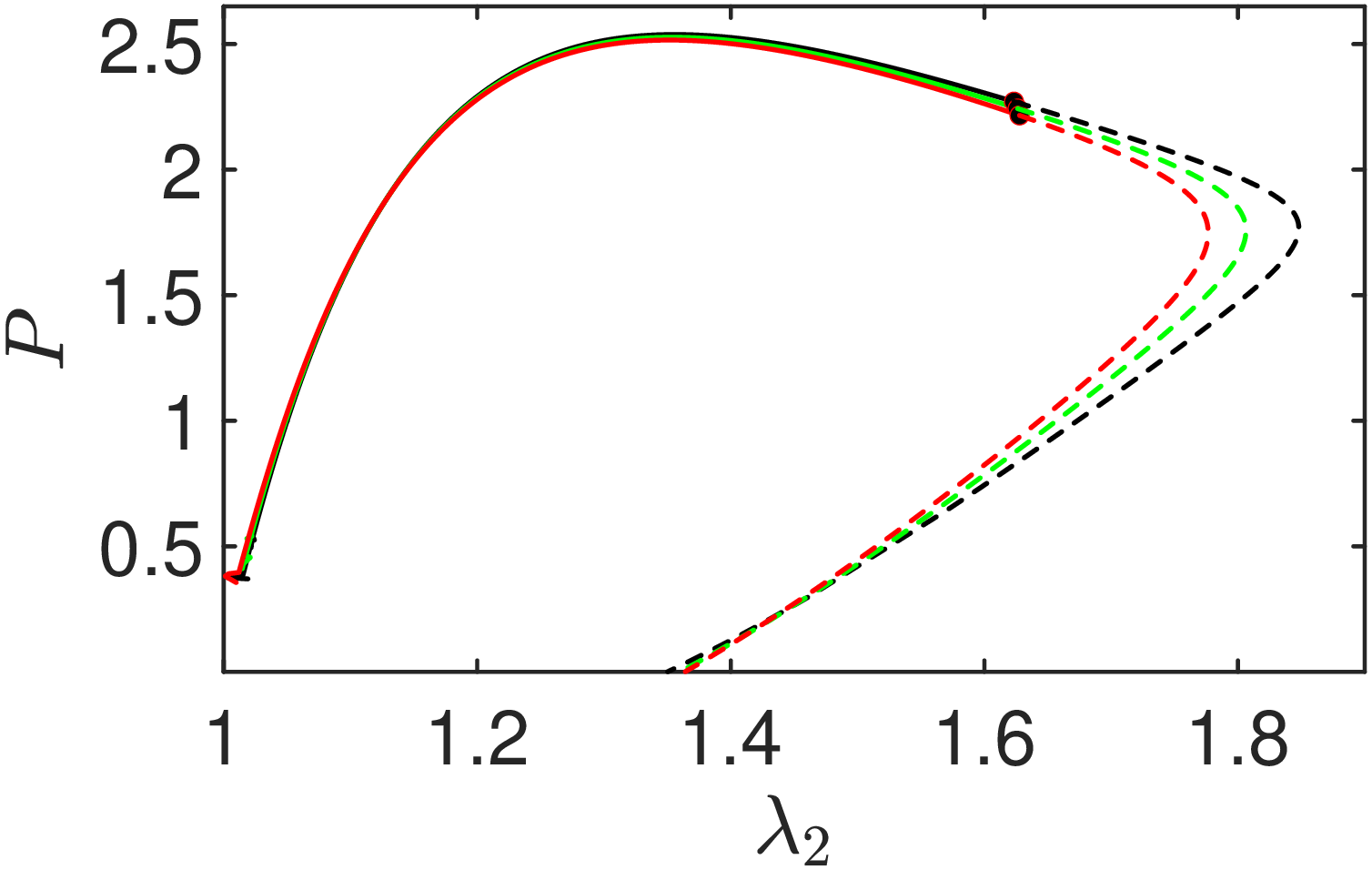}
\put(-3,62){\subcaptiontext*[3]{}}
\end{overpic}\hspace{0.1in}
\begin{overpic}[width=0.48\linewidth,trim=0in 0in 0in 0in]{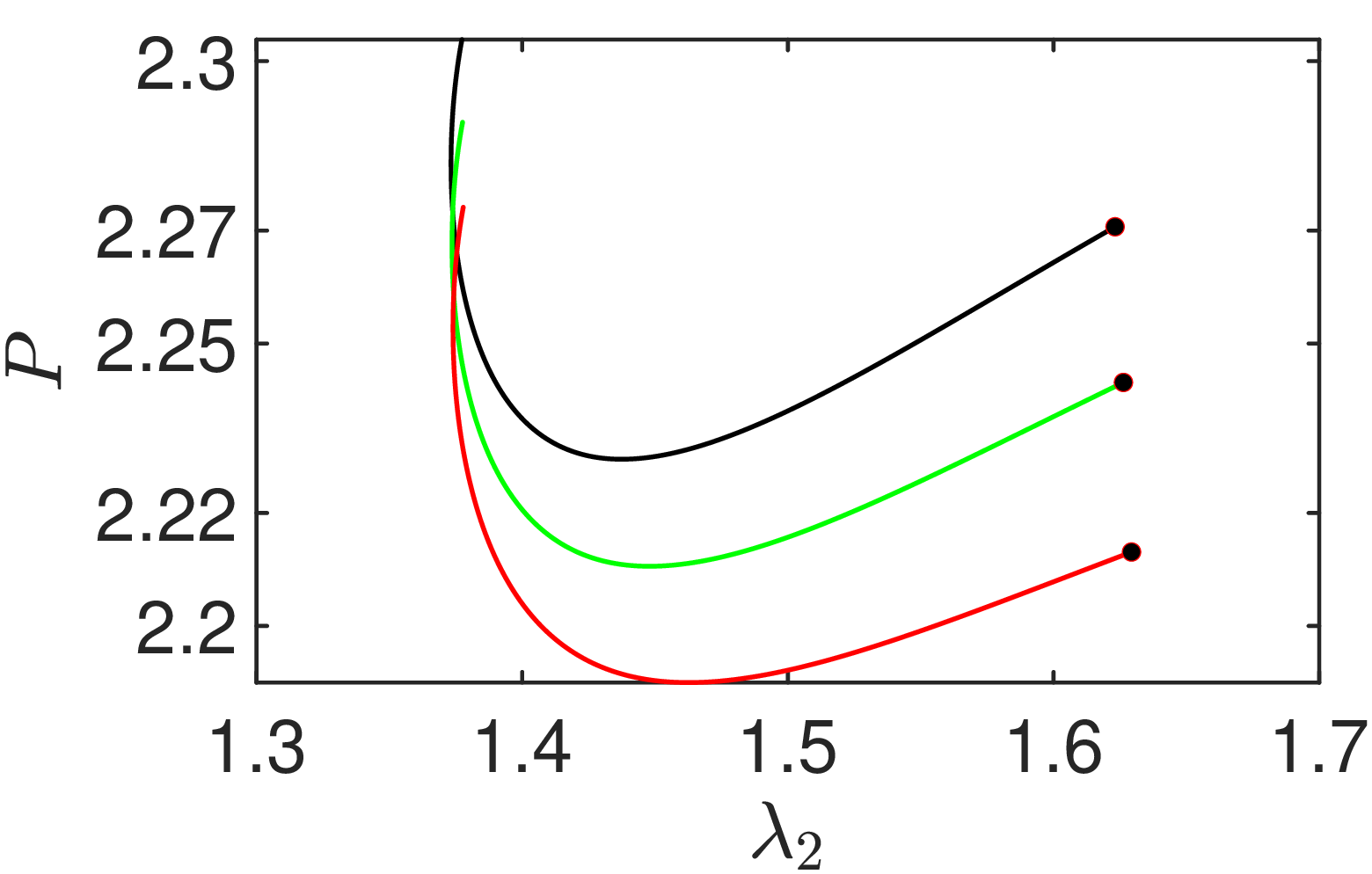}
\put(-3,62){\subcaptiontext*[4]{}}
\end{overpic}
\caption{ $P$ vs $\lambda_2$ at equator $\theta=\pi/2$ for two coil arrangement (a) SPOM1 plane of symmetry curve. (b) SPOM1 pear shaped curve. (c) TPOM1 plane of symmetry curve. (d) TPOM1 pear shaped curve. The solid line represents a stable equilibrium and the dotted line represents unstable equilibrium.} \label{fig:Magdiff_equator-l2_TC}
\end{figure}

Unlike the single coil arrangement, the SPOM1 material model exhibits single bifurcation point for the two coil arrangement, as illustrated in Fig.~\ref{fig:Magdiff_equator-l2_TC}(b). The critical pressure at which pear-shaped bifurcation occurs decreases as the magnetic field strength increases, as shown in Fig.~\ref{fig:Magdiff_equator-l2_TC}(b). The pressure vs. stretch curves for the primary and pear-shaped bifurcations in the TPOM1 model for the two-coil arrangement are similar to those observed in the single coil arrangement, as presented in Fig.~\ref{fig:Magdiff_equator-l2_TC}.


\begin{figure}[ht!]
\centering
\begin{overpic}[width=0.48\linewidth,trim=0in 0in 0in 0in]{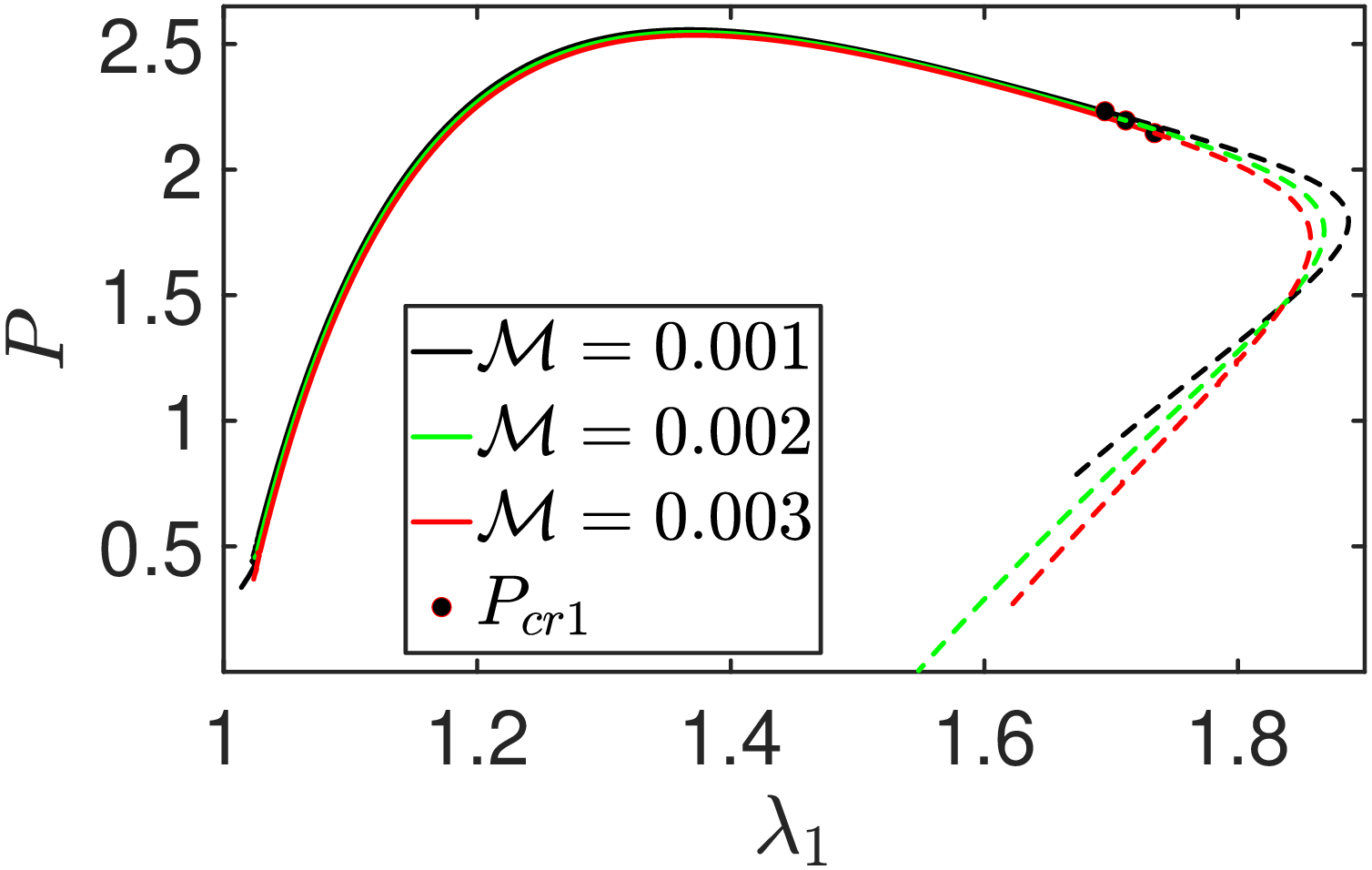}
\put(-3,62){\subcaptiontext*[1]{}}
\end{overpic}\hspace{0.1in}
\begin{overpic}[width=0.48\linewidth,trim=0in 0in 0in 0in]{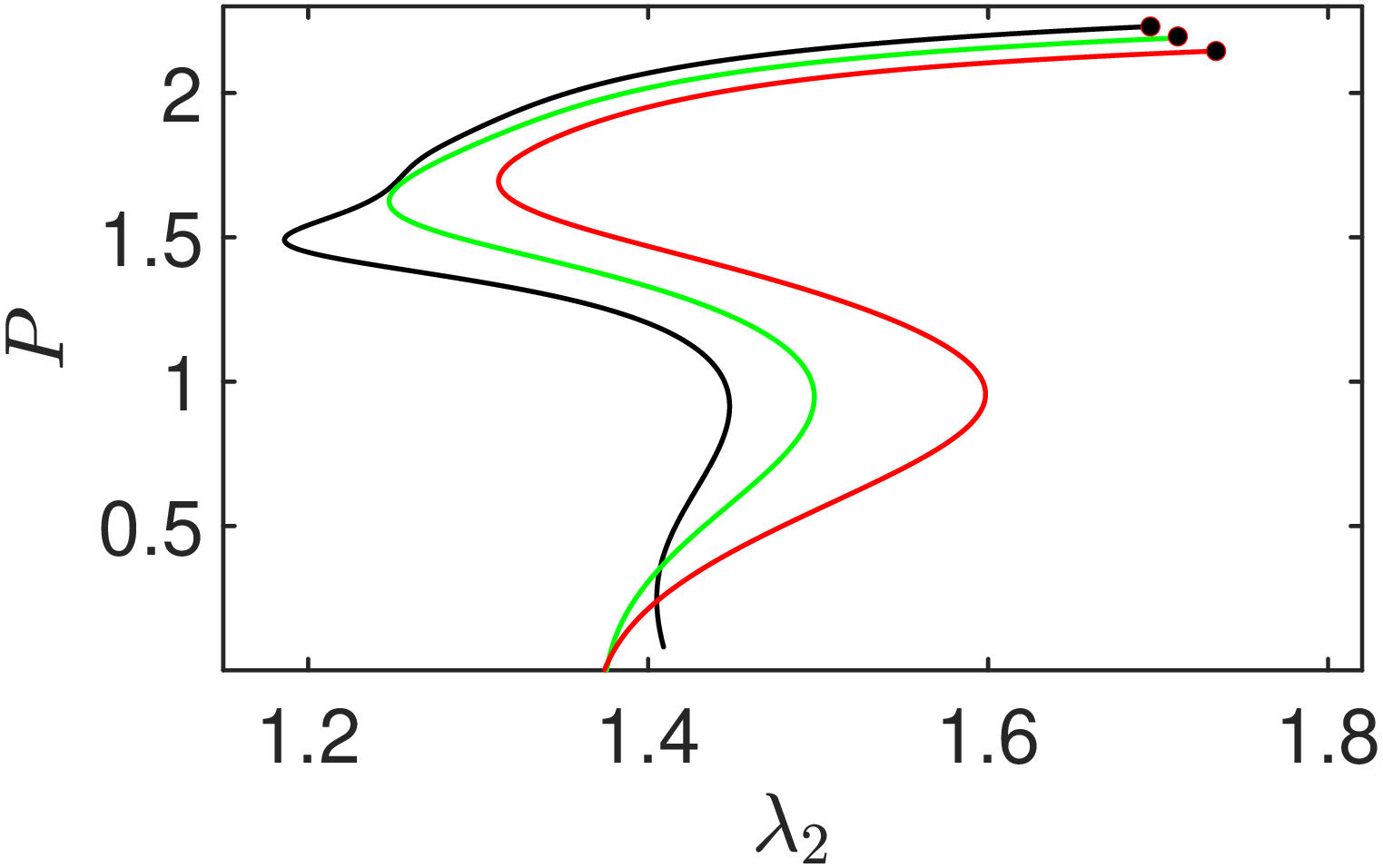}
\put(-3,62){\subcaptiontext*[2]{}}
\end{overpic} \\ 
\vspace{0.1in}
\begin{overpic}[width=0.48\linewidth,trim=0in 0in 0in 0in]{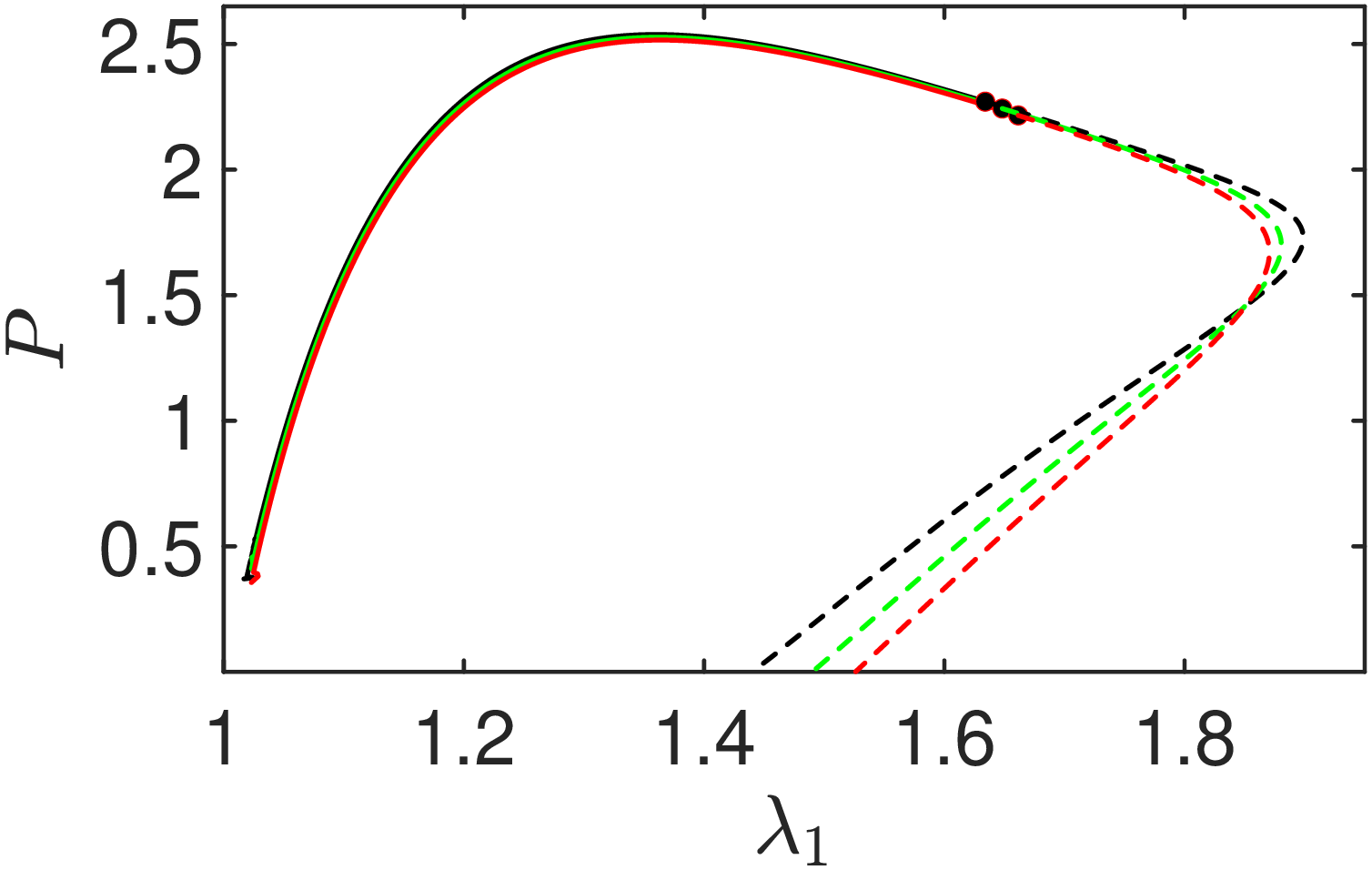}
\put(-3,62){\subcaptiontext*[3]{}}
\end{overpic}\hspace{0.1in}
\begin{overpic}[width=0.48\linewidth,trim=0in 0in 0in 0in]{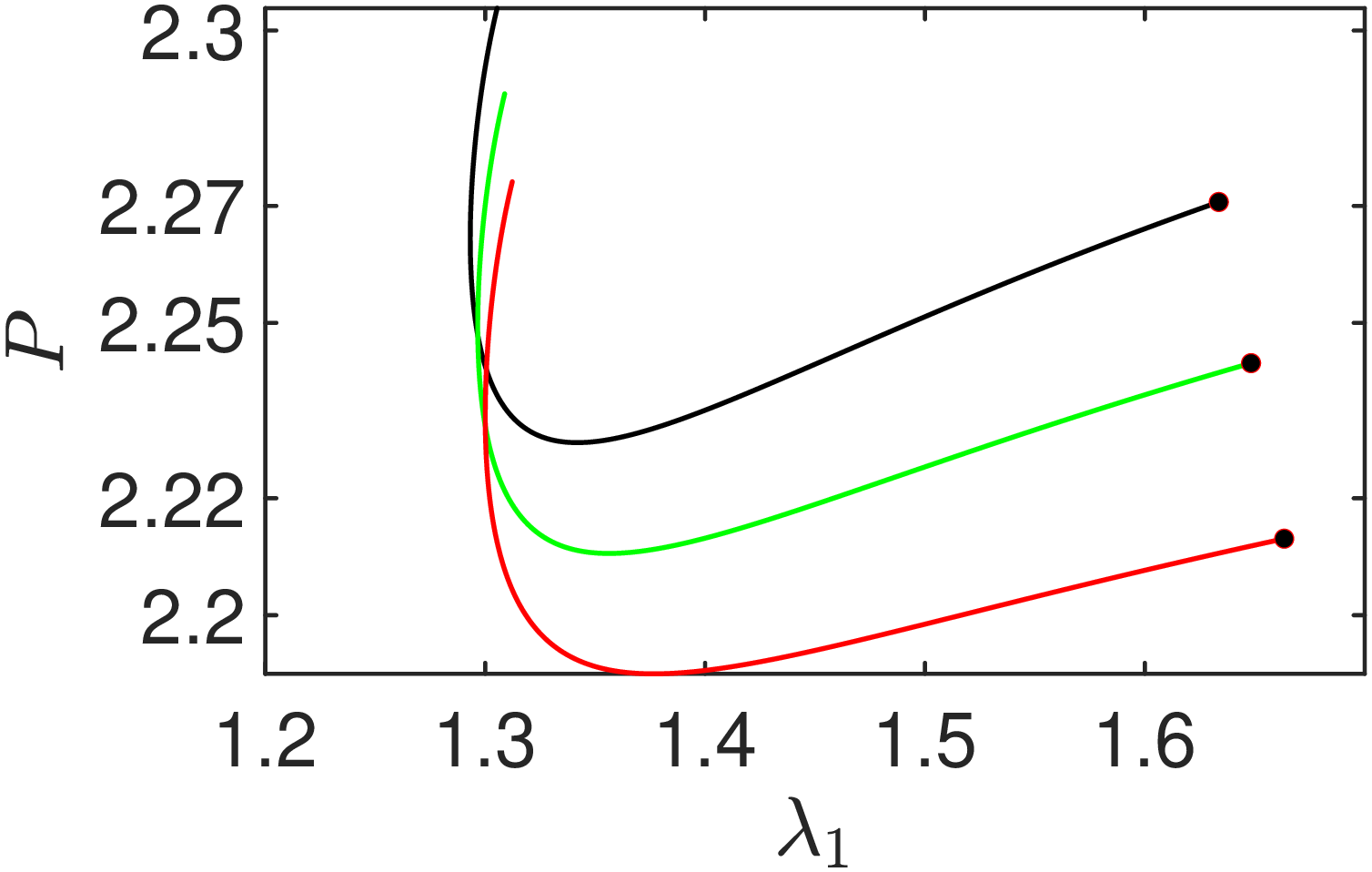}
\put(-3,62){\subcaptiontext*[4]{}}
\end{overpic}
\caption{ $P$ vs $\lambda_1$ at equator $\theta=\pi/2$ for two coil arrangement (a) SPOM1 plane of symmetry curve. (b) SPOM1 pear--shaped curve. (c) TPOM1 plane of symmetry curve. (d) TPOM1 pear--shaped curve. The solid line represents a stable equilibrium and the dotted line represents unstable equilibrium.} 
\label{fig:Magdiff_equator-l1_TC}
\end{figure}

Figs.~\ref{fig:Magdiff_equator-l2_TC} and \ref{fig:Magdiff_equator-l1_TC} show $P$ versus $\lambda_2$ and $\lambda_1$, respectively, at the equator for both axisymmetric and plane of symmetry breaking cases for two coil arrangement.  

\subsection{Deformed Configuration}


\begin{figure}[ht!]
\centering
\begin{overpic}[width=0.35\linewidth,trim=0in 0in 0in 0in]{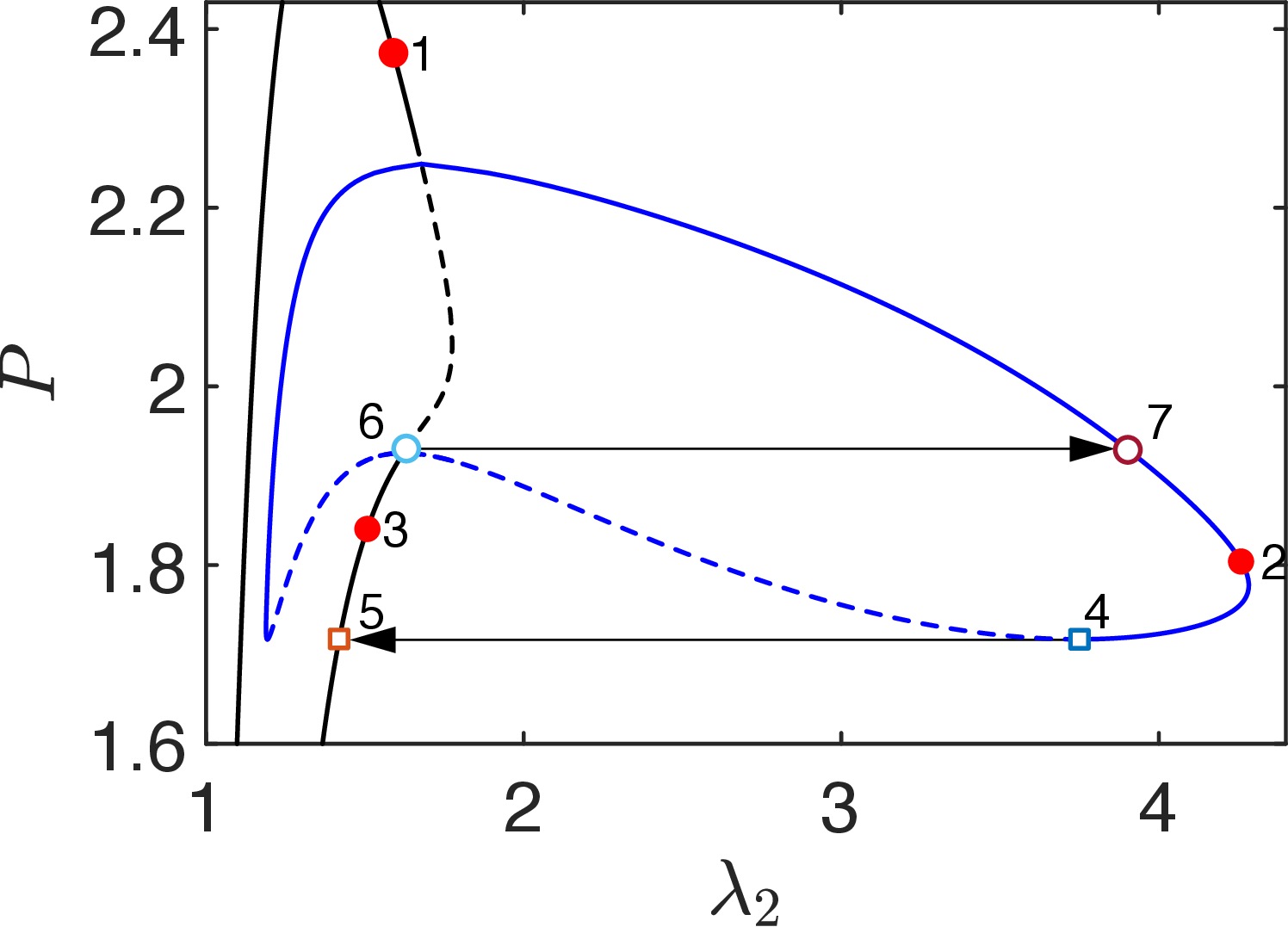}
\put(-4,72){\subcaptiontext*[1]{}}
\end{overpic}\hspace{0.2in}
\begin{overpic}[width=0.35\linewidth,trim=0in 0in 0in 0in]{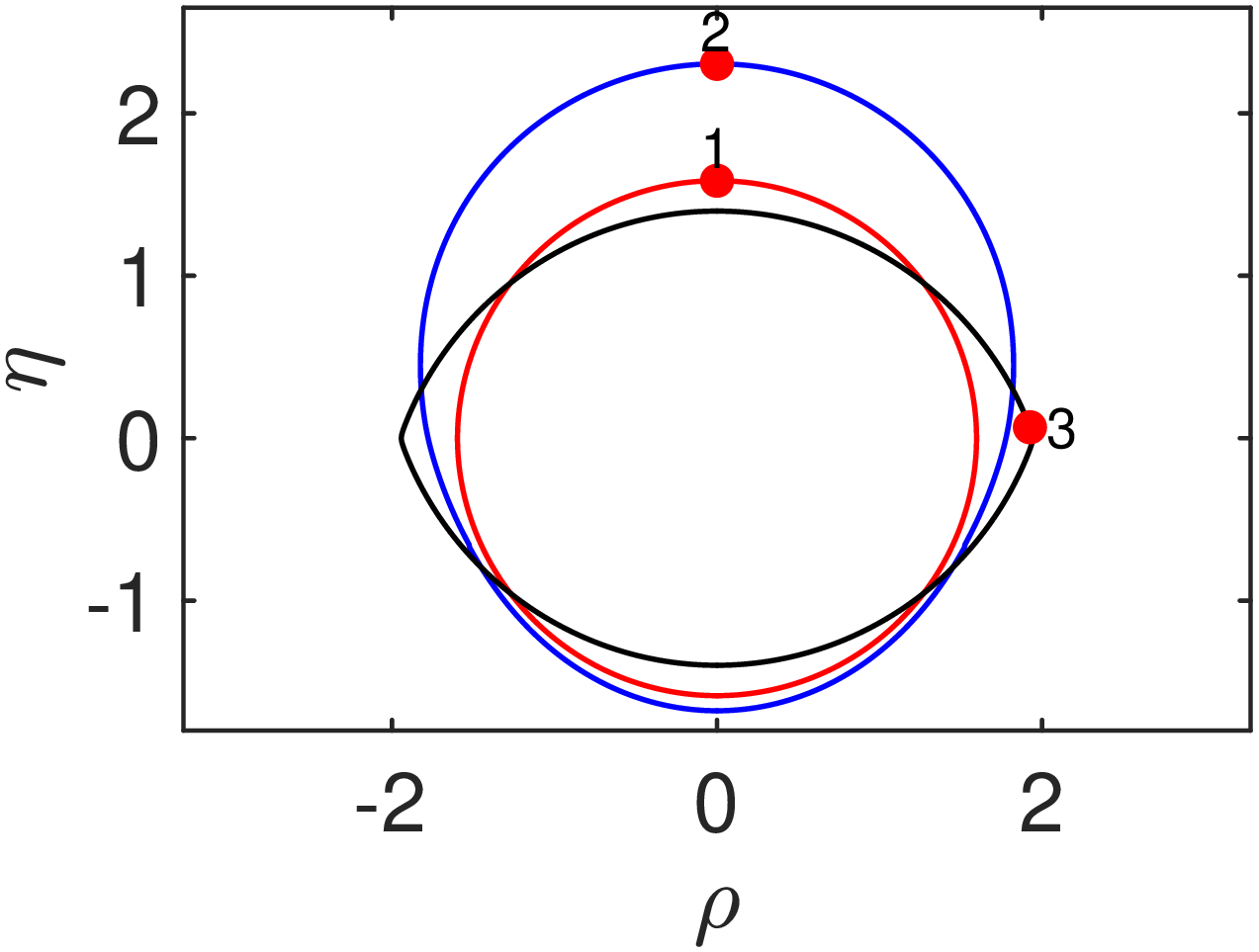}
\put(-4,72){\subcaptiontext*[2]{}}
\end{overpic} \\ 
\vspace{0.1in}
\begin{overpic}[width=0.35\linewidth,trim=0in 0in 0in 0in]{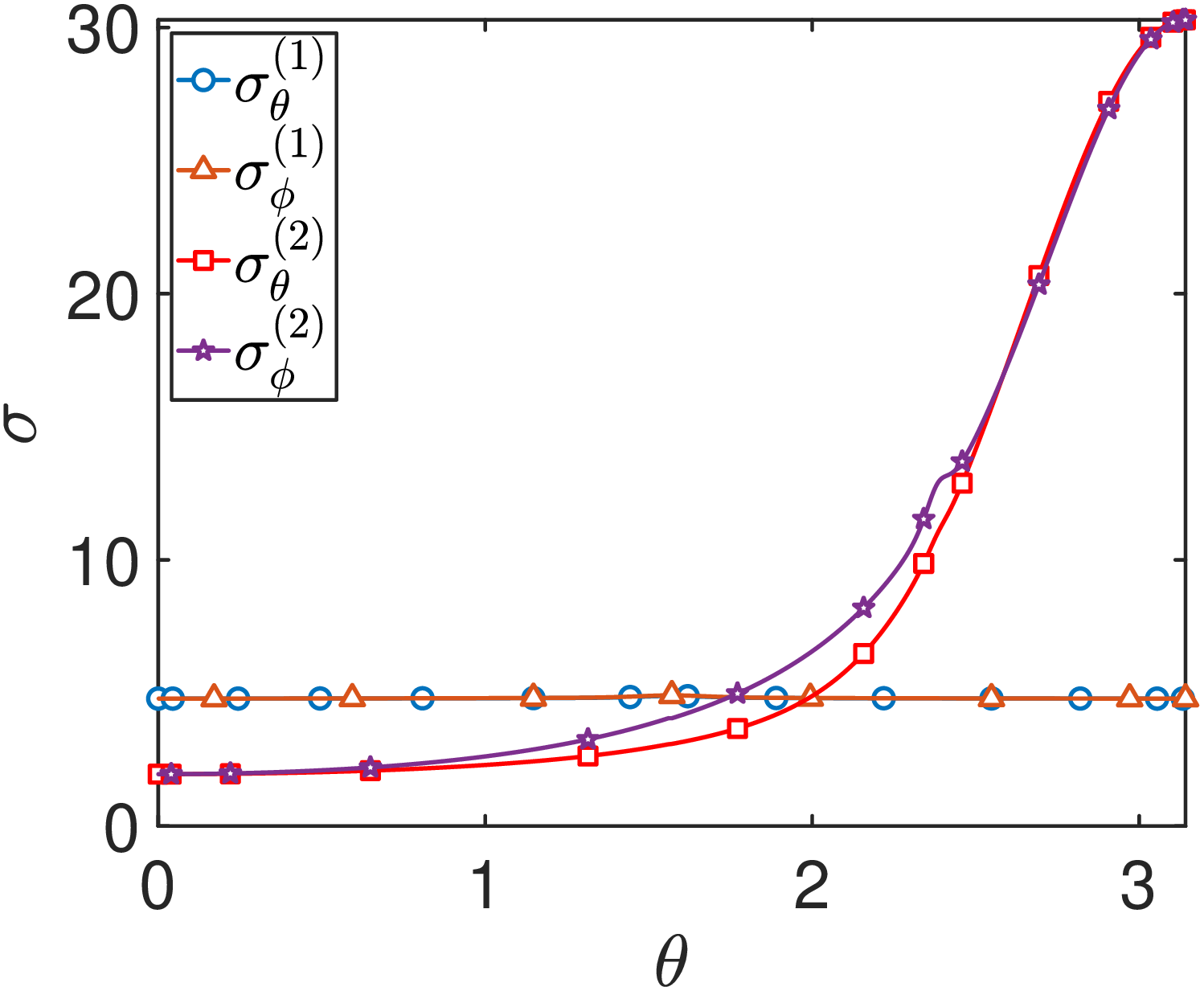}
\put(-4,80){\subcaptiontext*[3]{}}
\end{overpic}\hspace{0.2in}
\begin{overpic}[width=0.35\linewidth,trim=0in 0in 0in 0in]{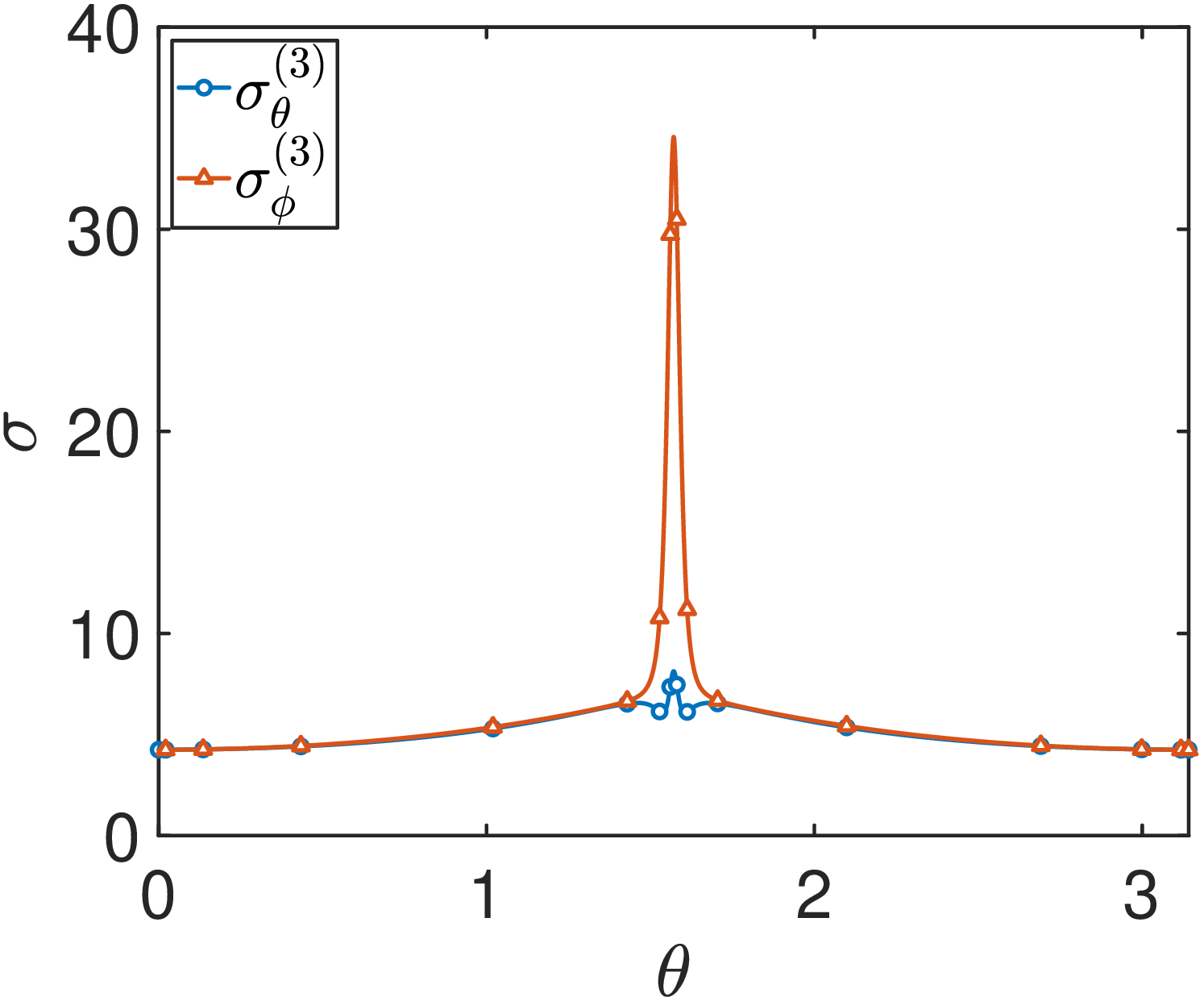}
\put(-4,80){\subcaptiontext*[4]{}}
\end{overpic}

\caption{ (a) $P$ vs $\lambda_2$ equilibrium curves for SPOM1 model and equilibrium configurations (b) shapes of equilibrium configurations $1$ (axisymmetric) $2$ (pear--shaped) $3$ (axisymmetric after turning point) (c) Stresses $\sigma_{\theta}$ and $\sigma_{\phi}$ vs $\theta$ for the configuration $1$ and $2$. (d) Stresses $\sigma_{\theta}$ and $\sigma_{\phi}$ vs $\theta$ for the configuration $3$. }
     \label{fig:og_shape_stress}
\end{figure}



\begin{figure}[ht!]
\centering
\begin{overpic}[width=0.325\linewidth,trim=0in 0in 0in 0in]{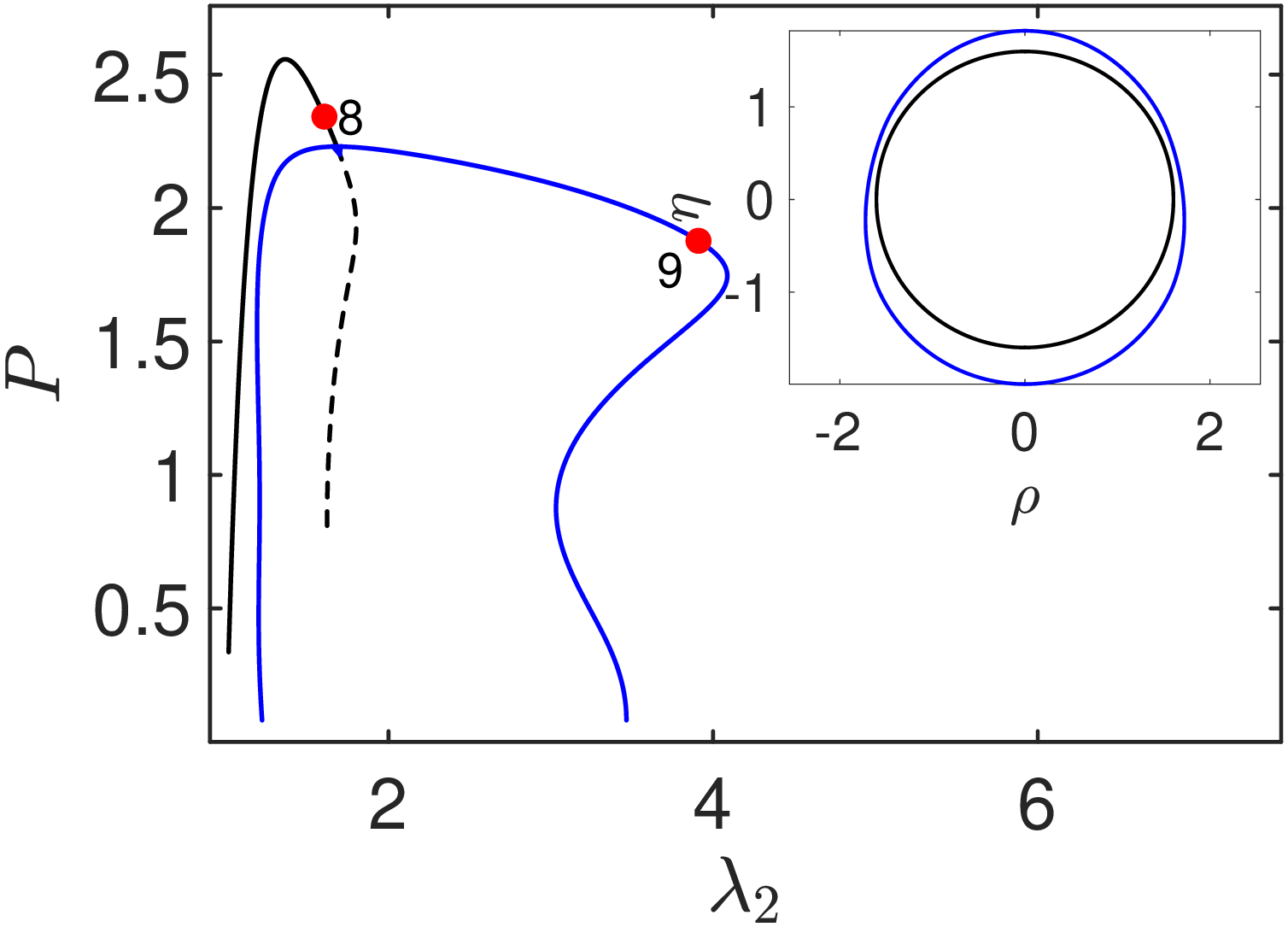}
\put(-2,70){\subcaptiontext*[1]{}}
\end{overpic}
\begin{overpic}[width=0.325\linewidth,trim=0in 0in 0in 0in]{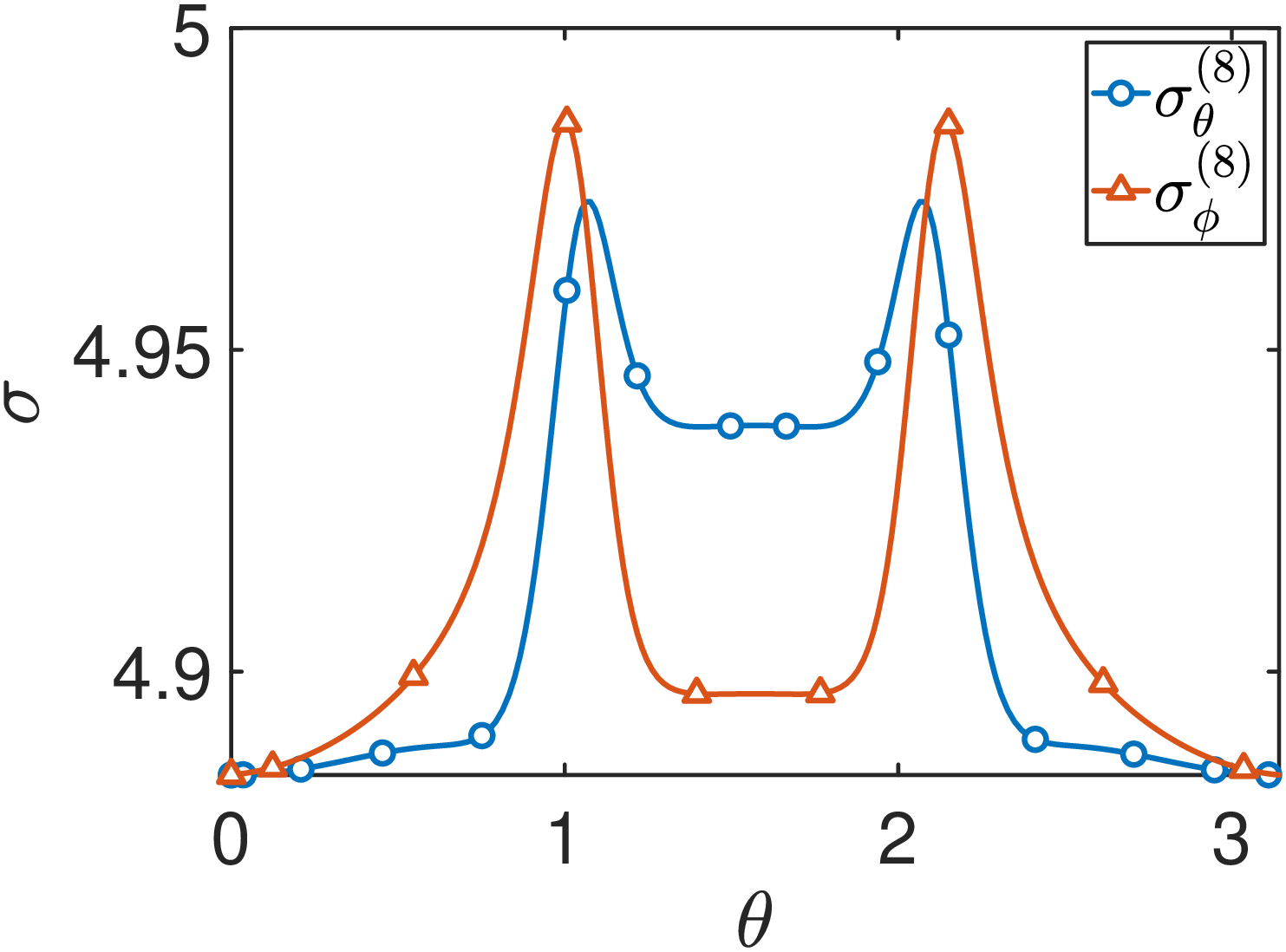}
\put(-2,70){\subcaptiontext*[2]{}}
\end{overpic} 
\begin{overpic}[width=0.325\linewidth,trim=0in 0in 0in 0in]{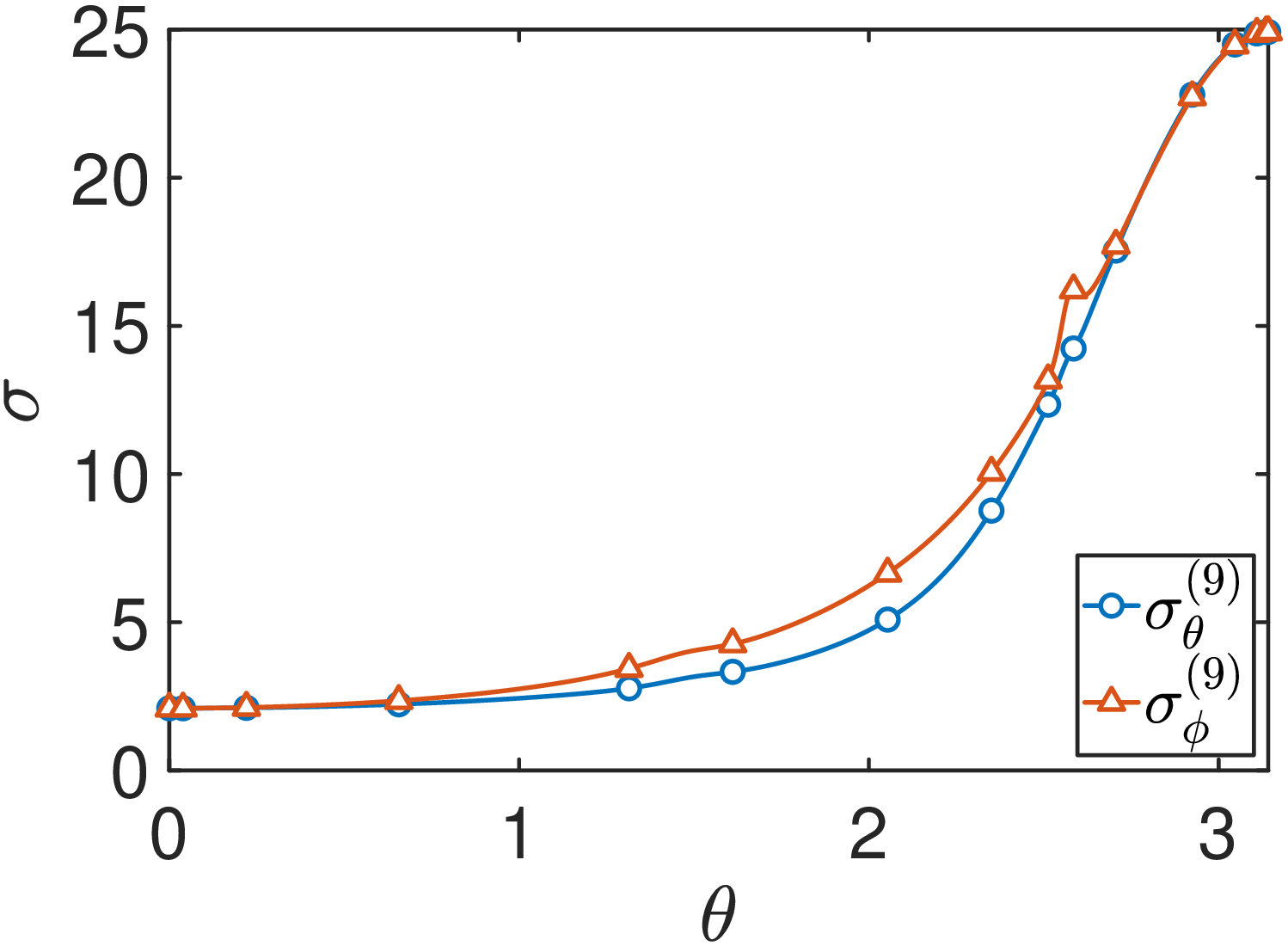}
\put(-2,72){\subcaptiontext*[3]{}}
\end{overpic}

\caption{ (a) $P$ vs $\lambda_2$ equilibrium curves with a two coil arrangement for SPOM1 model and shapes for configurations $8$ and $9$. (b) Stresses $\sigma_{\theta}$ and $\sigma_{\phi}$ vs $\theta$ for the configuration $8$. (c) Stresses $\sigma_{\theta}$ and $\sigma_{\phi}$ vs $\theta$ for the configuration $9$.}
\label{fig:og_shape_stress_2c}
\end{figure}

\begin{figure}[ht!]
\centering
\begin{overpic}[width=0.325\linewidth,trim=0in 0in 0in 0in]{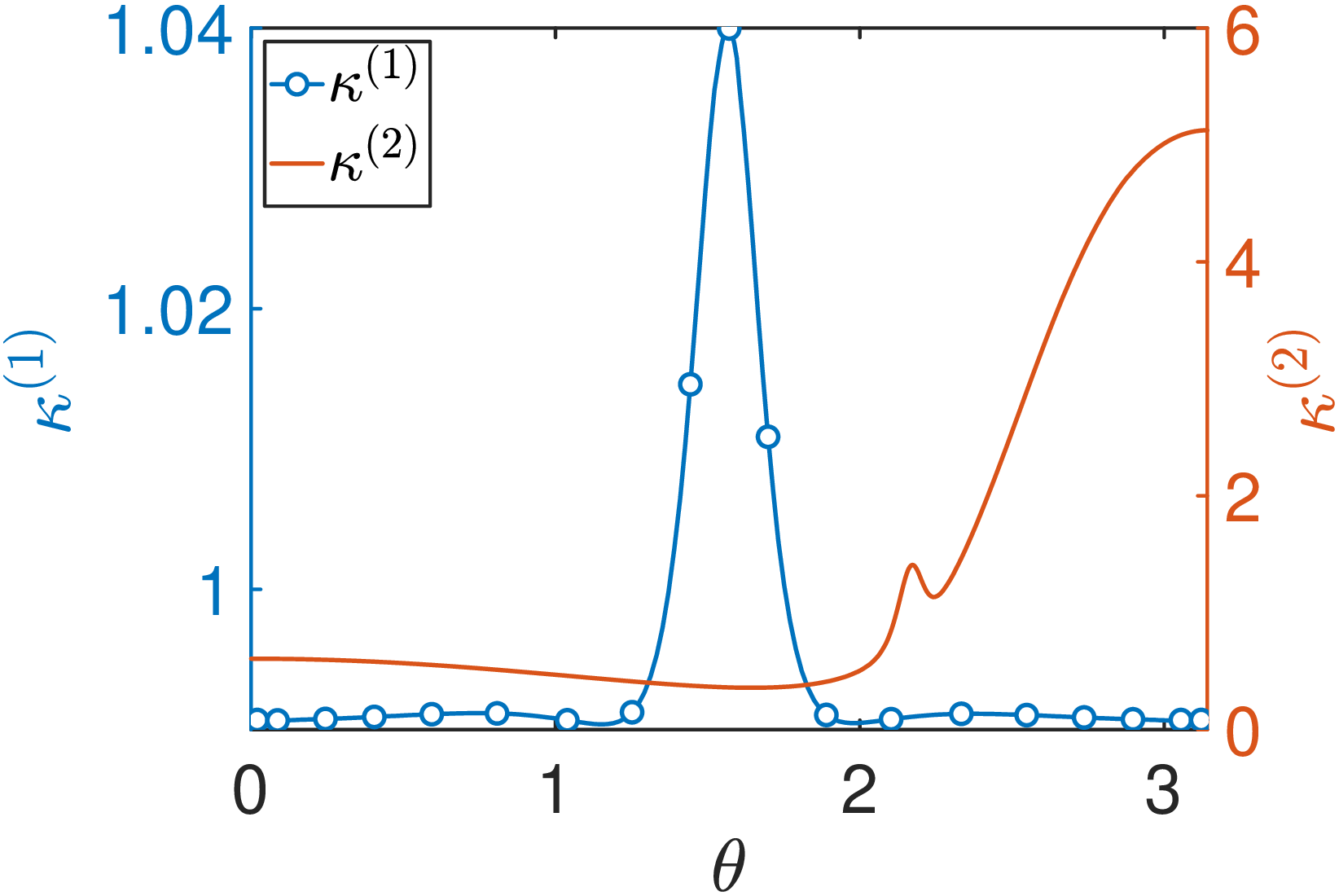}
\put(-2,67){\subcaptiontext*[1]{}}
\end{overpic}
\begin{overpic}[width=0.31\linewidth,trim=0in 0in 0in 0in]{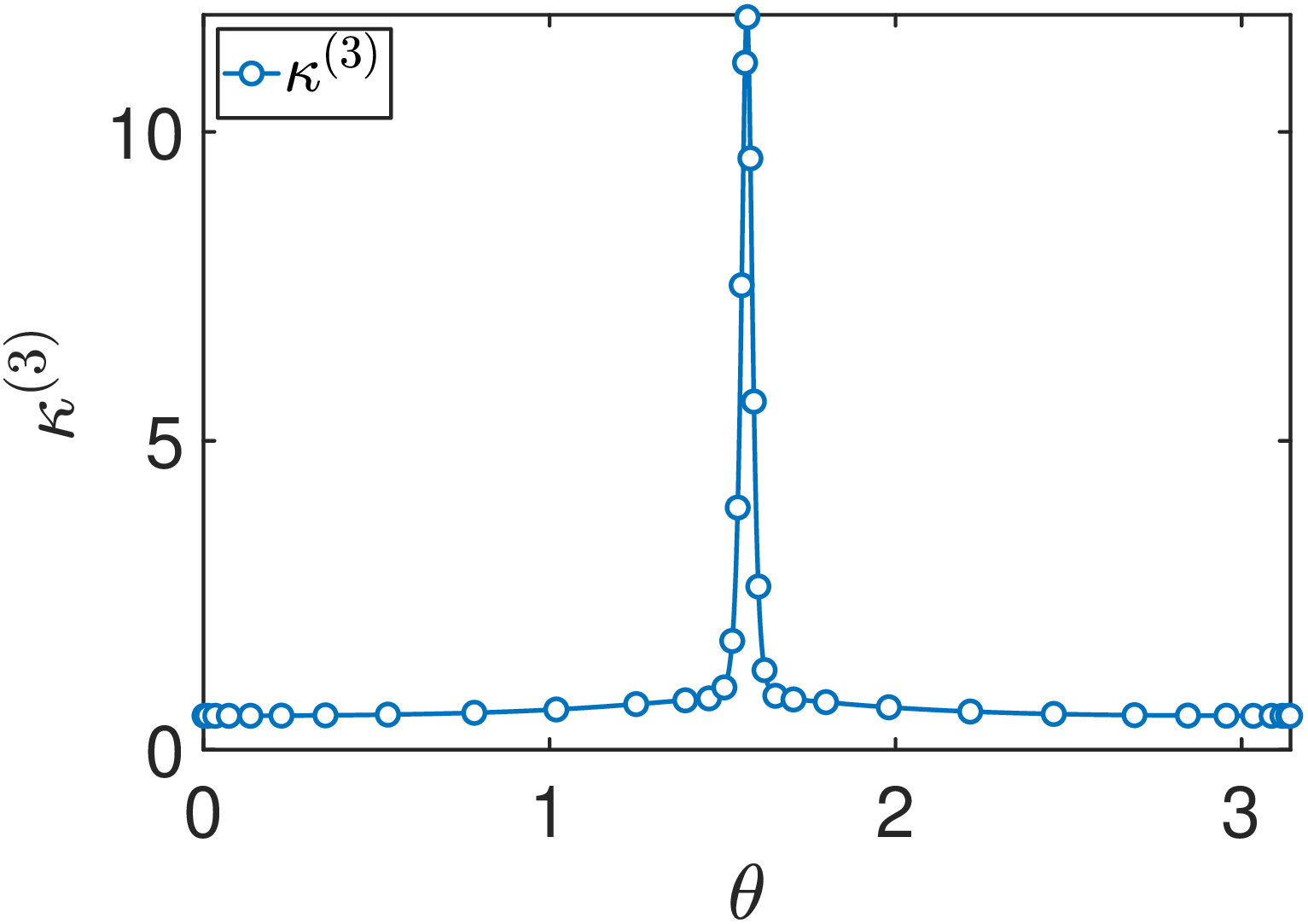}
\put(-2,67){\subcaptiontext*[2]{}}
\end{overpic} 
\begin{overpic}[width=0.325\linewidth,trim=0in 0in 0in 0in]{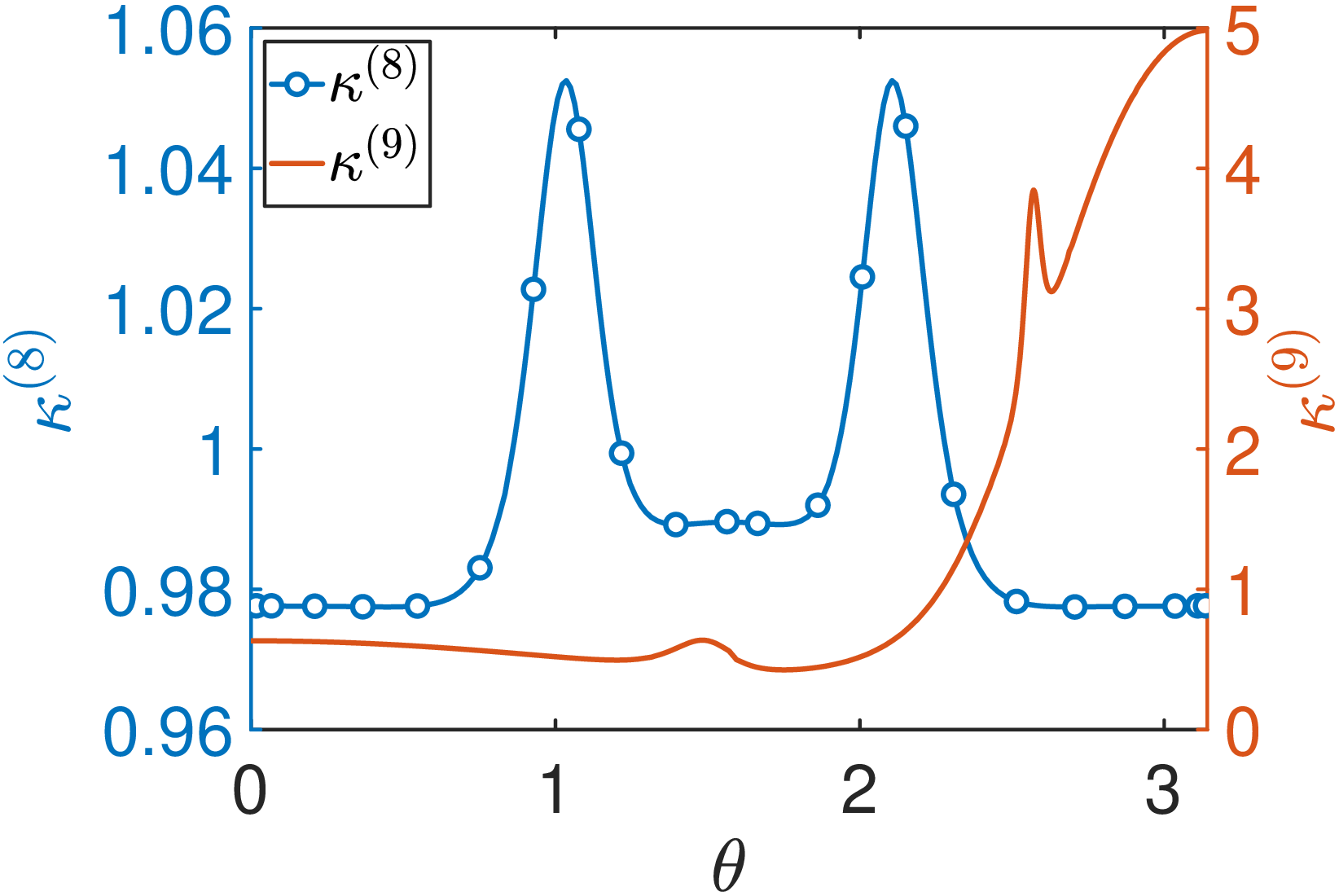}
\put(-2,67){\subcaptiontext*[3]{}}
\end{overpic}

\caption{Gaussian curvature $\kappa$ vs $\theta$ (a) for configuration  1 and 2  shown in Fig.~\ref{fig:og_shape_stress}(a). (b) for configuration 3 shown in Fig.~\ref{fig:og_shape_stress}(a). (c) for configuration 8 and 9 shown in Fig.~\ref{fig:og_shape_stress_2c}(a).}
    \label{fig:og_G_curv}
\end{figure}

Fig.~\ref{fig:og_shape_stress}(a) shows the equilibrium configurations on pressure stretch curve $1:$ (axisymmetric before turning point), $2:$ (pear--shaped), $3:$ (axisymmetric after turning point). Fig.~\ref{fig:og_shape_stress}(b) illustrates the deformed shapes for equilibrium configurations $1$, $2$ and $3$. The deformed shape along the axisymmetric curve (depicted by equilibrium configuration $1$) maintains axis symmetry and symmetry across the $Y^1-Y^2$ plane. In contrast, the deformed shape along the plane-symmetry-breaking curve retains axis symmetry but lacks symmetry across the $Y^1-Y^2$ plane (depicted by equilibrium configuration $2$) and resembles a top-heavy pear shape. For the shape after the turning point (depicted by equilibrium configuration $3$), a sharp change in the balloon's curvature occurs near the equator ($\theta = \pi/2$) as shown in Fig.~\ref{fig:og_G_curv}(b). This sharp change is caused by two factors: (1) the membrane near the equator is positioned very close to the current-carrying coil, resulting in a stronger magnetic field; and (2) there is a sharp variation in the radial component of the applied magnetic field around $\theta = \pi/2$. This abrupt curvature change near the equator is also reflected in the stress vs. $\theta$ plots in Fig.~\ref{fig:og_shape_stress}(d), where the stress component $\sigma_{\phi}$ shows a pronounced increase near the theta value where membrane comes very close to current carrying coil. For equilibrium points $1$ and $2$, change of curvature with respect to $\theta$ is shown in Fig.~\ref{fig:og_G_curv}(a).     

\begin{figure}[ht!]
\centering
\begin{overpic}[width=0.325\linewidth,trim=0in 0in 0in 0in]{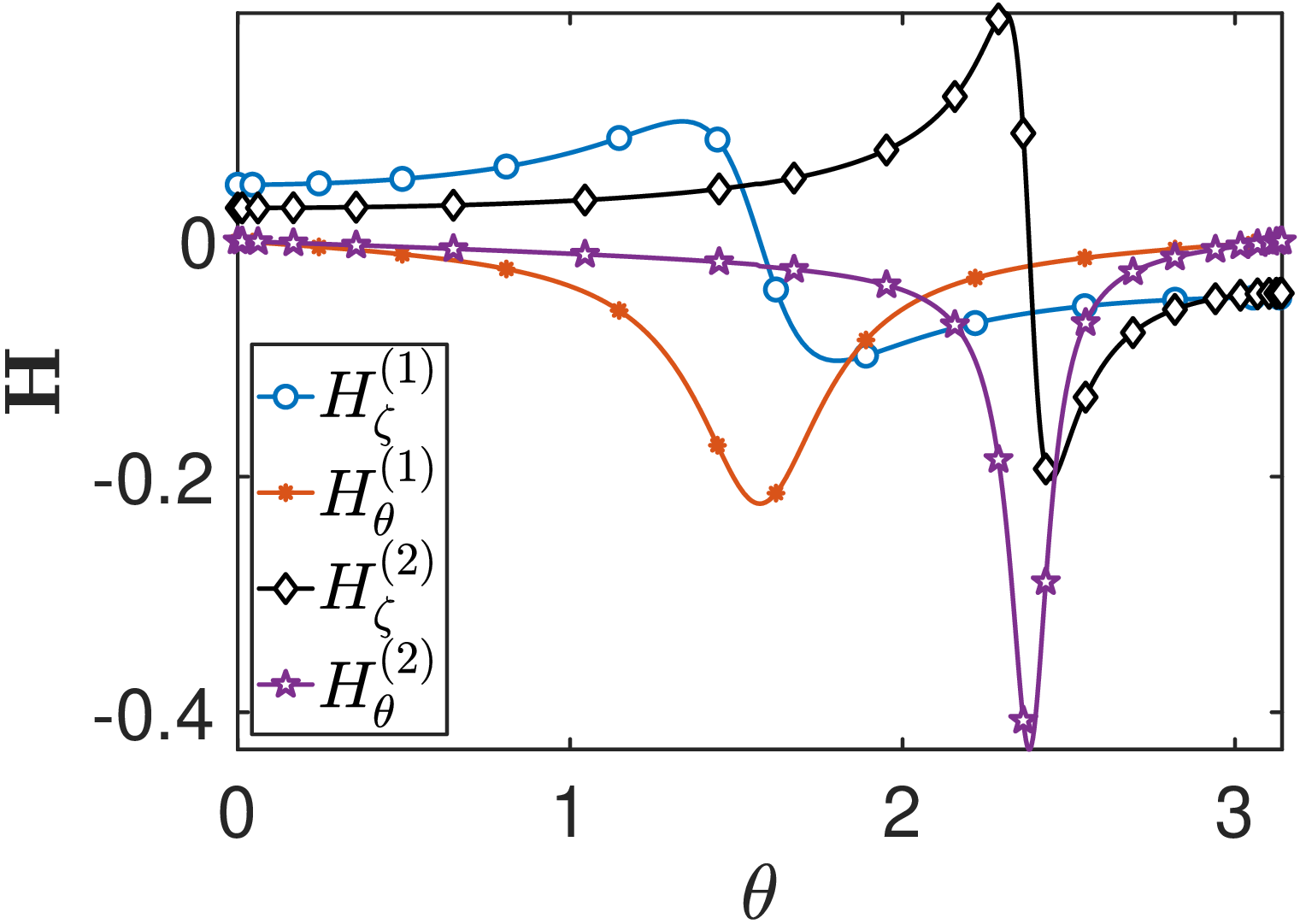}
\put(0,67){\subcaptiontext*[1]{}}
\end{overpic}
\begin{overpic}[width=0.325\linewidth,trim=0in 0in 0in 0in]{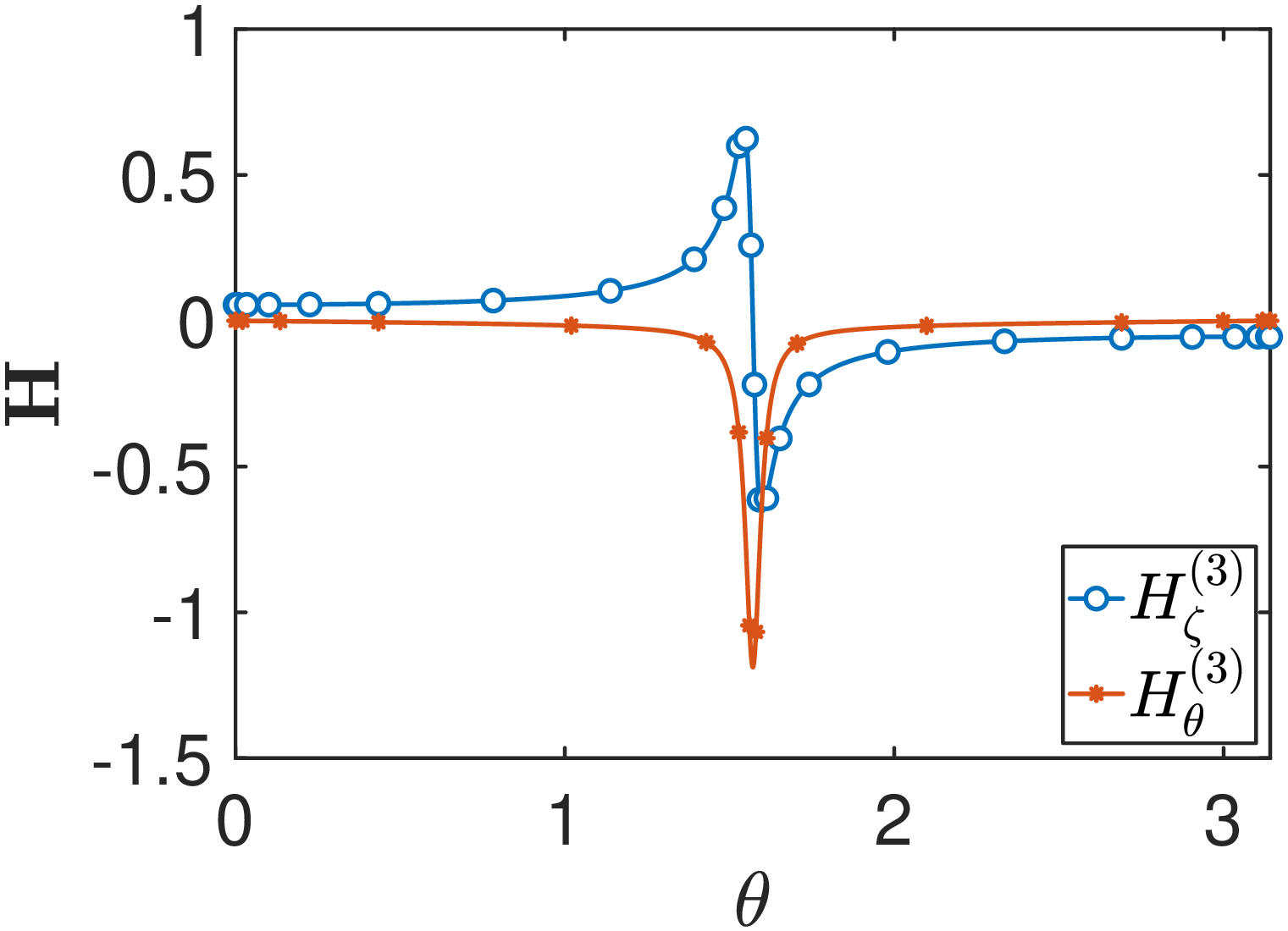}
\put(0,67){\subcaptiontext*[2]{}}
\end{overpic} 
\begin{overpic}[width=0.325\linewidth,trim=0in 0in 0in 0in]{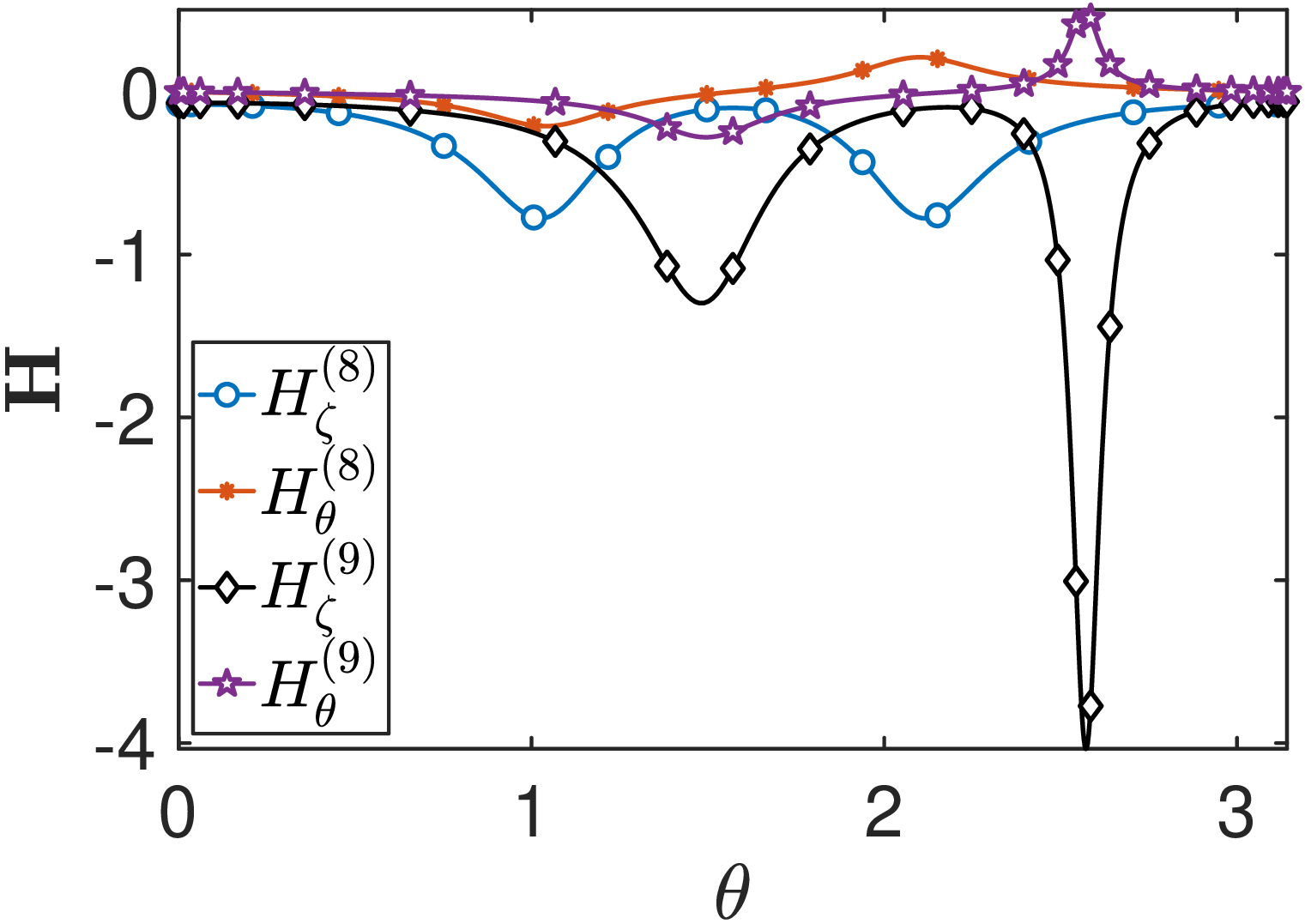}
\put(0,67){\subcaptiontext*[3]{}}
\end{overpic}

\caption{ Applied magnetic field components $H_{\zeta}$ and $H_{\theta}$ vs $\theta$ on the deformed membrane. (a) Configuration 1 and 2 shown in Fig.~\ref{fig:og_shape_stress}(a),  (b) Configuration 3, shown in Fig.~\ref{fig:og_shape_stress}(a). (c) Configurations 8 and 9, shown in Fig.~\ref{fig:og_shape_stress_2c}(a).}
     \label{fig:og_appMag}
\end{figure}

Figure \ref{fig:og_shape_stress_2c}(a) shows the deformed shapes of the axisymmetric equilibrium configuration 8 and the pear-shaped equilibrium configuration 9 for the two-coil arrangement. The stresses, $\sigma_\theta$ and $\sigma_\phi$ for configurations 8 and 9, are shown in Fig.~\ref{fig:og_shape_stress_2c}(b) and \ref{fig:og_shape_stress_2c}(c), respectively. The applied magnetic field and curvature for configurations 8 and 9 are presented in Fig.~\ref{fig:og_appMag}(c) and Fig.~\ref{fig:og_G_curv}(c), respectively. 

Equilibrium configurations for pressures below point 9 in Fig.~\ref{fig:og_shape_stress_2c}(a) exhibit sharp changes in curvature, similar to configuration 3 in Fig.~\ref{fig:og_shape_stress}(b). This phenomenon occurs as the balloon membrane approaches very close to the two coils located at heights $h=1$ and $h=-1$, where sharp variations in the radial component of the magnetic field are present. Consequently, the membrane experiences substantial stress concentrations, orders of magnitude higher than the ultimate tensile strength reported of magnetoelastic polymers or soft biological membranes (\cite{raz2019tensile,UTS_magnetic_polymer,UTS2_magnetic_polymer}).

\subsection{Stability} 
\cite{CHEN_IJNM1991} demonstrated the existence of an isola bifurcation in pressurized spherical membranes, forming a closed loop of non-spherical solutions during inflation. They observed that when pear-shaped and spherical configurations coexist, the pear-shaped configuration exhibits  lower strain energy. Furthermore, the spherical solution branch undergoes symmetry-breaking via supercritical pitchfork bifurcation and reconnects to the symmetric branch through reverse supercritical pitchfork bifurcation, thereby forming the isola bifurcation.

In this work, a supercritical pitchfork bifurcation occurs during the transition from a stable axisymmetric ellipsoidal configuration to a stable pear-shaped configuration. The isolated pear-shaped bifurcation curve, representing a closed loop of symmetry-breaking solutions, reconnects to the symmetry-preserving ellipsoidal shape curve through a reverse subcritical pitchfork bifurcation. Due to the subcritical nature of bifurcation, where higher order non-linearity play an important role, part of the pear-shaped solutions is found to be unstable.

In the absence of magnetic field, isola bifurcation is observed in the SPOM model but is absent in the TPOM model, as shown in Fig.~\ref{fig:M0}. When magnetic field is applied using a single coil, the closed loop of the isolated pear-shaped curve in the SPOM model shrinks toward its center (Fig.~\ref{fig:Magdiff}(b)) and eventually disappears at a critical magnetic field strength (not shown), similar to the behavior of isola bifurcation in the non-magnetic case. In the TPOM model (Fig.~\ref{fig:Magdiff}(d)), as the magnetic field increases, bifurcation occurs at progressively lower pressures and ultimately vanishes at a critical magnetic field strength (not shown).

Interestingly, when the magnetic field is applied using two coils, the isolated loop of pear-shaped solutions is absent in both the SPOM and TPOM models, as shown in Fig.~\ref{fig:Magdiff_pole-l2_TC}. However, with a single coil, the isolated loop of pear-shaped solutions appears but only persists below a certain critical magnetic field strength. 

As previously mentioned, stable pear-shaped and unstable axisymmetric configurations are possible only when the Hessian matrix in \ref{stability condition} is not positive definite, i.e., at least one of the eigenvalues of the matrix must cross zero and become negative. Let $\mathbf{v} = \Lambda \mathbf{u}$ represent a perturbation vector, where $\mathbf{u}$ is the unit eigenvector associated with the negative eigenvalue of the Hessian matrix and $\Lambda$ is the unknown (small) magnitude of the perturbation vector. Determining $\Lambda$ will fully characterize the pear-shaped configuration. The perturbed potential energy can then be expressed as, 
\begin{equation}
    \tilde{E}(\Lambda)  =  E + \Lambda E_1 + \Lambda^2 E_2 + \Lambda^3 E_3 +  \Lambda^4 E_4  + O(\Lambda^5).   
    \label{Energy_landscape}
\end{equation}
For equilibrium, we must have, 
\begin{equation}
    \frac{\partial \tilde{E}}{\partial \Lambda} = 0.  
    \label{Lambda sol}
\end{equation}
Solution of \ref{Lambda sol} gives the value of $\Lambda$. 

\begin{figure}[ht!]
\centering
\begin{overpic}[width=0.325\linewidth,trim=0in 0in 0in 0in]{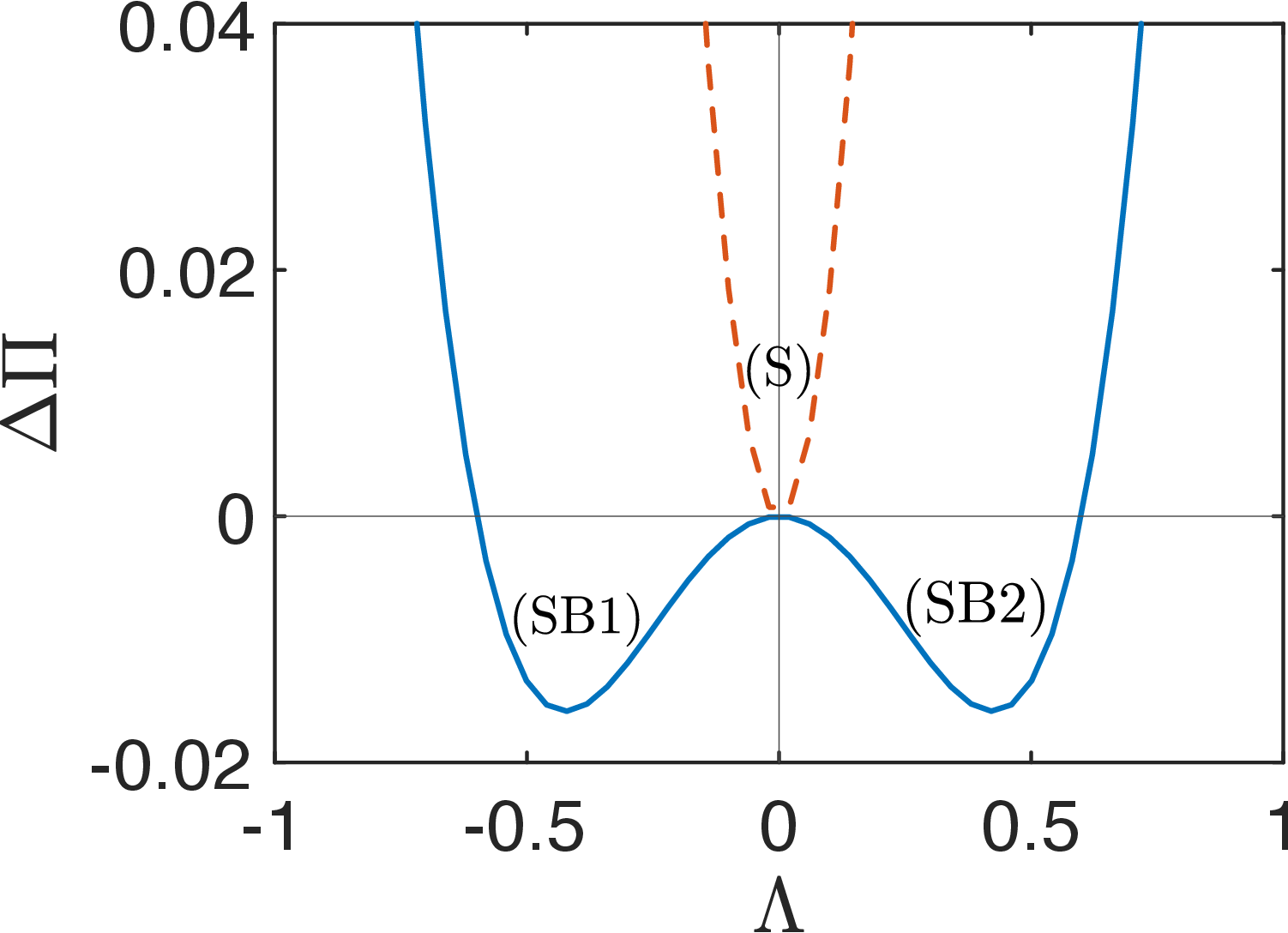}
\put(-2,67){\subcaptiontext*[1]{}}
\end{overpic}
\begin{overpic}[width=0.325\linewidth,trim=0in 0in 0in 0in]{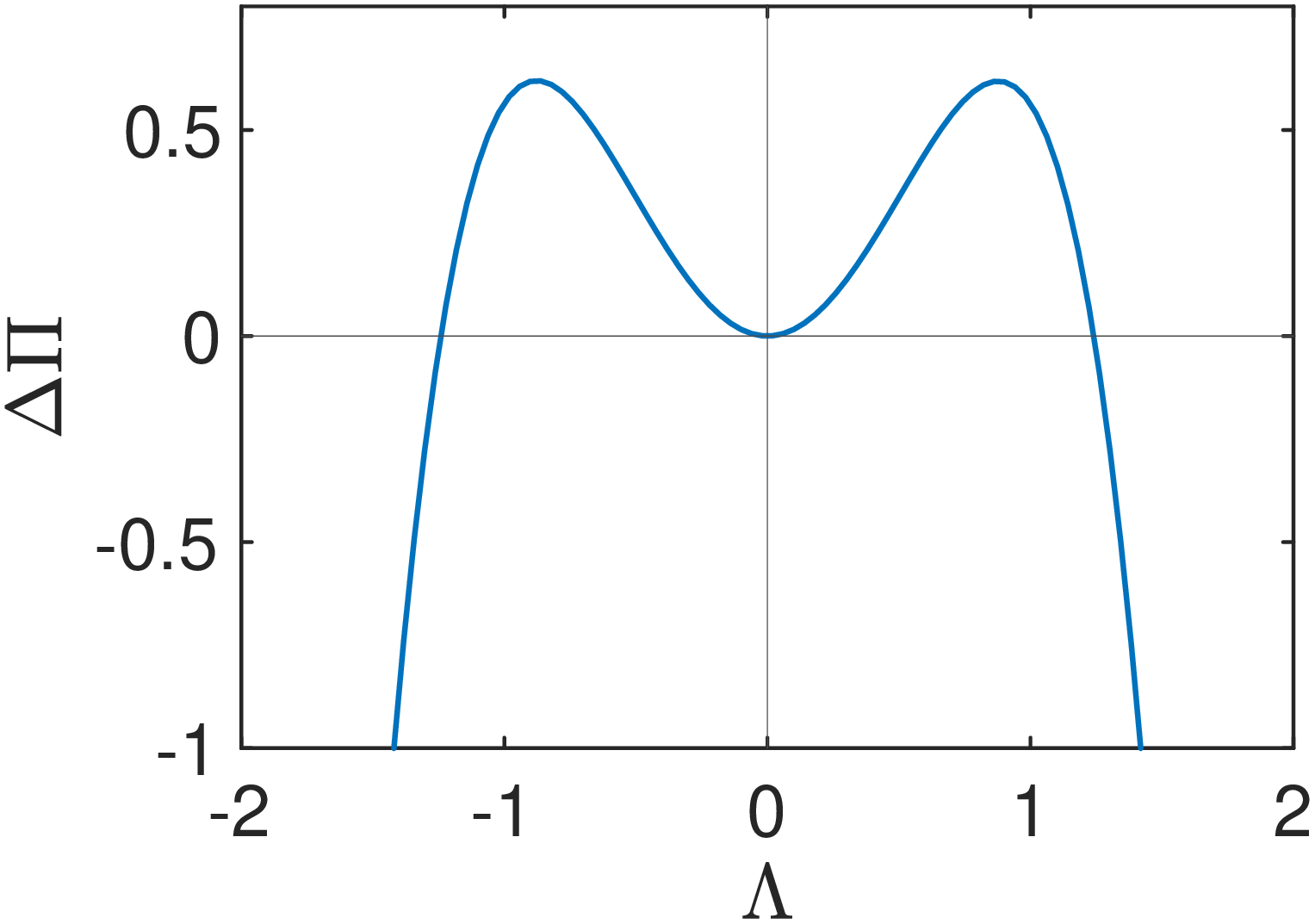}
\put(0,67){\subcaptiontext*[2]{}} 
\end{overpic} 
\begin{overpic}[width=0.3\linewidth,trim=0in 0in 0in 0in]{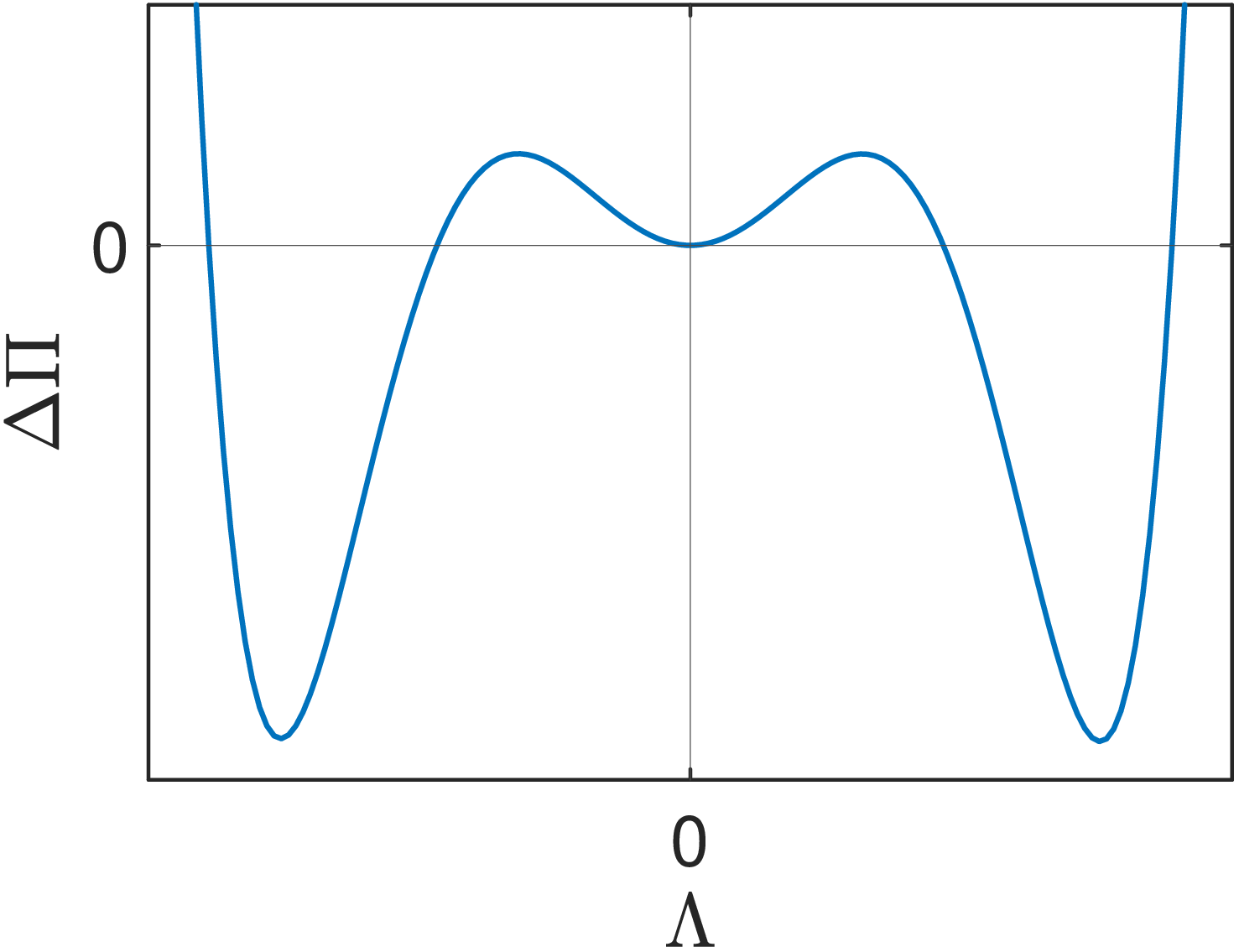}
\put(-2,74){\subcaptiontext*[3]{}} 
\end{overpic} 

\caption{ (a) Potential energy difference $\Delta\Pi$ vs $\Lambda$ showing axisymmetric equilibrium $(\text{S})$ and symmetry breaking configurations $(\text{SB}1)$ and $(\text{SB}2)$ near pressure $P_{cr1}$ for $\mathcal{M} = 1 \times 10^{-3}$. (b) $\Delta\Pi$ vs $\Lambda$ at $P_{cr2}$ for $\mathcal{M} = 1\times 10^{-3}$. (c) Qualitative potential energy difference $\Delta\Pi$ vs $\Lambda$ for reverse subcritical pitchfork bifurcation.} 
    \label{fig:Energy_plot}
\end{figure}

To visualize the potential energy for SPOM1 model and $\mathcal{M} = 1 \times 10^{-3}$ near the critical pressure points ($P_{cr1}$ and $P_{cr2}$), where the closed loop of the isolated pear-shaped solutions starts and merges, we plot the potential difference $ \Delta E = (\tilde{E} - E)$ against $\Lambda$ in Fig.~\ref{fig:Energy_plot}.

In Fig.~\ref{fig:Energy_plot}(a), the energy plot corresponds to configuration $\text{S}$, just above pressure value $P_{cr1}$. At this point, $\Lambda = 0$ corresponds to the axisymmetric solution, which is the only stable configuration. However, just below $P_{cr1}$, a supercritical pitchfork bifurcation occurs, giving rise to two stable pear-shaped configurations, corresponding to $\Lambda=0.4208$ and $\Lambda=-0.4208$  denoted as SB1 and SB2.  

Fig.~\ref{fig:Energy_plot}(b) illustrates the energy landscape just below $P_{cr2}$, where a reverse subcritical pitchfork bifurcation occurs. At this point, the axisymmetric configuration remains stable, while two unstable pear-shaped configurations, corresponding to $\Lambda=0.87723$ and $\Lambda=-0.87723$, bifurcate from it. However, Fig.~\ref{fig:Energy_plot}(b) provides only a partial representation of the reverse subcritical pitchfork bifurcation, as the energy expansion is considered only up to the $\Lambda^4$ term in (\ref{Energy_landscape}), revealing only two unstable solutions and one stable solution. A complete qualitative depiction of the energy landscape for the reverse subcritical pitchfork bifurcation is presented in Fig.~\ref{fig:Energy_plot}(c).

Fig.~\ref{fig:og_shape_stress}(a) shows the axisymmetric and pear-shaped equilibrium $P$ vs $\lambda_2$ curves at the poles ($\theta = 0 / \pi$) for the SPOM1 model with $\mathcal{M} = 1 \times 10^{-3}$. These curves illustrate the snap-through transition from the pear-shaped configuration $4$ to the axisymmetric configuration $5$, and from an axisymmetric configuration $6$ to a pear--shaped configuration $7$.

If we consider pressure as the control parameter for a balloon under the influence of a magnetic field with intensity $\mathcal{M} = 1 \times 10^{-3}$, the system initially reaches the pear-shaped configuration $4$. However, if the pressure is reduced even slightly further, the equilibrium configuration abruptly snaps through from the pear-shaped configuration to the axisymmetric configuration $5$.
This snap-through behavior occurs because the pear-shaped equilibrium curve (blue curve in Fig.~\ref{fig:og_shape_stress}(a)) becomes unstable for pressures below the equilibrium configuration point $4$. From axisymmetric configuration $5$ if we increase the pressure even very slightly the equilibrium configuration abruptly snaps through from the axisymmetric configuration to the pear--shaped configuration $6$. 

The stability of the plane of symmetry--preserving and pear-shaped equilibrium configurations was examined by introducing small asymmetric perturbations around the determined equilibrium solutions. However, no asymmetric bifurcation was observed and both configurations were found to remain stable under small asymmetric perturbations. 

\section{Conclusion} 
Axisymmetric inflation of an initially spherical magnetoelastic balloon is studied, focusing on stability under both non-magnetic and magnetic conditions. Stability was analyzed using axisymmetric and asymmetric perturbations represented by Legendre's polynomials of $\cos{\theta}$. In the presence of a magnetic field generated by a single coil, the axisymmetric curve bifurcates into a pear-shaped configuration through a supercritical pitchfork bifurcation, forming an isolated  bifurcation loop of pear--shaped solutions similar to the non-magnetic case. However, unlike the non-magnetic scenario where the isola curve rejoins via a reverse supercritical bifurcation, isolated  bifurcation loop of pear--shaped solutions rejoins through a reverse subcritical pitchfork bifurcation in the magnetic case. As the magnetic field intensity increases, the isolated  bifurcation loop of pear--shaped solutions shrinks and eventually disappears at a critical field value, mirroring the effect of increasing the material parameter $\alpha_1$ in the non-magnetic case. For a magnetic field generated by two coils, the isolated  bifurcation loop of pear--shaped solutions is absent for SPOM and FPOM models, though the pressure at which pear-shaped bifurcations occur decreases with increasing magnetic field intensity. These findings demonstrate that both the applied magnetic field and material parameters strongly influence the stability of axisymmetric configurations in magnetoelastic balloons. Additionally, snap-through phenomena between axisymmetric and pear-shaped configurations are observed under the influence of the magnetic field. 

In the present analysis, asymmetric perturbations have been considered for both symmetry-preserving and pear-shaped solutions. However, since the perturbations introduced were small, this linear perturbation approach can only capture asymmetric solutions that are very close to axisymmetric ones. As a result, no asymmetric solutions were identified for either the magnetic or non-magnetic cases. Extending the analysis to include more general perturbations would be valuable for determining such solutions.

Coupling the magnetoelastic inflation of a balloon with solvent diffusion gives rise to the problem of magnetic gel balloon inflation. Investigating the stability of such a multiphysics system involving magnetism, diffusion, and deformation interactions presents an intriguing challenge for future work. 

\justify{}
\textbf{Acknowledgement}\\
S.N.K. and G.T. acknowledge the Ministry of Education (MoE), Government of India, for providing research support. We also thank Arun Krishna B.J. for the helpful discussions.  


\appendix

\section{}
The total potential energy for inflation of a magneto-elastic balloon under the application of a magnetic field is given by \eqref{Total Energy nd}) as, 
\begin{equation} 
\label{Apendix: Total Energy nd}
    E_T = \int_0^{2\pi} \int_0^{\pi} \left[ \left( {\Phi} -\frac{1}{2} \mu_o \chi  \mathcal{M} \mathcal{H}^2 \right)\sin\theta + \frac{1}{2}{P}\lambda_2^2\sin^2\theta{{\eta}}_{,\theta}\right]   {\rm d}\theta \,{\rm d}\phi,
\end{equation} 

\begin{equation*}
\label{Apendix: Total Energy nd1}
    E_T = \int_0^{2\pi} \int_0^{\pi} \Pi \,  {\rm d}\theta \,{\rm d}\phi,
\end{equation*} 
From the principle of minimum potential energy, the equilibrium equations in terms of $\lambda_2$ and $\eta$ are $\delta_{\lambda_\theta}\hat{\Pi}=0$ and $\delta_{\eta}\hat{\Pi}=0$ are given as follows, 
\subsection{ Governing equations } 
\label{sec: l2 equation}
\begin{equation}
    \frac{\partial \Pi}{\partial \lambda_2} - \frac{\rm d}{{\rm d} \theta} \left(\frac{\partial \Pi}{\partial {\lambda_{2}}_{,\theta}}\right)=0,\quad \frac{\partial \Pi}{\partial \eta} - \frac{\rm d}{\rm d \theta} \frac{\partial \Pi}{\partial \eta_{,\theta}}=0. 
\label{e:euler}
\end{equation}
Using the substitution $v={\lambda_2}_{,\theta}\sin\theta$ and $\eta_{,\theta}=w$, \eqref{e:euler}$_1$ can be further expressed as   

\begin{equation*}
   \frac{\rm d}{{\rm d} \theta} \left[ \Phi_{,v} \sin^2\theta \right] =  \Phi_{,\lambda_2} \sin\theta - \mu_o \chi  \mathcal{M} \mathcal{H} \mathcal{H}_{,\lambda_2}\sin\theta  + P\lambda_2w\sin^2\theta 
\end{equation*}  
Expanding the above equation further with $(\cdot)_{,z}=\frac{\partial (\cdot)}{\partial z}$
\begin{align*}
\begin{split} 
    \sin^2\theta\left[ \Phi_{,vv}v_{,\theta} +\Phi_{,vw}w_{,\theta} +\Phi_{,v\lambda_2}\frac{v}{\sin\theta} + \Phi_{,v\cos\theta}\sin\theta \right] &= \Phi_{,\lambda_2} \sin\theta - \mu_o \chi  \mathcal{M} \mathcal{H} \mathcal{H}_{,\lambda_2} \sin\theta\\  
    &+ P\lambda_2w\sin^2\theta  - 2\sin\theta\cos\theta\,\Phi_{,v} 
\end{split} 
\end{align*}

\begin{align}
\begin{split} 
\label{lam2 eq1}
    \Phi_{,vv}v_{,\theta} +\Phi_{,vw}w_{,\theta} =\frac{\Phi_{,\lambda_2} - 2\cos\theta\,\Phi_{,v} }{\sin\theta}-\Phi_{,v\lambda_2}\frac{v}{\sin\theta} - \mu_o \chi  \mathcal{M} \mathcal{H} \mathcal{H}_{,\lambda_2} \frac{1}{\sin\theta} + P\lambda_2w   - \Phi_{v\cos\theta}\sin\theta
\end{split} 
\end{align}
The first and second terms on the right-hand side of the equation \eqref{lam2 eq1} are of form $\frac{0}{0}$ at poles of the balloon, i.e. at $\theta=0$ and $\theta=\pi$ due to division by $\sin\theta$. These terms at poles are calculated using L'Hôpital's rule. In the third term $\sin\theta$ in denominator gets cancelled by $\sin\theta$ term in the numerator of the term $\frac{\partial\mathcal{H}}{\partial \lambda_2}$. The presence of a mathematical form $\frac{0}{0}$ which is resolved using L'Hôpital's rule is called the removable singularity.    

Similarly, \eqref{e:euler}$_2$ can be expressed as 

\begin{equation*}
    \frac{\rm d}{\rm d \theta} \left[\Phi_{,w}\sin\theta + \frac{1}{2}P\lambda_2^2\sin^2\theta \right] = -\mu_0 \chi  \mathcal{M} \mathcal{H} {\mathcal{H}}_{,\eta} \sin\theta 
\end{equation*} 


\begin{equation*}
    \frac{{\rm d} \Phi_{,w}}{\rm d \theta} = -\mu_0 \chi  \mathcal{M} \mathcal{H} \mathcal{H}_{,\eta}   - \cot\theta\,\Phi_{,w} - P\lambda_2v  -P\cos\theta\lambda_2^2  
\end{equation*}  
\begin{align}
\begin{split} 
\label{eta eq2}
    \Phi_{,wv}v_{,\theta} +\Phi_{,ww}w_{,\theta} =-\mu_o \chi  \mathcal{M} \mathcal{H} \mathcal{H}_{,\eta} -\Phi_{,w\lambda_2}\frac{v}{\sin\theta}  - \cot\theta\,\Phi_{,w}- P\lambda_2v  -P\lambda_2^2\cos\theta  -\Phi_{,w\cos\theta}\sin\theta
\end{split} 
\end{align}
Similar to the removable singularity case discussed in the section (\ref{sec: l2 equation}) the second and third terms on the right-hand side of the equation (\ref{eta eq2}) are of form $\frac{0}{0}$ at poles of the balloon, i.e. at $\theta=0$ and $\theta=\pi$ due to division by $\sin\theta$. These terms at poles are calculated using L'Hôpital's rule.

\subsection{ Applied magnetic field}  
It is convenient to express the applied magnetic field $\mathbf{h}_a$ in both coordinate systems i.e. Cartesian as well as curvilinear coordinate systems associated with deformed configuration. From equations \eqref{Position vec}, \eqref{tangentC vec}  tangent vectors i.e. bases for curvilinear coordinates ($\theta$, $\phi$, $\zeta$), can be written in terms of bases vectors of Cartesian coordinate system as, 
\begin{equation}
   \mathbf{\hat{g}}_i = Q_{ij} \mathbf{\hat{e}}_j 
    \label{curvilinear to cartesian} 
\end{equation} 
\begin{align}
    \begin{split}
        &\text{where} \qquad Q_{11} = \frac{\rho_{,\theta}\cos \phi }{\sqrt{ {\rho_{,\theta}}^2 +{\eta_{,\theta}^2} }  }, \quad Q_{12} = -\sin \phi,  \\ 
        &Q_{21} = -\sin \phi, \quad Q_{22} = \cos\phi, \quad Q_{23} = 0, \quad Q_{31} = -  \frac{\eta_{,\theta}\cos \phi }{\sqrt{ {\rho_{,\theta}}^2 +{\eta_{,\theta}}^2 }  }, \\ 
        &Q_{32} = -  \frac{\eta_{,\theta}\sin \phi }{\sqrt{ {\rho_{,\theta}}^2 +{\eta_{,\theta}}^2 }  }, \quad Q_{33} = -  \frac{ \rho_{,\theta}}{\sqrt{ {\rho_{,\theta}}^2 +{\eta_{,\theta}}^2 }  },
    \end{split}
\end{align}
here $\mathbf{\hat{g}}_3 = \mathbf{n}$ the outward surface normal as mentioned in  \eqref{current normal vec}. Inverting equation \eqref{curvilinear to cartesian}, Cartesian bases can be written in terms of curvilinear bases as follows, 

\begin{equation}
    \mathbf{\hat{e}}_i = \bar{Q}_{ij} \mathbf{\hat{g}}_j,   
    \label{cartesian to curvilinear}
\end{equation}
where $\bar{Q}_{ij} = Q^{-1}_{ij} =Q_{ji}$. Using equations \eqref{cartesian to curvilinear}, \eqref{curvilinear to cartesian} and \eqref{ha calc} the applied magnetic field $\mathbf{h}_a$ can be written in the curvilinear bases associated with deformed configuration as follows, 

\begin{align}
    \begin{split}
        \mathbf{h}_a &= \frac{I}{4\pi R_0} \int_0^{2\pi} \mathbf{\hat{g}}_1  \frac{\eta_{,\theta}a^2 + \left( \rho_{,\theta} \left(\eta -h \right)a - \rho \eta_{,\theta} a  \right) \cos \left( \phi_i - \phi \right) }{ \left( \sqrt{\rho_{,\theta}^2 + \eta_{,\theta}^2}  \right)  \left( {\rho}^2 + {\left(\eta-h\right)}^2 + {a}^2 - 2{\rho}{a}\cos\left( \phi_i - \phi \right) \right)^{3/2} } \,{\rm d} \phi_i  \\ 
        &+\frac{I}{4\pi R_0} \int_0^{2\pi} \mathbf{\hat{g}}_3  \frac{\rho_{,\theta}a^2 - \left( \eta_{,\theta}\left(\eta -h\right)a + \rho_{,\theta}\rho a  \right) \cos \left( \phi_i - \phi \right) }{ \left( \sqrt{\rho_{,\theta}^2 + \eta_{,\theta}^2}  \right)  \left( {\rho}^2 + {\left(\eta - h\right)}^2 + {a}^2 - 2{\rho}{a}\cos\left( \phi_i - \phi \right) \right)^{3/2} }\, {\rm d} \phi_i
    \end{split}
    \label{ha curvi} 
\end{align}
Expressions for the applied magnetic field $\mathbf{h}_a$ given in equations \eqref{ha calc} and \eqref{ha curvi} can be written shortly as, 
\begin{align}
    \begin{split}
       \mathbf{h}_a &= \frac{I}{4\pi R_0} \left( \mathbf{\hat{e}}_1 (h_a)_1 + \mathbf{\hat{e}}_2 (h_a)_2 + \mathbf{\hat{e}}_3 (h_a)_3  \right) \\ 
       &=\frac{I}{4\pi R_0} \left( \mathbf{\hat{g}}_1 (h_a)_\theta + \mathbf{\hat{g}}_3 (h_a)_\zeta   \right)
    \end{split}
\end{align}

  \bibliographystyle{abbrvnatX2}    
  \bibliography{cas-refsX-2}

\begin{thebibliography}{35}
\providecommand{\natexlab}[1]{#1}
\providecommand{\url}[1]{\texttt{#1}}
\expandafter\ifx\csname urlstyle\endcsname\relax
  \providecommand{\doi}[1]{doi: #1}\else
  \providecommand{\doi}{doi: \begingroup \urlstyle{rm}\Url}\fi

\bibitem[Barham et~al.(2007)Barham, Steigmann, McElfresh, and Rudd]{Barham_ActaM2007}
M.~Barham, D.~J. Steigmann, M.~McElfresh, and R.~E. Rudd.
\newblock 2007.
\newblock Finite deformation of a pressurized magnetoelastic membrane in a stationary dipole field.
\newblock \emph{Acta Mech.}, 191\penalty0 (1):\penalty0 1--19.

\bibitem[Barham et~al.(2008)Barham, Steigmann, McElfresh, and Rudd]{Barham_2008}
M.~Barham, D.~J. Steigmann, M.~McElfresh, and R.~E. Rudd.
\newblock 2008.
\newblock Limit-point instability of a magnetoelastic membrane in a stationary magnetic field.
\newblock \emph{Smart Mater. Struct.}, 17\penalty0 (5):\penalty0 055003.

\bibitem[Bartlett et~al.(2015)Bartlett, Tolley, Overvelde, Weaver, Mosadegh, Bertoldi, Whitesides, and Wood]{Combution_robot}
N.~W. Bartlett, M.~T. Tolley, J.~T.~B. Overvelde, J.~C. Weaver, B.~Mosadegh, K.~Bertoldi, G.~M. Whitesides, and R.~J. Wood.
\newblock 2015.
\newblock A 3d-printed, functionally graded soft robot powered by combustion.
\newblock \emph{Science}, 349\penalty0 (6244):\penalty0 161--165.

\bibitem[Bauer et~al.(1970)Bauer, Reiss, and Keller]{hemisphere_buckling2}
L.~Bauer, E.~L. Reiss, and H.~B. Keller.
\newblock 1970.
\newblock Axisymmetric buckling of hollow spheres and hemispheres.
\newblock \emph{Comm. Pure Appl. Math.}, 23\penalty0 (4):\penalty0 529--568.

\bibitem[Chen and Healey(1991)]{CHEN_IJNM1991}
Y.-C. Chen and T.~J. Healey.
\newblock 1991.
\newblock Bifurcation to pear-shaped equilibria of pressurized spherical membranes.
\newblock \emph{Int. J. Non-Linear Mech.}, 26\penalty0 (3):\penalty0 279--291.

\bibitem[Cheng et~al.(2019)Cheng, Jia, Guo, Nie, and Li]{TengLi_JMPS2019}
J.~Cheng, Z.~Jia, H.~Guo, Z.~Nie, and T.~Li.
\newblock 2019.
\newblock Delayed burst of a gel balloon.
\newblock \emph{J. Mech. Phys. Solids}, 124:\penalty0 143--158.

\bibitem[Dellwo et~al.(1982)Dellwo, Keller, Matkowsky, and Reiss]{birth_of_isola}
D.~Dellwo, H.~B. Keller, B.~J. Matkowsky, and E.~L. Reiss.
\newblock 1982.
\newblock On the birth of isolas.
\newblock \emph{SIAM J. Appl. Math.}, 42\penalty0 (5):\penalty0 956--963.

\bibitem[Fu and Xie(2014)]{FU201433}
Y.~Fu and Y.~Xie.
\newblock 2014.
\newblock Stability of pear-shaped configurations bifurcated from a pressurized spherical balloon.
\newblock \emph{J. Mech. Phys. Solids}, 68:\penalty0 33--44.

\bibitem[Ghosh and Basu(2021)]{GHOSH_JMPS2021}
A.~Ghosh and S.~Basu.
\newblock 2021.
\newblock Soft dielectric elastomer tubes in an electric field.
\newblock \emph{J. Mech. Phys. Solids}, 150:\penalty0 104371.

\bibitem[Heidarian and Kouzani(2023)]{UTS_magnetic_polymer}
P.~Heidarian and A.~Z. Kouzani.
\newblock 2023.
\newblock Starch-g-acrylic acid/magnetic nanochitin self-healing ferrogels as flexible soft strain sensors.
\newblock \emph{Sens.}, 23\penalty0 (3):\penalty0 1138.

\bibitem[Humphrey(2003)]{Membranes_in_Bio}
J.~Humphrey.
\newblock 2003.
\newblock Review paper: Continuum biomechanics of soft biological tissues.
\newblock \emph{Proc. R. Soc. A}, 459\penalty0 (2029):\penalty0 3--46.

\bibitem[Jenkins(2001)]{space_book}
C.~Jenkins.
\newblock 01 2001.
\newblock Gossamer spacecraft: Membrane and inflatable structures technology for space applications.
\newblock \emph{AIAA}.

\bibitem[Keller and Wolfe(1965)]{shperical_shell_buckling1}
H.~B. Keller and A.~W. Wolfe.
\newblock 1965.
\newblock On the nonunique equilibrium states and buckling mechanism of spherical shells.
\newblock \emph{J. Soc. Ind. Appl. Math.}, 13\penalty0 (3):\penalty0 674--705.

\bibitem[Kerr and El-Bayoumy(1968)]{arch_buckling3}
A.~D. Kerr and L.~El-Bayoumy.
\newblock \emph{Equilibrium and stability of a shallow arch subjected to a uniform lateral load}.
\newblock Q. Appl. Math., 1968.

\bibitem[Kim et~al.(2011)Kim, Lu, Ghaffari, Kim, Lee, Xu, Wu, Kim, Song, Liu, Viventi, de~Graff, Elolampi, Mansour, Slepian, Hwang, Moss, Won, Huang, Litt, and Rogers]{surgical_tools}
D.-H. Kim, N.~Lu, R.~Ghaffari, Y.-S. Kim, S.~P. Lee, L.~Xu, J.~Wu, R.-H. Kim, J.~Song, Z.~Liu, J.~Viventi, B.~de~Graff, B.~Elolampi, M.~Mansour, M.~J. Slepian, S.~Hwang, J.~D. Moss, S.-M. Won, Y.~Huang, B.~Litt, and J.~A. Rogers.
\newblock 2011.
\newblock Materials for multifunctional balloon catheters with capabilities in cardiac electrophysiological mapping and ablation therapy.
\newblock \emph{Nat. Mater.}, 10\penalty0 (4):\penalty0 316--323.

\bibitem[Liang and Cai(2015)]{Cai_JAM2015}
X.~Liang and S.~Cai.
\newblock 2015.
\newblock {Shape Bifurcation of a Spherical Dielectric Elastomer Balloon Under the Actions of Internal Pressure and Electric Voltage}.
\newblock \emph{J. Appl. Mech.}, 82\penalty0 (10):\penalty0 101002.

\bibitem[Liao et~al.(2023)Liao, Zoumhani, and Boutry]{UTS2_magnetic_polymer}
Z.~Liao, O.~Zoumhani, and C.~M. Boutry.
\newblock 2023.
\newblock Recent advances in magnetic polymer composites for biomems: A review.
\newblock \emph{Mater.}, 16\penalty0 (10):\penalty0 3802.

\bibitem[Nelder and Mead(1965)]{nelder_simplex}
J.~A. Nelder and R.~Mead.
\newblock 1965.
\newblock A simplex method for function minimization.
\newblock \emph{Comput. J.}, 7\penalty0 (4):\penalty0 308--313.

\bibitem[Patil et~al.(2015)Patil, Nordmark, and Eriksson]{AMITPATIL_Royal_soc}
A.~Patil, A.~Nordmark, and A.~Eriksson.
\newblock 2015.
\newblock Instability investigation on fluid-loaded pre-stretched cylindrical membranes.
\newblock \emph{Proc. R. Soc. A}, 471\penalty0 (2177):\penalty0 20150016.

\bibitem[Patil et~al.(2016)Patil, Nordmark, and Eriksson]{AMITPATIL_JMPS}
A.~Patil, A.~Nordmark, and A.~Eriksson.
\newblock 2016.
\newblock Instabilities of wrinkled membranes with pressure loadings.
\newblock \emph{J. Mech. Phys. Solids}, 94:\penalty0 298--315.

\bibitem[Raz et~al.(2019)Raz, Brosh, Ronen, and Tal]{raz2019tensile}
P.~Raz, T.~Brosh, G.~Ronen, and H.~Tal.
\newblock 2019.
\newblock Tensile properties of three selected collagen membranes.
\newblock \emph{Biomed Res. Int.}, 2019\penalty0 (1):\penalty0 5163603.

\bibitem[Reddy and Saxena(2017)]{REDDY_IJNM2017}
N.~H. Reddy and P.~Saxena.
\newblock 2017.
\newblock Limit points in the free inflation of a magnetoelastic toroidal membrane.
\newblock \emph{Int. J. Non-Linear Mech.}, 95:\penalty0 248--263.

\bibitem[Reddy and Saxena(2018)]{REDDY_IJSS2018}
N.~H. Reddy and P.~Saxena.
\newblock 2018.
\newblock Instabilities in the axisymmetric magnetoelastic deformation of a cylindrical membrane.
\newblock \emph{Int. J. Solids Struct.}, 136-137:\penalty0 203--219.

\bibitem[Roychowdhury and DasGupta(2015{\natexlab{a}})]{roychowdhury2015inflating}
S.~Roychowdhury and A.~DasGupta.
\newblock 2015{\natexlab{a}}.
\newblock Inflating a flat toroidal membrane.
\newblock \emph{Int. J. Solids Struct.}, 67:\penalty0 182--191.

\bibitem[Roychowdhury and DasGupta(2015{\natexlab{b}})]{roychowdhury2015response}
S.~Roychowdhury and A.~DasGupta.
\newblock 2015{\natexlab{b}}.
\newblock On the response and stability of an inflated toroidal membrane under radial loading.
\newblock \emph{Int. J. Non-Linear Mech.}, 77:\penalty0 254--264.

\bibitem[Roychowdhury and DasGupta(2018)]{ROYCHOWDHURY2018328}
S.~Roychowdhury and A.~DasGupta.
\newblock 2018.
\newblock Symmetry breaking during inflation of a toroidal membrane.
\newblock \emph{J. Mech. Phys. Solids}, 121:\penalty0 328--340.

\bibitem[Saxena et~al.(2019)Saxena, Reddy, and Pradhan]{SAXENA_IJNM2019}
P.~Saxena, N.~H. Reddy, and S.~P. Pradhan.
\newblock 2019.
\newblock Magnetoelastic deformation of a circular membrane: Wrinkling and limit point instabilities.
\newblock \emph{Int. J. Non-Linear Mech.}, 116:\penalty0 250--261.

\bibitem[Tamadapu(2022)]{TAMADAPU_IJNM2022}
G.~Tamadapu.
\newblock 2022.
\newblock Swelling and inflation of a toroidal gel balloon.
\newblock \emph{Int. J. Non-Linear Mech.}, 138:\penalty0 103838.

\bibitem[Tamadapu and DasGupta(2014{\natexlab{a}})]{TAMADAPU_Euro_A}
G.~Tamadapu and A.~DasGupta.
\newblock 2014{\natexlab{a}}.
\newblock Effect of curvature and anisotropy on the finite inflation of a hyperelastic toroidal membrane.
\newblock \emph{Euro. J. Mech. A/Solids.}, 46:\penalty0 106--114.

\bibitem[Tamadapu and DasGupta(2014{\natexlab{b}})]{TAMADAPU_IJSS_rim}
G.~Tamadapu and A.~DasGupta.
\newblock 2014{\natexlab{b}}.
\newblock Finite inflation of a hyperelastic toroidal membrane over a cylindrical rim.
\newblock \emph{Int. J. Solids Struct.}, 51\penalty0 (2):\penalty0 430--439.

\bibitem[Tamadapu et~al.(2013)Tamadapu, Dhavale, and DasGupta]{GT_PhysRevE}
G.~Tamadapu, N.~N. Dhavale, and A.~DasGupta.
\newblock 2013.
\newblock Geometrical feature of the scaling behavior of the limit-point pressure of inflated hyperelastic membranes.
\newblock \emph{Phys. Rev. E}, 88:\penalty0 053201.

\bibitem[Truesdell et~al.(2004)Truesdell, Noll, and Antman]{Truesdell}
C.~Truesdell, W.~Noll, and S.~Antman.
\newblock 2004.
\newblock The non-linear field theory of mechanics.
\newblock \emph{Handb. Phys.}, 3.

\bibitem[Wang et~al.(2018)Wang, Xu, Huo, and Potier-Ferry]{Wang_IJNM2018}
T.~Wang, F.~Xu, Y.~Huo, and M.~Potier-Ferry.
\newblock 2018.
\newblock Snap-through instabilities of pressurized balloons: Pear-shaped bifurcation and localized bulging.
\newblock \emph{Int. J. Non-Linear Mech.}, 98:\penalty0 137--144.

\bibitem[Yao et~al.(2011)Yao, McDowell, Ryu, Wu, Liu, Hu, Nix, and Cui]{Yao2011}
Y.~Yao, M.~T. McDowell, I.~Ryu, H.~Wu, N.~Liu, L.~Hu, W.~D. Nix, and Y.~Cui.
\newblock 2011.
\newblock Interconnected silicon hollow nanospheres for lithium-ion battery anodes with long cycle life.
\newblock \emph{Nano Lett.}, 11\penalty0 (7):\penalty0 2949--2954.

\bibitem[Yi et~al.(2023)Yi, Zou, Huang, Ren, Tian, Yu, Wang, Yuan, Tang, Wang, Liang, Cao, Li, Yu, Jiang, Zhang, Yang, Li, Wang, Luo, Loh, Li, Hu, Liu, Gao, and Chen]{Bio-elect-application}
J.~Yi, G.~Zou, J.~Huang, X.~Ren, Q.~Tian, Q.~Yu, P.~Wang, Y.~Yuan, W.~Tang, C.~Wang, L.~Liang, Z.~Cao, Y.~Li, M.~Yu, Y.~Jiang, F.~Zhang, X.~Yang, W.~Li, X.~Wang, Y.~Luo, X.~J. Loh, G.~Li, B.~Hu, Z.~Liu, H.~Gao, and X.~Chen.
\newblock 2023.
\newblock Water-responsive supercontractile polymer films for bioelectronic interfaces.
\newblock \emph{Nature}, 624\penalty0 (7991):\penalty0 295--302.

\end{thebibliography}






\end{document}